\documentclass[twocolumn]{aastex62}

\def\nsings{45}
\def\ngalsings{36}
\def\ngal{38}
\def\nssfr{21}
\def\nx{4442}
\def\nxrb{2478}
\def\nmod{537}
\def\nlmxb{1230}
\def\nhmxb{710}

\def\chandra{{\itshape Chandra\/}}

\def\spitzer{{\itshape Spitzer\/}}
\def\herschel{{\itshape Herschel\/}}
\def\galex{{\itshape GALEX\/}}

\def\xmm{{\itshape XMM-Newton\/}}

\def\xray{\hbox{X-ray}}

\def\etal{{et\,al.}}

\def\ltsima{$\; \buildrel < \over \sim \;$}
\def\simlt{\lower.5ex\hbox{\ltsima}}
\def\gtsima{$\; \buildrel > \over \sim \;$}
\def\simgt{\lower.5ex\hbox{\gtsima}}
\def\kms{\ifmmode{~{\rm km~s^{-1}}}\else{~km s$^{-1}$}\fi}
\def\lsim{\lower0.3em\hbox{$\,\buildrel <\over\sim\,$}}
\def\gsim{\lower0.3em\hbox{$\,\buildrel >\over\sim\,$}}

\def\msol{$M_\odot$}
\def\h2{H$_2$}
\def\flux{erg~cm$^{-2}$~s$^{-1}$}
\def\lum{erg~s$^{-1}$}

\def\sfr{$M_{\odot}$~yr$^{-1}$}

\def\aap{A\&A}
\def\apj{ApJ}
\def\apjl{ApJL}
\def\apjs{ApJS}
\def\aj{AJ}
\def\mnras{MNRAS}
\def\araa{ARA\&A}

\def\nat{Nature}

\usepackage{underscore}
\usepackage{longtable}
\usepackage{amsmath}
\shorttitle{XRB Local Scaling Relations}
\shortauthors{Lehmer et al.}
\begin{document}

%
%%%%%%%%%%%%%%%%%%%%%%%%%%%%%%%%%%%%%%%%%%%%%%%%%%%%%%%%%%%%%%%%%%%%%%%%
\title{X-ray Binary Luminosity Function Scaling Relations for Local Galaxies Based on Subgalactic Modeling}

\correspondingauthor{Bret Lehmer}
\email{lehmer@uark.edu}

\author{Bret~D.~Lehmer}
\affil{Department of Physics, University of Arkansas, 226 Physics Building, 825 West Dickson Street, Fayetteville, AR 72701, USA}

\author{Rafael~T.~Eufrasio}
\affil{Department of Physics, University of Arkansas, 226 Physics Building, 825 West Dickson Street, Fayetteville, AR 72701, USA}

\author{Panayiotis Tzanavaris}
\affiliation{NASA Goddard Space Flight Center, Code 662, Greenbelt, MD 20771, USA}
\affiliation{Center for Space Science and Technology, University of
Maryland Baltimore County, 1000 Hilltop Circle, Baltimore, MD 21250, USA}

\author{Antara Basu-Zych}
\affiliation{NASA Goddard Space Flight Center, Code 662, Greenbelt, MD 20771, USA}
\affiliation{Center for Space Science and Technology, University of
Maryland Baltimore County, 1000 Hilltop Circle, Baltimore, MD 21250, USA}

\author{Tassos Fragos}
\affiliation{Geneva Observatory, Geneva University, Chemin des Maillettes
51, 1290 Sauverny, Switzerland}

\author{Andrea Prestwich}
\affiliation{Harvard-Smithsonian Center for Astrophysics, 60 Garden Street, Cambridge, MA 02138, USA}

\author{Mihoko Yukita}
\affiliation{The Johns Hopkins University, Homewood Campus, Baltimore, MD
21218, USA}

\author{Andreas Zezas}
\affiliation{Harvard-Smithsonian Center for Astrophysics, 60 Garden Street, Cambridge, MA 02138, USA}
\affiliation{Foundation for Research and Technology-Hellas, 100 Nikolaou Plastira Street, 71110 Heraklion, Crete, Greece}
\affiliation{Physics Department \& Institute of Theoretical \& Computational Physics, P.O. Box 2208, 71003 Heraklion, Crete, Greece}

\author{Ann~E. Hornschemeier}
\affiliation{NASA Goddard Space Flight Center, Code 662, Greenbelt, MD 20771, USA}
\affiliation{The Johns Hopkins University, Homewood Campus, Baltimore, MD
21218, USA}

\author{Andrew Ptak}
\affiliation{NASA Goddard Space Flight Center, Code 662, Greenbelt, MD 20771, USA}
\affiliation{The Johns Hopkins University, Homewood Campus, Baltimore, MD
21218, USA}

%
%%%%%%%%%%%%%%%%%%%%%%%%%%%%%%%%%%%%%%%%%%%%%%%%%%%%%%%%%%%%%%%%%%%%%%%%
\begin{abstract}
%%%%%%%%%%%%%%%%%%%%%%%%%%%%%%%%%%%%%%%%%%%%%%%%%%%%%%%%%%%%%%%%%%%%%%%%
%

We present new \chandra\ constraints on the \xray\ luminosity functions (XLFs)
of \xray\ binary (XRB) populations, and their scaling relations, for a sample
of \ngal\ nearby galaxies ($D =$~3.4--29~Mpc).  Our galaxy sample is drawn
primarily from the \spitzer\ infrared nearby galaxy survey (SINGS), and
contains a wealth of \chandra\ (5.8~Ms total) and multiwavelength data,
allowing for star-formation rates (SFRs) and stellar masses ($M_\star$) to be
measured on subgalactic scales.
We divided the \nxrb\ \xray\ detected sources into \nssfr\ subsamples in bins
of specific-SFR (sSFR~$\equiv$~SFR/$M_\star$) and constructed XLFs.  To model
the XLF dependence on sSFR, we fit a global XLF model, containing contributions
from high-mass XRBs (HMXBs), low-mass XRBs (LMXBs), and background sources from
the cosmic \xray\ background (CXB) that respectively scale with SFR, $M_\star$,
and sky area.  We find an HMXB XLF that is more complex in shape than
previously reported and an LMXB XLF that likely varies with sSFR, potentially
due to an age dependence.
When applying our global model to XLF data for each individual galaxy, we
discover a few galaxy XLFs that significantly deviate from our model beyond
statistical scatter.  Most notably, relatively low-metallicity galaxies have an
excess of HMXBs above $\approx$$10^{38}$~\lum\ and elliptical galaxies that
have relatively rich populations of globular clusters (GCs) show excesses of
LMXBs compared to the global model.  Additional modeling of how the XRB XLF
depends on stellar age, metallicity, and GC specific frequency is required to
sufficiently characterize the XLFs of galaxies.

%
%%%%%%%%%%%%%%%%%%%%%%%%%%%%%%%%%%%%%%%%%%%%%%%%%%%%%%%%%%%%%%%%%%%%%%%%
\end{abstract}
%%%%%%%%%%%%%%%%%%%%%%%%%%%%%%%%%%%%%%%%%%%%%%%%%%%%%%%%%%%%%%%%%%%%%%%%
%


%
%%%%%%%%%%%%%%%%%%%%%%%%%%%%%%%%%%%%%%%%%%%%%%%%%%%%%%%%%%%%%%%%%%%%%%%%
\section{Introduction} \label{sec:intro}
%%%%%%%%%%%%%%%%%%%%%%%%%%%%%%%%%%%%%%%%%%%%%%%%%%%%%%%%%%%%%%%%%%%%%%%%
%

X-ray binaries (XRBs) provide a direct probe of compact object (i.e., black
hole [BH] and neutron star [NS]) populations and close binary systems in
galaxies.  The XRB phase of close-binary evolution results when mass is
transferred from a normal star (secondary) to an accreting compact-object
remnant (primary), via Roche-lobe overflow or stellar-wind mass transfer.
Depending on the binary parameters, subsequent evolution beyond the XRB phase
is expected to result in a variety of astrophysical systems, including, e.g.,
gravitational wave (GW) mergers, millisecond pulsars, and short gamma-ray
bursts (GRBs).  Recent discoveries of gravitational waves (GWs) from merging
BHs and NSs from LIGO (e.g., Abbott \etal\ 2016, 2017) have prompted a
resurgence in efforts to self-consistently model close binary populations and
their evolution (e.g., Belczynski \etal\ 2016, 2018; Mandel \& de~Mink~2016;
Marchant \etal\ 2017; Kruckow \etal\ 2018; Mapelli \& Giacobbo~2018).  As such,
statistically meaningful constraints on XRB populations are critical to such
efforts.   

Thanks largely to data collected over the last two decades by \chandra\ and
\xmm, substantial insight has been gained into how the XRB phase is manifested
within a variety of galactic environments beyond the Milky Way and Magellanic
Clouds.  Several studies of XRB emission from galaxies in the nearby Universe
($D\simlt50$~Mpc) have established that the \xray\ luminosity functions (XLFs),
and population-integrated luminosities of high-mass XRBs (HMXBs) and low-mass
XRBs (LMXBs) scale with star-formation rate (SFR) and stellar mass ($M_\star$),
respectively (e.g., Grimm \etal\ 2003; Ranalli \etal\ 2003; Colbert \etal\
2004; Gilfanov~2004; Lehmer \etal\ 2010; Boroson \etal\ 2011; Mineo \etal\
2012a, 2012b, Zhang \etal\ 2012).  These scaling relations have been assumed to
be ``universal'' in applications outside of studies focused on XRBs.  For
example, studies of distant active galactic nuclei (AGN) routinely utilize
local scaling relations when assessing the levels of XRB emission in distant
populations.  (see, e.g., $\S$2.2 of Hickox \& Alexander~2018).  

However, more recently, it has
been suggested that the scatter in basic XRB scaling relations is larger than
expected if the correlations were universal.  XRB population synthesis models
have indicated that universal scaling relations are unrealistic on physical
grounds (e.g., Fragos \etal\ 2008, 2013a, 2013b; Zuo \etal\ 2014).  For example, the population
synthesis models from Fragos \etal\ (2013b) predict order-of-magnitude variations of $L_{\rm
X}$(HMXB)/SFR and $L_{\rm X}$(LMXB)/$M_\star$ with metallicity and
stellar age, respectively, over ranges of these quantities present in the observable Universe.

Since the ranges of metallicities and mean stellar ages for typical galaxies in
the local Universe are relatively narrow, empirically measuring the predicted
deviations of the scaling relations with these parameters has been challenging.
Nonetheless, targeted observations of relatively rare, low metallicity
late-type galaxies (e.g., Basu-Zych \etal\ 2013a, 2016; Douna \etal\ 2015; Brorby \etal\ 2016; Tzanavaris \etal\ 2016)
and early-type galaxies with a range of stellar ages (e.g., Kim \& Fabbiano
2010; Lehmer \etal\ 2014), have provided tantalizing evidence of variations in
the scaling relations in line with those predicted by population synthesis
models.  New studies of XRB formation rates within very nearby galaxies (e.g., Magellanic Clouds, M33, M51, NGC~3310, and NGC~2276) have revealed similar variations with physical properties on subgalactic scales (e.g., Antoniou \& Zezas~2016; Lehmer \etal\ 2017; Garofali \etal\ 2018; Anastasopoulou \etal\ 2018; Antoniou \etal\ 2019).
Furthermore, \xray\ stacking analyses of distant galaxy populations in deep
\chandra\ surveys (e.g., the \chandra\ Deep Fields and \chandra\ COSMOS
surveys) have claimed that there is redshift evolution in the scaling
relations, potentially due to the corresponding decline in mean stellar
population age and metallicity with lookback time (e.g., Lehmer \etal\ 2007,
2016; Basu-Zych \etal\ 2013b; Kaaret~2014; Aird \etal\ 2017).  

The measured evolution of $L_{\rm X}$(HMXB)/SFR~$\propto (1+z)$ and $L_{\rm
X}$(LMXB)/$M_\star$~$\propto (1+z)^{2-3}$ out to $z \approx$~2--4 (Lehmer
\etal\ 2016; Aird \etal\ 2017) is only loosely constrained, but consistent with
the population synthesis predictions from Fragos \etal\ (2013a); however, see
Fornasini \etal\ (2018) for caveats.  Extrapolation of the theoretical
predictions into the very early Universe at $z \simgt 10$, when the Universe
was of very low metallicity ($\simlt$1/10~$Z_\odot$; e.g., based on the
Millenium~II simulations; Guo \etal\ 2011), indicate that XRBs were likely the
most luminous \xray\ emitting population in the Universe (e.g., Fragos \etal\ 2013b; Lehmer \etal\
2016; Madau \& Fragos 2017).  In fact, emission from XRBs is
thought to play a dominant role in heating the IGM at $z \approx$~10--20 (e.g., Mirabel \etal\ 2011; Mesinger \etal\
2013; Pacucci \etal\ 2014; Das \etal\ 2017; Grieg \& Mesinger~2018).

%
%%%%%%%%%%%%%%%%%%%%%%%%%%%%%%%%%%%%%%%%%%%%%%%%%%%%%%%%%%%%%%%%%%%%%%%%%%%%%%%%%%
% Figure 1
%%%%%%%%%%%%%%%%%%%%%%%%%%%%%%%%%%%%%%%%%%%%%%%%%%%%%%%%%%%%%%%%%%%%%%%%%%%%%%%%%%
%
\begin{figure*}[t!]
\figurenum{1}
\centerline{
\includegraphics[width=18cm]{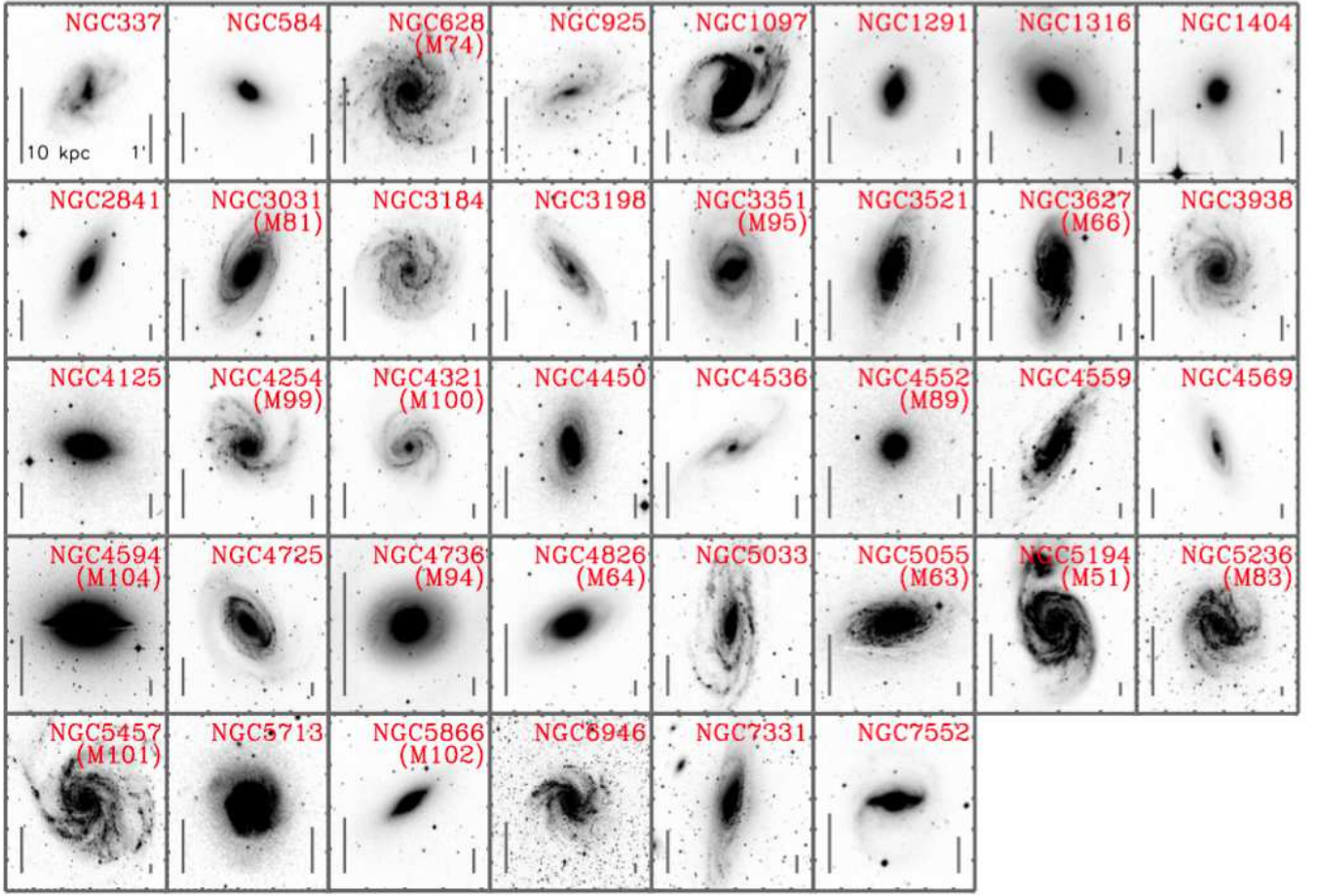}
}
\caption{
%%%
Digitized sky survey (DSS) images of the \ngal\ galaxies in our sample.  All images
have square dimensions with the length of each side being equal to the 1.1 times the total
$K_s$-band major axis (as reported by Jarrett \etal\ 2003).  For reference,
vertical bars of size 10~kpc and 1~arcmin are provided in the lower left and
right of each panel, respectively.
%%%
}
\end{figure*}
%%%%%%%%%%%%%%%%%%%%%%%%%%%%%%%%%%%%%%%%%%%%%%%%%%%%%%%%%%%%%%%%%%%%%%%%%%%%%%%%%%

The studies outlined above indicate that XRBs play an important role in a
variety of astrophysical systems and that the XRB scaling relations have
non-negligible dependencies on galaxy physical properties.  Although we now have
some indications of how the XRB emission and scaling relations vary with
important physical properties, there is still large uncertainty in how the {\em
distributions} of XRB populations (i.e., XLFs) vary with these physical
properties.  In particular, we do not know precisely how the XRB XLFs vary with
age and metallicity. There are some indications that the HMXB XLF in
low-metallicity galaxies contains an excess of ultraluminous \xray\ sources
(ULXs) above $10^{39}$~\lum\ (Mapelli \etal\ 2010; Kaaret \etal\ 201; Prestwich
\etal\ 2013; Basu-Zych \etal\ 2016) and the bright-end of the LMXB XLF for
young elliptical galaxies contains more LMXBs with $\simgt$$10^{39}$~\lum\ than
older ellipticals (e.g., Kim \& Fabbiano~2010; Lehmer \etal\ 2014, 2017).  But
for both HMXBs and LMXBs, it is not clear whether there is an excess of XRBs
over the full range of luminosities that are important to the galaxy-wide
global \xray\ power output, and to what extent these populations are elevated (due to small
number statistics).  These details are powerful constraints for population
synthesis models, as they provide several additional degrees of freedom for
modeling XRB populations, beyond scalings with integrated $L_{\rm X}$.

The most recent large-scale measurements of the XRB XLFs and their scalings
with galaxy properties have employed a strategy of selecting galaxy samples
with high specific-SFR (sSFR~$\equiv$~SFR/$M_\star$) to isolate HMXB
populations (Mineo \etal\ 2012, hereafter, M12; Sazonov \& Khabibullin~2017a,
2017b) and elliptical galaxy populations that lack HMXBs to isolate LMXB
populations (Zhang \etal\ 2012; hereafter, Z12; Peacock \etal\ 2017).  By
design, such a strategy excludes data from more representative populations of
galaxies that are likely to have a mix of populations and has the potential to
yield misleading results for a number of physical reasons.  For example,
late-type galaxies generally have younger mean stellar ages, and could have
larger contributions from LMXBs than elliptical galaxies, since the LMXB
emission per unit mass is expected to decline with increasing age (e.g., Fragos
\etal\ 2008).  Similarly, massive elliptical galaxies, which dominate studies
of LMXB scaling relations, tend to have larger numbers of globular clusters
(GCs) per unit mass than lower-mass late-type galaxies (e.g., Brodie \& Strader
2006).  GCs very efficiently produce LMXBs through dynamical interactions
(Clark~1975; Fabian \etal\ 1975; Sivakoff \etal\ 2007; Cheng \etal\ 2018a,
2018b) and can even dominate the LMXB population of massive ellipticals (e.g.,
Irwin~2005; Kim \etal\ 2009; Voss \etal\ 2009; Lehmer \etal\ 2014) and produce
XLFs that are different in shape to those of the LMXB population found in the
galactic field. 

In this paper, we delve into the \chandra\
archive of local ($D \simlt 30$~Mpc) galaxies to establish XRB XLF correlations
with physical properties that are representative of the local galaxy population
that makes up most of the mass of the local Universe (e.g., Blanton \& Moustakas~2009).  We make use of 5.8~Ms of
\chandra\ ACIS imaging data across \ngal~galaxies to simultaneously
constrain the HMXB and LMXB XLF shapes and scalings with SFR and
$M_\star$, respectively.  We employ a galaxy decomposition technique, developed
in Lehmer \etal\ (2017), to statistically extract the contributions from HMXBs,
LMXBs, and unrelated background sources (e.g., AGN and Galactic stars).  This
technique uses spatially-resolved maps of SFR and $M_\star$ for the galaxies in
our sample to extract XRB population statistics from a range of local
specific-SFRs, and then self-consistently models the XRB XLFs across the entire
sSFR range.  

Our goal here is to establish a baseline XLF model, for which we can compare
observed XLFs of other galaxies and identify outliers to study in more detail.
Furthermore, in subsequent studies, we will expand our sample and will
investigate quantitatively how metallicity, stellar age, and GC populations
influence the XRB XLFs.  Our paper is organized as follows.  In $\S$2, we
discuss the galaxy sample selection.  In $\S$3 we outline our analysis
procedures for constructing maps of SFR and $M_\star$, as well as our detailed
\xray\ data reduction and point-source cataloging procedure.  In $\S$4, we
present the XLFs for our galaxies and culled regions selected by sSFR, and
provide model fits to the XLFs.  In $\S$5, we make comparisons of our HMXB and
LMXB XLFs with past observational estimates and XRB population synthesis
models, identify interesting galaxies with XRB populations that are outliers to
the average, and discuss possible physical trends that explain these
deviations.  We also characterize the galaxy-to-galaxy scatter of the
integrated XRB luminosity implied by our XLFs.  Finally, we summarize our
results in $\S$6.  Full catalogs of the \chandra\ sources, \chandra\ images, as
well as our SFR and $M_\star$ maps, are provided publicly at {\ttfamily
https://lehmer.uark.edu/downloads/}.

%
%%%%%%%%%%%%%%%%%%%%%%%%%%%%%%%%%%%%%%%%%%%%%%%%%%%%%%%%%%%%%%%%%%%%%%%%
\section{Galaxy Sample Selection and Properties}
%%%%%%%%%%%%%%%%%%%%%%%%%%%%%%%%%%%%%%%%%%%%%%%%%%%%%%%%%%%%%%%%%%%%%%%%
%

We started by selecting a sample of nearby galaxies with \chandra\ coverage, as
well as far-UV--to--IR multiwavelength data that was sufficient for measuring
accurate SFR and $M_\star$ values on subgalactic scales.  To this end, we
searched for galaxies in the \spitzer\ Infrared Nearby Galaxies Survey (SINGS;
Kennicutt \etal\ 2003) that also contained \chandra\ ACIS imaging data in the
archive.  The SINGS sample itself contains 75 nearby ($\simlt$30~Mpc) galaxies,
which were selected to be diverse in properties, and well resolved and
efficiently observed by \spitzer\ and other multiwavelength facilities
(covering angular sizes of 5--15~arcmin).  We first limited our search to
galaxies with $B$-band absolute magnitudes of $M_B < -19$~mag (as provided by
Moustakas \etal\ 2010), which includes galaxies that are $\approx$1~mag below
the knee of the $B$-band luminosity function and are in the range of galaxies
that dominate the stellar mass density of the local universe (e.g., Blanton
\etal\ 2003).  We further restricted our sample to galaxies with inclinations
to our line of sight that are $\simlt$70~deg. Inclination, $i$,  was estimated
as $\sin (i) = \sqrt{1-(b/a)^2}$, where $a$ and $b$ are the semi-major and
semi-minor axes, as defined in the $K_s$-band by Jarrett \etal\ (2003).  This
criteria is motivated by the fact that extinction due to a thin disk rapidly
increases for inclinations above this value (e.g., Tuffs \etal\ 2004).  
Since
we are unable to accurately correct for intrinsic extinction for the point
sources, and expect that this extinction could have substantial effects on the observed XLFs, we have elected to exclude these galaxies.

The above selection resulted in \nsings\ SINGS galaxies, with \ngalsings\ of
them having sufficient \chandra\ data.  In addition to these galaxies, we
elected to add to our sample NGC~5236 (M83) and NGC~5474 (M101), both of which
have properties consistent with those selected in the SINGS galaxy sample and
also have outstanding \xray\ coverage due to large \chandra\ campaigns (Kuntz
\& Snowden~2010; Long \etal\ 2014).   We note that the overall selection of
galaxies is driven by the presence of excellent multiwavelength data mainly
available through SINGS.  The SINGS sample has 80\% \chandra\ completeness,
with many of the galaxies being observed due to their SINGS coverage (e.g., via
the XSINGS program; PI: L.~Jenkins; Tzanavaris \etal\ 2013), suggesting that
our sample is not significantly biased towards \xray\ bright galaxies.  In
total, our final sample contains \ngal\ nearby galaxies.  

%%%%%%%%%%%%%%%%%%%%%%%%%%%%%%%%%%%%%%%%%%%%%%%%%%%%%%%%%%%%%%%%%%%%%%%%%%%%%%%%%%
% Table 1
%%%%%%%%%%%%%%%%%%%%%%%%%%%%%%%%%%%%%%%%%%%%%%%%%%%%%%%%%%%%%%%%%%%%%%%%%%%%%%%%%%
\begin{longrotatetable}
\begin{deluxetable*}{lllcccccccccccc}
\tablewidth{1.0\columnwidth}
\tabletypesize{\footnotesize}
\tablecaption{Nearby Galaxy Sample and Properties}
\tablehead{
\multicolumn{1}{c}{\sc Galaxy}  & \colhead{} & \colhead{} & \colhead{} &\colhead{} & \colhead{} & \multicolumn{4}{c}{\sc Size Parameters} & \colhead{} & \colhead{} & \colhead{} & \colhead{} & \colhead{}\\
\vspace{-0.25in} \\
\multicolumn{1}{c}{\sc Name} & \multicolumn{1}{c}{\sc Alt.} &  \multicolumn{1}{c}{\sc Morph.} & \multicolumn{2}{c}{\sc Central Position} & \colhead{$D$} & \colhead{$a$} & \colhead{$b$} & \colhead{PA} & \colhead{$r_{\rm remove}$} & \colhead{{\sc SFR}} & \colhead{$\log M_\star$} & \colhead{$\log$ sSFR} & \colhead{12 + $\log$~[O/H]} & \colhead{} \\ 
\vspace{-0.25in} \\
\multicolumn{1}{c}{(NGC)} & \multicolumn{1}{c}{\sc Name} & \multicolumn{1}{c}{\sc Type} & \colhead{$\alpha_{\rm J2000}$} & \colhead{$\delta_{\rm J2000}$} & \colhead{(Mpc)} & \colhead{(arcmin)} & \colhead{(arcmin)} & \colhead{(deg)} & \colhead{(arcsec)} & \colhead{($M_\odot$~yr$^{-1}$)} & \colhead{($M_\odot$)}  & \colhead{(${\rm yr}^{-1}$)} & \colhead{(dex)} & \colhead{$S_N$} \\ 
\vspace{-0.25in} \\
\multicolumn{1}{c}{(1)} & \multicolumn{1}{c}{(2)} & \multicolumn{1}{c}{(3)} & \colhead{(4)} & \colhead{(5)} & \colhead{(6)} & \colhead{(7)} & \colhead{(8)} & \colhead{(9)} & \colhead{(10)} & \colhead{(11)} & \colhead{(12)} & \colhead{(13)} & \colhead{(14)} & \colhead{(15)}
}
\startdata
       337 &            &        SBd &     00 59 50.1 & $-$07 34 40.7 &   22.40$\pm$2.30 &   0.87 &   0.49 &    $-$22.5 &  0 &  1.09 &  9.32 &                   $-$9.28 &             8.44$\pm$0.07 &          \ldots \\
       584 &            &         E4 &     01 31 20.8 & $-$06 52 05.0 &   20.10$\pm$1.90 &   1.47 &   0.91 &       62.5 &  0 &  0.05 & 10.48 &                  $-$11.77 &               8.75$^\ast$ &   1.69$\pm$0.67 \\
       628 &        M74 &        SAc &       01 36 41.8 & +15 47 00.5 &    7.30$\pm$1.40 &   2.10 &   1.80 &       87.5 &  3 &  0.33 &  9.48 &                   $-$9.96 &             8.54$\pm$0.15 &          \ldots \\
       925 &            &       SABd &       02 27 16.9 & +33 34 44.0 &    9.12$\pm$0.17 &   1.87 &   0.82 &    $-$75.0 &  0 &  0.18 &  9.03 &                   $-$9.78 &             8.38$\pm$0.15 &          \ldots \\
      1097 &            &        SBb &     02 46 19.1 & $-$30 16 29.7 &   17.10$\pm$2.30 &   2.63 &   1.44 &    $-$35.0 &  5 &  4.51 & 10.76 &                  $-$10.11 &             8.83$\pm$0.05 &          \ldots \\
\\
      1291 &            &      SB0/a &     03 17 18.6 & $-$41 06 29.1 &   10.80$\pm$2.30 &   2.39 &   1.70 &    $-$10.0 &  3 &  0.08 & 10.81 &                  $-$11.89 &               9.20$^\ast$ &          \ldots \\
      1316 &            &       SAB0 &     03 22 41.8 & $-$37 12 29.5 &   21.50$\pm$1.70 &   2.77 &   1.99 &       47.5 &  3 &  0.49 & 11.48 &                  $-$11.79 &               9.52$^\ast$ &   0.54$\pm$0.27 \\
      1404 &            &         E1 &     03 38 51.9 & $-$35 35 39.8 &   20.80$\pm$1.70 &   1.38 &   1.24 &    $-$17.5 &  3 &  0.10 & 10.98 &                  $-$11.99 &               9.21$^\ast$ &   1.78$\pm$0.32 \\
      2841 &            &        SAb &       09 22 02.7 & +50 58 35.3 &   14.10$\pm$1.50 &   3.02 &   1.36 &    $-$30.0 &  0 &  0.61 & 10.67 &                  $-$10.89 &             8.89$\pm$0.05 &          \ldots \\
      3031 &        M81 &       SAab &       09 55 33.2 & +69 03 54.9 &    3.55$\pm$0.13 &   8.13 &   4.14 &    $-$31.0 & 12 &  0.25 & 10.39 &                  $-$10.98 &             8.60$\pm$0.09 &   1.11$\pm$0.37 \\
\\
      3184 &            &      SABcd &       10 18 17.0 & +41 25 27.8 &   11.10$\pm$1.90 &   1.91 &   1.62 &      117.5 &  0 &  0.48 &  9.68 &                  $-$10.00 &             8.75$\pm$0.12 &          \ldots \\
      3198 &            &        SBc &       10 19 55.0 & +45 32 58.9 &   13.68$\pm$0.50 &   1.91 &   0.67 &       40.0 &  0 &  0.55 &  9.70 &                   $-$9.96 &             8.43$\pm$0.15 &          \ldots \\
      3351 &        M95 &        SBb &       10 43 57.7 & +11 42 13.0 &    9.33$\pm$0.39 &   1.94 &   1.71 &    $-$17.0 &  0 &  0.57 &  9.95 &                  $-$10.19 &             9.21$\pm$0.05 &          \ldots \\
      3521 &            &      SABbc &     11 05 48.6 & $-$00 02 09.2 &   10.10$\pm$2.30 &   2.74 &   1.40 &    $-$14.5 &  0 &  1.43 & 10.41 &                  $-$10.25 &             8.74$\pm$0.09 &          \ldots \\
      3627 &        M66 &       SABb &       11 20 15.0 & +12 59 28.6 &    9.38$\pm$0.35 &   3.08 &   1.70 &        6.5 &  3 &  1.83 & 10.30 &                  $-$10.04 &             8.66$\pm$0.11 &          \ldots \\
\\
      3938 &            &        SAc &       11 52 49.5 & +44 07 14.6 &   13.40$\pm$2.30 &   1.30 &   1.23 &       28.5 &  0 &  0.58 &  9.64 &                   $-$9.88 &               8.74$^\ast$ &          \ldots \\
      4125 &            &     E6 pec &       12 08 06.0 & +65 10 26.9 &   23.90$\pm$2.80 &   1.76 &   1.11 &       82.5 &  0 &  0.13 & 10.84 &                  $-$11.73 &               9.30$^\ast$ &          \ldots \\
      4254 &        M99 &        SAc &       12 18 49.6 & +14 24 59.4 &   16.50$\pm$0.60 &   1.70 &   1.62 &       23.5 &  0 &  3.17 & 10.21 &                   $-$9.71 &             8.77$\pm$0.11 &          \ldots \\
      4321 &       M100 &      SABbc &       12 22 54.9 & +15 49 20.6 &   14.32$\pm$0.46 &   2.51 &   1.96 &    $-$72.5 &  0 &  2.04 & 10.24 &                   $-$9.93 &             8.81$\pm$0.07 &          \ldots \\
      4450 &            &       SAab &       12 28 29.6 & +17 05 05.3 &   16.50$\pm$0.60 &   1.87 &   1.18 &        2.5 &  3 &  0.19 & 10.40 &                  $-$11.12 &               8.82$^\ast$ &          \ldots \\
\\
      4536 &            &      SABbc &       12 34 27.1 & +02 11 16.4 &   14.45$\pm$0.27 &   1.89 &   0.98 &    $-$85.0 &  0 &  1.88 & 10.13 &                   $-$9.86 &             8.45$\pm$0.23 &          \ldots \\
      4552 &        M89 &          E &       12 35 39.9 & +12 33 21.7 &   15.92$\pm$0.81 &   1.48 &   1.39 &    $-$30.0 &  3 &  0.08 & 10.54 &                  $-$11.66 &               8.83$^\ast$ &   7.68$\pm$1.40 \\
      4559 &            &      SABcd &       12 35 57.7 & +27 57 35.1 &   10.30$\pm$2.30 &   2.04 &   0.96 &    $-$32.5 &  0 &  0.45 &  9.34 &                   $-$9.68 &             8.40$\pm$0.13 &          \ldots \\
      4569 &            &      SABab &       12 36 49.8 & +13 09 46.3 &   16.50$\pm$0.60 &   2.75 &   1.10 &       15.0 &  2 &  1.06 & 10.48 &                  $-$10.45 &               9.26$^\ast$ &          \ldots \\
      4594 &       M104 &        SAa &     12 39 59.5 & $-$11 37 23.1 &    9.33$\pm$0.34 &   3.36 &   1.82 &       87.5 &  3 &  0.18 & 10.86 &                  $-$11.59 &               9.22$^\ast$ &   2.70$\pm$0.28 \\
\\
      4725 &            &  SABab pec &       12 50 26.6 & +25 30 02.7 &   11.91$\pm$0.33 &   2.91 &   1.51 &       50.0 &  0 &  0.37 & 10.38 &                  $-$10.81 &             8.79$\pm$0.08 &          \ldots \\
      4736 &        M94 &       SAab &       12 50 53.1 & +41 07 12.5 &    5.20$\pm$0.43 &   2.87 &   2.27 &       85.0 &  0 &  0.50 & 10.13 &                  $-$10.43 &             8.72$\pm$0.04 &          \ldots \\
      4826 &        M64 &       SAab &       12 56 43.7 & +21 40 57.6 &    7.48$\pm$0.69 &   3.58 &   2.04 &    $-$70.0 &  0 &  0.42 & 10.41 &                  $-$10.79 &             9.24$\pm$0.04 &          \ldots \\
      5033 &            &        SAc &       13 13 27.5 & +36 35 37.1 &   14.80$\pm$2.30 &   1.79 &   0.80 &     $-$5.0 &  5 &  0.84 & 10.37 &                  $-$10.44 &             8.55$\pm$0.13 &          \ldots \\
      5055 &        M63 &       SAbc &       13 15 49.3 & +42 01 45.4 &    7.80$\pm$2.30 &   3.40 &   1.97 &    $-$82.5 &  0 &  0.94 & 10.26 &                  $-$10.29 &             8.80$\pm$0.10 &          \ldots \\
\\
      5194 &        M51 &  SABbc pec &       13 29 52.7 & +47 11 42.9 &    8.58$\pm$0.10 &   3.29 &   2.24 &       57.5 &  3 &  2.61 & 10.24 &                   $-$9.83 &             8.87$\pm$0.11 &   0.76$\pm$0.15 \\
      5236 &        M83 &       SABc &     13 37 00.9 & $-$29 51 56.7 &    4.66$\pm$0.33 &   5.21 &   4.01 &       45.0 &  0 &  2.48 & 10.33 &                   $-$9.94 &  8.95$\pm$0.03$^\ddagger$ &   0.17$\pm$0.05 \\
      5457 &       M101 &      SABcd &       14 03 12.5 & +54 20 55.5 &    6.81$\pm$0.03 &   3.94 &   3.90 &       28.5 &  0 &  1.07 &  9.91 &                   $-$9.88 &  9.10$\pm$0.08$^\ddagger$ &   0.43$\pm$0.11 \\
      5713 &            &  SABbc pec &     14 40 11.5 & $-$00 17 21.2 &   29.40$\pm$2.30 &   0.90 &   0.89 &    $-$20.0 &  0 &  5.48 & 10.15 &                   $-$9.41 &             8.63$\pm$0.06 &          \ldots \\
      5866 &       M102 &         S0 &       15 06 29.6 & +55 45 47.9 &   15.42$\pm$0.85 &   1.86 &   0.78 &    $-$57.0 &  0 &  0.14 & 10.46 &                  $-$11.32 &               8.81$^\ast$ &   1.37$\pm$0.26 \\
\\
      6946 &            &      SABcd &       20 34 52.3 & +60 09 13.2 &    6.80$\pm$1.70 &   4.21 &   2.95 &       52.5 &  0 &  2.46 & 10.01 &                   $-$9.61 &             8.66$\pm$0.11 &   0.29$\pm$0.13 \\
      7331 &            &        SAb &       22 37 04.1 & +34 24 57.3 &   14.52$\pm$0.60 &   2.60 &   1.27 &    $-$12.5 &  0 &  2.12 & 10.75 &                  $-$10.42 &             8.73$\pm$0.05 &   0.43$\pm$0.27 \\
      7552 &            &       SBab &     23 16 10.8 & $-$42 35 05.4 &   21.00$\pm$2.30 &   1.27 &   0.75 &    $-$85.0 & 15 &  3.58 & 10.04 &                   $-$9.48 &             8.85$\pm$0.01 &          \ldots \\
\hline
{\bf Total} & \ldots & \ldots & \ldots & \ldots & \ldots & \ldots & \ldots & \ldots & \ldots & {\bf 45.4} & {\bf 12.10} & {\bf $-$10.44} & \ldots & \ldots \\
\enddata
\tablecomments{Col.(1): NGC number of galaxy. Col.(2): Alternative Messier designation, if applicable. Col.(3): Morphological type as provided in the Third Reference Catalog of Bright Galaxies (RC3; de Vaucouleurs \etal\ 1991).  Col.(4) and (5): Right ascension and declination of the galactic center based on the 2 Micron All Sky Survey (2MASS) positions derived by Jarrett \etal\ (2003). Col.(6): Adopted distance and 1$\sigma$ error in units of Mpc.  Distances were adopted from the SINGS values provided from Col.(9) of Table~1 in Moustakas \etal\ (2010), except for NGC~5194, 5236, and 5457, which were provided by McQuinn \etal\ (2016), Tully \etal\ (2013), and Nataf \etal\ (2015), respectively.  Col.(7)--(9): $K_s$-band isophotal ellipse parameters, including, respectively, semi-major axis, $a$, semi-minor axis, $b$, and position angle east from north, PA.  The ellipses estimate the 20~mag~arcsec$^{-2}$ surface brightness contour of each galaxy (derived by Jarrett \etal\ 2003).  Col.(10): Radius of central region removed from the galaxy due to either the presence of an AGN or extreme crowding.  Col.(11)--(13): SFR, $M_\star$, and sSFR values derived using the maps described in $\S$3.1, and correspond to areal coverage within the regions defined by Col.(7)--(10) (i.e., with contributions from $r_{\rm remove}$ excluded).  Col.(14): Estimated average oxygen abundances, 12+$\log$~[O/H], from Moustakas \etal\ (2010), except for M83 and M101, which are based on the central metallicities from Bresolin \etal\ (2009) and Hu \etal\ (2018), respectively (denoted as $\ddagger$).  Most oxygen abundances are based on strong line indicators, with the exception of those denoted with asterisks, which are from the optical luminosity--metallicity correlation.  For consistency with other studies of XRB scaling relations that include metallicity, we have converted the Moustakas \etal\ (2010) abundances based on the Kobulnicky \& Kewley~(2004; KK04) calibration to the Pettini \& Pagel~(2004; PP04) calibration following the prescriptions in Kewley \& Ellison~(2008). Col.(15): GC specific frequency, $S_N$, as reported by Harris \etal\ (2013).\\
}
\end{deluxetable*}
\end{longrotatetable}

In Figure~1, we show cut-out optical images of the galaxy sample, and in
Table~1 we summarize the basic properties of each galaxy.  
Here we are interested in XLF scaling relations with the basic
properties: SFR and $M_\star$.  Calculations of galaxy-wide SFR and $M_\star$
values for our sample are detailed in $\S$3.1 below, and in Figure~2$a$ we graphically show
their values on the SFR--$M_\star$ plane.  Our sample spans 2.5~dex
in SFR and $M_\star$, and by design, these galaxies were
chosen to be diverse and do not strictly follow the galaxy ``main sequence''
(e.g., Elbaz \etal\ 2007; Noeske \etal\ 2007; Karim \etal\
2011; Whitaker \etal\ 2014).

Since we expect that the HMXB-to-LMXB ratio will be dependent on sSFR, this
quantity is of particular interest.  In Figure~2$b$, we show the distribution
of galaxy-wide sSFR (i.e., total galaxy SFR/$M_\star$) values for the \ngal\
galaxies in our sample.  Past studies have shown that around sSFR~$\approx
10^{-10}$~yr$^{-1}$ the relative \xray\ luminosities from HMXBs and LMXBs is
nearly equal, while at higher and lower sSFR values HMXBs and LMXBs,
respectively, dominate the XRB population luminosities (see, e.g., Colbert
\etal\ 2004; Lehmer \etal\ 2010; M12).  Our galaxy sample
contains 15 and 23 galaxies, respectively, above and below this threshold,
with the most extreme cases being NGC~337 (sSFR~$\approx 5 \times 10^{-10}$~yr$^{-1}$) and 
 NGC~1404 (sSFR~$\approx 10^{-12}$~yr$^{-1}$).
As we will show below, we can quantify the HMXB and LMXB contributions to the
XLFs of all late-type galaxies based on a self-consistent ``global'' model of
the HMXB and LMXB XLF scaling with SFR and $M_\star$, respectively.

%
%%%%%%%%%%%%%%%%%%%%%%%%%%%%%%%%%%%%%%%%%%%%%%%%%%%%%%%%%%%%%%%%%%%%%%%%
\section{Data Analysis and Products}
%%%%%%%%%%%%%%%%%%%%%%%%%%%%%%%%%%%%%%%%%%%%%%%%%%%%%%%%%%%%%%%%%%%%%%%%
%

\subsection{Multiwavelength Tracer Maps}

For each galaxy in our sample, we generated SFR and $M_\star$ maps, using
multiwavelength tracers of these quantities.  For SFR, we made use of FUV
\galex\ and 24~$\mu$m \spitzer\ maps, and for $M_\star$, we utilized $K$-band
data from the Two Micron All Sky Survey (2MASS) combined with optical $g$ and
$i$ band data from the Sloan Digital Sky Survey (SDSS), when available.  In the
absence of SDSS, we utilized $B$ and $V$ band data available from the SINGS
collaboration,\footnote{https://irsa.ipac.caltech.edu/data/SPITZER/SINGS/doc/sings\_fifth\_delivery\_v2.pdf}
which originated from either the Kitt Peak National Observatory (KPNO) or Cerro
Tololo Inter-American Observatory (CTIO), or in the case of NGC~6946, we made use of $B$ and $V$ band data from {\it Swift}.  Our data preparation procedure,
including the identification and subtraction of foreground Galactic stars,
background subtraction, and convolution techniques followed closely that
outlined in $\S\S$2.1--2.4 of Eufrasio \etal\ (2017) with a few minor
differences.  All images were convolved to a common Gaussian point-spread
function (PSF) with a 15~arcsec full-width at half maximum (FWHM), which is
significantly larger than the 24$\mu$m PSF to comfortably remove all PSF
features and produce a Gaussian PSF.  The images were projected to a common
pixel scale of 3~arcsec~pixel$^{-1}$.  For a galaxy at 30~Mpc, just beyond the
most distant galaxy in our sample, this pixel scale results in a physical size
of 436~pc~pixel$^{-1}$.

To calculate SFRs, we made use of the Hao \etal\ (2011) relation (implied by
their Table~3):
\begin{equation}
\left( \frac{{\rm SFR}}{M_\odot~{\rm yr}^{-1}} \right) = 1.6 \times 10^{-10}
\left[ \left(\frac{L_{\rm FUV}^{\rm obs}}{L_\odot} \right) + 3.89 \left(
\frac{L_{24~\mu{\rm m}}^{\rm obs}}{L_\odot} \right)  \right],
\end{equation}
where $L_{\rm FUV}^{\rm obs}$ and $L_{24~\mu{\rm m}}^{\rm obs}$ are the
observed (i.e., corrected only for Galactic extinction and not intrinsic
extinction) monochromatic luminosities (e.g., $\nu L_\nu$) at 1528~\AA\ and
24~$\mu$m, respectively.  For each pixel, values of $L_{\rm FUV}^{\rm obs}$ and
$L_{24~\mu{\rm m}}^{\rm obs}$ are determined from the \galex\ FUV and \spitzer\
24~$\mu$m maps, respectively.  In the case of NGC~7552, \herschel\ 70~$\mu$m
data was used instead of the \spitzer\ 24$\mu$m data, due to strong PSF
contributions from the 24$\mu$m-bright nuclear starburst at large
galactocentric radii.  For this galaxy, we converted $L_{70}^{\rm obs}$ to
$L_{24}^{\rm obs}$, using scaling relations from Kennicutt \& Evans (2012) and
Galametz \etal\ (2013).  These SFRs are based on an assumed constant
star-formation history with duration of 100~Myr, a Kroupa~(2001) initial mass
function (IMF), and solar metallicity (i.e., $Z = 0.02$).  The calibration is
reported to have a 1$\sigma$ uncertainty of 0.1~dex.  

Stellar masses
($M_\star$) were computed following the relations in Zibetti \etal\ (2009; see their Table~B1):
\begin{equation}
\log (M_\star/M_\odot) = \log (L_K/L_{K,\odot}) - 1.321 + 0.754 (g -i),
\end{equation}
\begin{equation}
\log (M_\star/M_\odot) = \log (L_K/L_{K,\odot}) - 1.390 + 1.176 (B -V).
\end{equation}
We utilized Eqn.~(2) for 30 of our galaxies, and this was our preferred
calibration.  Eqn.~(3) was applied for the remaining 8 galaxies in our sample.
Both equations are reported to have 1$\sigma$ calibration uncertainties of
$\approx$0.13~dex.  For 17 of the galaxies, both $g-i$ and $B-V$ colors were
available.  We generated maps based on both calibrations and found good
agreement between tracers and consistent with the uncertainty in the Zibetti
\etal\ (2009) calibration.

%
%%%%%%%%%%%%%%%%%%%%%%%%%%%%%%%%%%%%%%%%%%%%%%%%%%%%%%%%%%%%%%%%%%%%%%%%%%%%%%%%%%
% Figure 2
%%%%%%%%%%%%%%%%%%%%%%%%%%%%%%%%%%%%%%%%%%%%%%%%%%%%%%%%%%%%%%%%%%%%%%%%%%%%%%%%%%
%
\begin{figure*}[t!]
\figurenum{2}
\centerline{
\includegraphics[width=9cm]{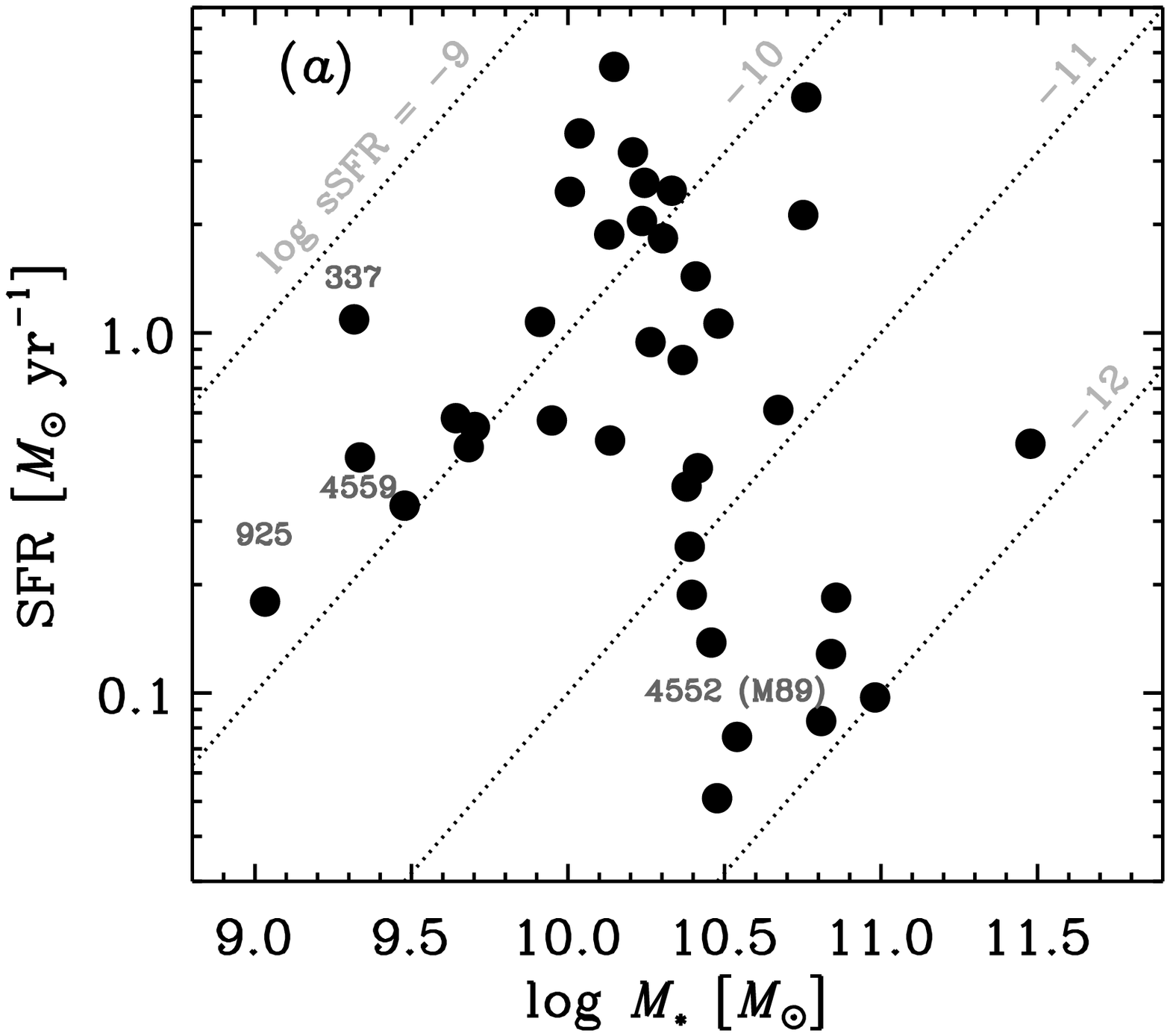}
\hfill
\includegraphics[width=9cm]{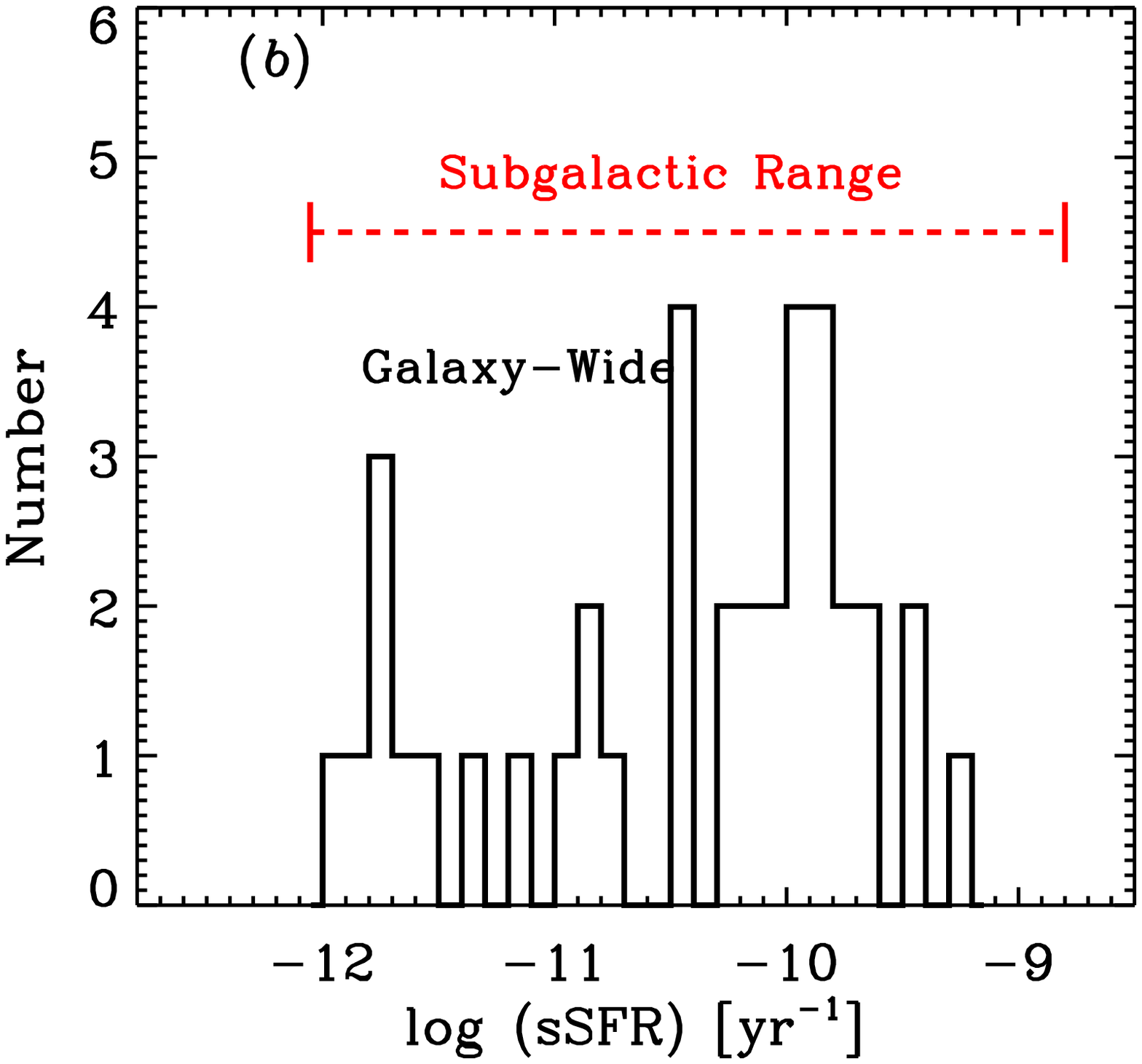}
}
\caption{
%%%
($a$) Galaxy-wide SFR versus $M_\star$ for the \ngal\ galaxies in our sample.  
Dotted lines show locations of objects with $\log$~sSFR (yr$^{-1}$)~=~$-9$, $-10$, $-11$, and $-12$
(see annotations).
%%%
($b$) Distribution of sSFR values for whole galaxies ({\it solid black
histogram\/}; see $\S$2) and range of subgalactic regions used ({\it dashed
horizontal bar\/}; see $\S$4.2).  As expected, the range of environments is
broader for the subgalactic regions, allowing us to more cleanly probe how the
XRB luminosity function varies with sSFR.  
}
\end{figure*}
%%%%%%%%%%%%%%%%%%%%%%%%%%%%%%%%%%%%%%%%%%%%%%%%%%%%%%%%%%%%%%%%%%%%%%%%%%%%%%%%%%

%%%%%%%%%%%%%%%%%%%%%%%%%%%%%%%%%%%%%%%%%%%%%%%%%%%%%%%%%%%%%%%%%%%%%%%%%%%%%%%%%%
% Table 2
%%%%%%%%%%%%%%%%%%%%%%%%%%%%%%%%%%%%%%%%%%%%%%%%%%%%%%%%%%%%%%%%%%%%%%%%%%%%%%%%%%
\begin{table*}
{\small
\begin{center}
\caption{\chandra\ Advanced CCD Imaging Spectrometer (ACIS) Observation Log}
\begin{tabular}{lcccccccc}
\hline\hline
& \multicolumn{2}{c}{\sc Aim Point} & {\sc Obs. Start} & {\sc Exposure}$^a$ & {\sc Flaring}$^b$ &  $\Delta \alpha$ & $\Delta \delta$  & {\sc Obs.} \\
\multicolumn{1}{c}{\sc Obs. ID} & $\alpha_{\rm J2000}$ & $\delta_{\rm J2000}$ & (UT) & (ks) & {\sc Intervals} & (arcsec) & (arcsec)  & {\sc Mode}$^c$ \\
\hline\hline
\multicolumn{9}{c}{{\bf NGC0337}} \\
\hline
12979$^d$ & 00 59 49.29 & $-$07 34 28.15 & 2011-07-19T23:07:02 & 10 & \ldots & \ldots & \ldots & F \\
\hline
\multicolumn{9}{c}{{\bf NGC0584}} \\
\hline
12175$^d$ & 01 31 20.38 & $-$06 51 38.45 & 2010-09-07T01:40:53 & 10 & \ldots & \ldots & \ldots & V \\
\hline
\multicolumn{9}{c}{{\bf NGC0628}} \\
\hline
14801  & 01 36 47.41 & +15 45 32.58 & 2013-08-21T15:40:51 & 10 & \ldots & $+$0.05 & $+$0.01 & V \\
16000  & 01 36 47.37 & +15 45 31.61 & 2013-09-21T06:40:27 & 40 & \ldots & $+$0.56 & $-$0.24 & V \\
16001  & 01 36 47.39 & +15 45 29.57 & 2013-10-07T23:56:17 & 15 & \ldots & $+$0.24 & $-$0.07 & V \\
16002  & 01 36 48.85 & +15 45 26.66 & 2013-11-14T20:10:48 & 38 & \ldots & $+$0.08 & $+$0.16 & V \\
16003  & 01 36 48.89 & +15 45 28.36 & 2013-12-15T15:55:42 & 40 & \ldots & $+$0.04 & $-$0.11 & V \\
16484  & 01 36 47.38 & +15 45 29.36 & 2013-10-10T14:31:23 & 15 & \ldots & $+$0.45 & $+$0.14 & V \\
16485  & 01 36 47.39 & +15 45 29.44 & 2013-10-11T11:13:35 & 9 & \ldots & $+$0.32 & $+$0.06 & V \\
2057  & 01 36 40.35 & +15 48 17.73 & 2001-06-19T19:03:09 & 46 & 1, 0.5 & $-$0.05 & $-$0.05 & F \\
2058$^d$ & 01 36 36.11 & +15 46 51.99 & 2001-10-19T04:08:30 & 46 & \ldots & \ldots & \ldots & F \\
4753  & 01 36 51.21 & +15 45 12.44 & 2003-11-20T04:14:02 & 5 & \ldots & $-$0.10 & $-$0.03 & F \\
4754  & 01 36 51.51 & +15 45 12.89 & 2003-12-29T13:07:58 & 5 & \ldots & $+$0.09 & $+$0.07 & F \\
Merged$^e$  &01 36 44.82 & +15 46 11.67 & & 269 & 1, 0.5 &  \ldots & \ldots & \ldots \\
\hline
\end{tabular}
\end{center}
Note.---The full version of this table contains entries for all \ngal\ galaxies and 164 ObsIDs, and is available in the electronic edition.  An abbreviated version of the table is displayed here to illustrate its form and content.  \\
$^a$ All observations were continuous. These times have been corrected for removed data that were affected by high background; see $\S$~3.2.\\
$^b$ Number of flaring intervals and their combined duration.  These intervals were rejected from further analyses. \\
$^c$ The observing mode (F=Faint mode; V=Very Faint mode).\\
$^d$ Indicates Obs.~ID by which all other observations are reprojected to for alignment purposes.  This Obs.~ID was chosen for reprojection as it had the longest initial exposure time, before flaring intervals were removed.\\
$^e$ Aim point represents exposure-time weighted value.
}
\end{table*}

\subsection{{\it Chandra} Data Reduction and Catalog Production}

For our \xray\ point-source measurements, we use \chandra\ ACIS imaging data
(both ACIS-S and ACIS-I) of the galaxies in our sample.  In Table~2, we
tabulate the full \chandra\ observing log used in this paper.  We restricted
our analyses to \chandra\ data sets that had aim points within 5~arcmin of the
central coordinates of the galaxy.  This restriction ensures that the ObsID
combined images reach deep limits with a sharp PSF ($\simlt$1.5~arcsec 90\%
encircled-counts fraction radii) in the central nuclear regions of the
galaxies, where source confusion could potentially be problematic.  Some of the
galaxies in our sample have much more extensive archives than we utilize here.
For example, for M81, we make use of only 18 of the 27~ObsIDs that were available
in the archive, as a result of us excluding observations from a large program
to observe the periphery of the galaxy (PI: D.~Swartz).

Our \chandra\ data reduction was carried out using CIAO~v.~4.8 with {\ttfamily
CALDB}~v.~4.7.1,\footnote{http://cxc.harvard.edu/ciao/} and our procedure
followed closely the methods outlined in $\S$2.2 of Lehmer \etal\ (2017).
Briefly, we (1) reprocessed pipeline products using the {\ttfamily
chandra\_repro} script; (2) removed bad pixels and columns, and filtered the
events list to include only good time intervals without significant ($>$3
$\sigma$) flares above the background level; (3) when applicable, aligned
events lists and aspect histograms, via {\ttfamily wcs\_match} and {\ttfamily
wcs\_update}, to the deepest \chandra\ ObsID for a given galaxy, using small
translations (median shifts and 1$\sigma$ standard deviations of
$\delta$R.A.~=~$0.16 \pm 0.14$~arcsec and $\delta$decl.~=~$0.16 \pm
0.18$~arcsec);  (4) constructed merged events lists and astrometric solutions
using the {\ttfamily merge\_obs} script; and (5) created additional products,
including images, exposure maps, and exposure-weighted PSF maps with a 90\%
enclosed-count fraction, appropriate for the 0.5--2~keV, 2--7~keV, and
0.5--7~keV bands, which we hereafter refer to as the soft band (SB), hard band
(HB), and full band (FB), respectively.

%
%%%%%%%%%%%%%%%%%%%%%%%%%%%%%%%%%%%%%%%%%%%%%%%%%%%%%%%%%%%%%%%%%%%%%%%%%%%%%%%%%%
% Figure 3
%%%%%%%%%%%%%%%%%%%%%%%%%%%%%%%%%%%%%%%%%%%%%%%%%%%%%%%%%%%%%%%%%%%%%%%%%%%%%%%%%%
%
\begin{figure*}[t!]
\figurenum{3}
\centerline{
\includegraphics[width=18cm]{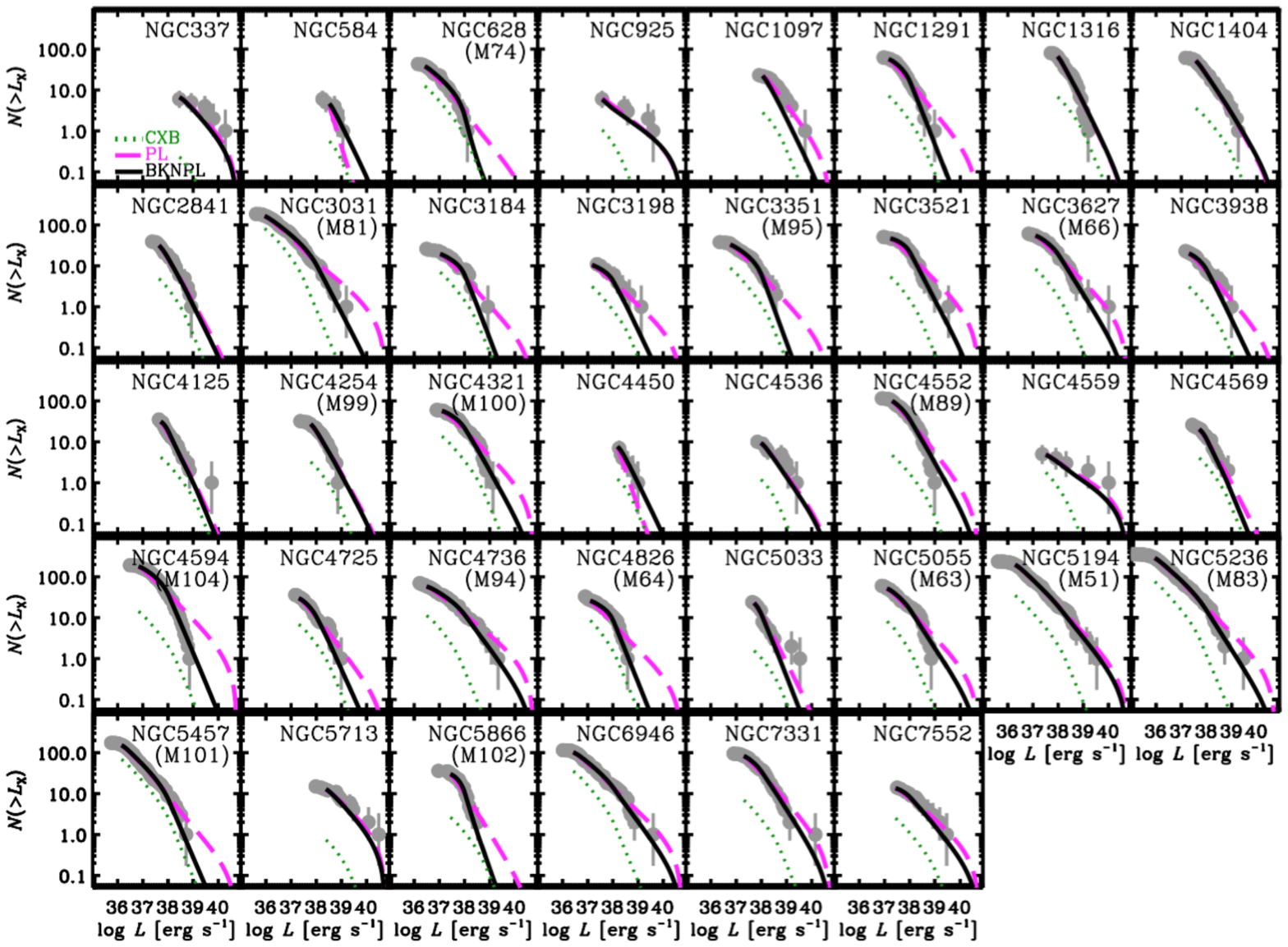}
}
\vspace{-0.1in}
\caption{
%%%
Observed cumulative XLFs for all galaxies in our sample ({\it gray circles with
1$\sigma$ error bars\/}).  These XLFs are not corrected for incompleteness,
explaining the perceptible turnovers at the lowest luminosity values.  Model
fits, which include contributions from the CXB ({\it green dotted curves\/})
and intrinsic point sources, are shown for single ({\it dashed magenta curves\/}) 
and broken ({\it black solid curves\/}) power-law models.  Displayed models
(and CXB contributions) include the effects of incompleteness for the purposes
of fitting the observed data (see $\S$4.1 for details).  All data above the
50\% completeness limits, $L_{50}$, were used in our fits, and the plotted model
curves are displayed going down to these limits.
%%%
}
\end{figure*}
%%%%%%%%%%%%%%%%%%%%%%%%%%%%%%%%%%%%%%%%%%%%%%%%%%%%%%%%%%%%%%%%%%%%%%%%%%%%%%%%%%

Merged 0.5--7~keV images were searched using {\ttfamily wavdetect} at a
false-positive probability threshold of $1 \times 10^{-6}$ over seven wavelet
scales from 1--8 pixels in a $\sqrt{2}$ sequence (i.e., 1, $\sqrt{2}$, 2,
2$\sqrt{2}$, 4, 4$\sqrt{2}$, and 8 pixels).  We ran {\ttfamily wavdetect} using
the merged exposure maps and 90\% enclosed-count fraction PSF maps, which
resulted in an initial source catalog with properties (e.g., positions and counts)
appropriate for point sources.  We inspected images from the three bands (i.e.,
SB, HB, and FB) by eye with source candidates indicated to ensure this process
produced sensible source candidates.  We found in the case of M81 that
several sources were identified along read-out streaks associated with the 
piled-up central AGN.  Unless the sources were obviously real (based on having
spatial count distributions consistent with the PSF and clear multi-band detections),
the sources along these streaks were removed from further consideration.  
Finally, for 14 galaxies, we found that point-source crowding in the
central region of the galaxy (near the galactic nuclei) was prohibitively large
(e.g., NGC~7552), or the central AGN was bright (e.g., M81).  In such cases, we
identified circular regions around these sources, within which we excluded the
sources, as well as the SFR and $M_\star$ contributions, from our \xray\ luminosity
function analyses (see Col.~10 of Table~1).  For completeness, these \xray\ sources are included in our
catalogs with a flag indicating that the source was excluded from our analyses
for the above reasons.

Source photometry was computed for all sources using the {\ttfamily ACIS
Extract} ({\ttfamily AE}) v.~2016sep22 software package (Broos \etal\ 2010,
2012).\footnote{ The {\ttfamily ACIS Extract} software package and User's Guide
are available at http://www.astro.psu.edu/xray/acis/acis\_analysis.html.}
{\ttfamily AE} extracts source events and exposure times from all pixels that
have exposure within polygonal regions that nominally trace the $\approx$90\%
encircled counts fraction (ECF).  These polygonal contours are constructed by
{\ttfamily AE}, for each source, using 1.497~keV PSFs generated by the
{\ttfamily MARX}~v.~5.3.2\footnote{http://space.mit.edu/ASC/MARX/} ray-tracing
code.  In a number of cases, the 90\% polygonal regions overlapped, and
{\ttfamily AE} iteratively generated non-overlapping polygonal regions that
encompassed a smaller fraction of the PSF, and kept track of those PSF
fractions.  Local background events files were extracted by {\ttfamily AE} by
first masking the source events within a circular masking region that is 1.1$\times$
the size of the 99.9\% ECF at 1.497~keV and then extracting events from a larger
circular aperture centered around the sources.  The larger circular aperture
size is determined by requiring that the summed exposure map value of the
background pixels (i.e., those not masked), $T_{\rm bkg}$, is 5--10~times that
determined for the source extraction pixels, $T_{\rm src}$, and also contains a
minimum number of 5 counts.  The latter criterion is generally met for $T_{\rm
bkg} = 5 \times T_{\rm src}$, but if it is not, then the background aperture is
increased up until $T_{\rm bkg} = 10 \times T_{\rm src}$, regardless of whether
the aperture contains 5 counts or more.

For sources near the {\ttfamily wavdetect} threshold, we found that the
{\ttfamily AE} photometry would sometimes provide negative counts in the
detection bandpass.  Instead of re-evaluating the significance of these sources
with {\ttfamily AE}, and culling low-significance sources from the catalog, we
chose to include them and utilize the {\ttfamily wavdetect} photometry.  The
primary reason for such discrepancies is likely due to the fact that {\ttfamily
AE} evaluates photometry based on events within the 90\% ECF, while {\ttfamily
wavdetect} uses wavelets of various scales to identify sources (sometimes based
on scales smaller than the 90\% ECF) and reconstructs a model of the source
counts.  Thus, {\ttfamily wavdetect} will be somewhat more sensitive than
{\ttfamily AE} in identifying sources when only the core of the PSF is
significant compared to the background.  Our choice to keep the
low-significance sources is also motivated by our later use of {\ttfamily
wavdetect} in calculating the completeness of a given galaxy's detected sources
as a function of counts and location, using large simulations of fake sources
(see $\S$3.3 for details).  Such completeness calculations are not feasible using
the computationally intensive {\ttfamily AE} photometry procedure.

For sources with $>$20 net counts, we performed basic spectral modeling of the
data within {\ttfamily AE}, using {\ttfamily xspec} v.~12.9.1 (Arnaud~1996).
We adopted an absorbed power-law model with both a fixed component of Galactic
absorption and a free variable intrinsic absorption component ({\ttfamily TBABS
$\times$ TBABS $\times$ POW} in {\ttfamily xspec}).  The free parameters
include the intrinsic column density, $N_{\rm H, int}$, and photon index, $\Gamma$.
The Galactic absorption column, $N_{\rm H, gal}$, for each source was fixed to
the value appropriate for the location of each galaxy, as derived by Dickey \&
Lockman~(1990).\footnote{Galactic column density values were extracted using
the {\ttfamily colden} tool at {\ttfamily
http://cxc.harvard.edu/toolkit/colden.jsp}}  All spectral fits were derived by
minimizing the C-statistic within {\ttfamily xspec} (Cash~1979), with both the on-source
events (i.e., those within the {\ttfamily AE} extraction regions discussed
above) and background events supplied.  {\ttfamily AE} simultaneously fits the
background spectrum, using a piecewise linear model, and the on-source spectrum
including the background spectrum model plus the physical source model (i.e., the
absorbed power law).

For the subsample of sources where spectral fitting was possible, we found
median and interquartile ranges of $\log N_{\rm H, int} = 21.3^{+0.5}_{-0.7}$
and $\Gamma = 1.7^{+0.3}_{-0.5}$.  Whenever possible, we computed 0.5--8~keV
\xray\ fluxes and corresponding luminosities using these best fit models.  For
sources where spectral fitting was not possible, we converted the 0.5--7~keV
count rates to 0.5--8~keV fluxes using the median model (i.e., $\log N_{\rm H, int} = 21.3$ and $\Gamma = 1.7$).  

In the Appendix, we provide the properties of \nx\ \xray\ point sources in all
\ngal\ galaxies in our sample.  Of these \xray\ sources, \nxrb\ had $L >
10^{35}$~\lum\ and were determined to lie within the galactic footprints of our sample.
The galactic footprints were taken to be the ellipses that trace the
$K_s \approx 20$~mag~arcsec$^{-2}$ galactic surface brightness (see Jarrett
\etal\ 2003), with some central regions excised due to the presence of AGN or
substantial source crowding. These detailed regions, including exclusion region
radii, $r_{\rm remove}$, are provided in Table~1.  The remaining sources were
either located outside the $K$-band based regions or within the central regions
removed from further analysis (i.e., AGN and clearly crowded sources).  We note
that a substantial number of sources that we have excluded from our XLF
analyses are outside the designated $K_s \approx 20$~mag~arcsec$^{-2}$ region,
yet within the larger ``total'' $K_s$-band ellipse, defined by Jarrett \etal\
(2003), or the generally larger RC3 regions, defined by de Vaucouleurs~(1991).
Such sources still have some reasonable probability of being associated with
the galaxy, so we report them in our \xray\ point-source catalogs; however,
their numbers are expected to be small compared with the number of CXB sources
in those areas and are therefore not included in our XLF analyses.  For
convenience, we flag sources in our \xray\ catalog that lie within the total
$K_s$-band ellipse, but outside the 20~mag~arcsec$^{-2}$ ellipse (Flag = 3). 

\subsection{Catalog Completeness Functions}

Since our \xray\ data sets span a broad range of \chandra\ depths, in terms of
intrinsic \xray\ point-source luminosity, it is essential to understand well
the completeness of each of our data sets when fitting XLF models.
To address this, we first derive radially-dependent completeness functions for
each galaxy using simulations, in which fake sources are added to the FB images
and searched for using {\ttfamily wavdetect} following the prescription adopted
in $\S$3.2.  For a given galaxy, we generated 700 simulated images in total.  Each
image consisted of our original 0.5--7~keV \chandra\ image plus 400 fake \xray\
point sources, each of which contained a fixed number of source counts.  Each
fake \xray\ source was placed randomly within the boundaries of a single box in
a $20 \times 20$ grid of boxes that spanned the image in equal intervals of
R.A. and decl.  A given simulated image would thus contain 400 fake \xray\
sources with one source per box and an equal number of \xray\ counts per
source.  Fifty simulated images were created for each of 14 different choices
of simulated source counts with nearly logarithmic spacing (spanning 3--500
source counts).  Source counts were probabilistically placed
onto the base image using the nearest {\ttfamily MARX}-based, exposure-weighted
PSF that was generated in the {\ttfamily AE} runs (see $\S$3.2) for the
original source catalog.  This method was adopted as a practical compromise
between running very accurate time-consuming PSF models for a small number of
simulated sources and having a robust characterization of the completeness
functions based on many sources with slightly inaccurate local PSFs.

To construct the completeness functions themselves we (1) repeated the
source detection procedure described in $\S$3.2 for all 700 mock images and (2)
compared the mock catalogs with the input catalogs to determine whether a given source was recovered.  In a general sense, the
completeness functions, for a given galaxy, vary with off-axis angle with
respect to the mean aim point and local background and point-source density. In
$\S$4 below, we describe how we use our completeness functions when measuring
XRB XLFs.

%
%%%%%%%%%%%%%%%%%%%%%%%%%%%%%%%%%%%%%%%%%%%%%%%%%%%%%%%%%%%%%%%%%%%%%%%%
\section{X-ray Luminosity Function Measurements}
%%%%%%%%%%%%%%%%%%%%%%%%%%%%%%%%%%%%%%%%%%%%%%%%%%%%%%%%%%%%%%%%%%%%%%%%
%

\subsection{Galaxy-Wide X-ray Luminosity Function Properties}

We began our XLF analyses by fitting the galaxy-wide 0.5--8~keV XLFs for each
of the galaxies.  As discussed above, we utilized only \xray\ point sources and
galaxy properties that are appropriate for the regions defined in Table~1,
which in some cases means excluding central regions (due to source crowding and
AGN).  In Figure~3, we display the galaxy-wide {\it observed} cumulative XLFs
({\it gray filled circles with 1$\sigma$ Poisson error bars\/}) for the
galaxies in our sample.  The data used here are simply raw counts, and not
corrected for incompleteness.  Furthermore, the \xray\ point sources will
contain contributions from objects that are intrinsic to the galaxies, but also
background \xray\ point sources from the cosmic \xray\ background (CXB; e.g.,
Kim \etal\ 2007; Georgakakis \etal\ 2008) and occasionally foreground stars
that are \xray\ detected.

We fit the observed galaxy-wide XLFs following a forward-fitting approach, in
which we include contributions from the intrinsic \xray\ sources (the vast
majority of which we expect to be XRBs) and CXB sources, with incompleteness
folded into our models.  For the {\it intrinsic} point-source XLF, we began by
fitting the data to single and broken power law models of the respective forms:
\begin{equation}
\frac{dN}{dL} = K_{\rm PL} \left \{ \begin{array}{lr} 
L^{-\alpha}, & \;\;\;\;\;\;\;\;(L < L_c) \\
0, & (L \ge L_c) 
\end{array}
  \right.
\end{equation}

\begin{equation}
\frac{dN}{dL} = K_{\rm BKNPL}  \left \{ \begin{array}{lr} L^{-\alpha_1}  &
\;\;\;\;\;\;\;\;(L < L_b) \\ 
L_b^{\alpha_2-\alpha_1} L^{-\alpha_2},  & (L_b \le L < L_c) \\ 
0,  & (L \ge L_c) \\ 
\end{array}
  \right.
\end{equation}
where $K_{\rm PL}$ and $\alpha$ are the single power-law normalization and
slope, respectively, and $K_{\rm BKNPL}$, $\alpha_1$, $L_b$, and $\alpha_2$ are
the broken power-law normalization, low-luminosity slope, break luminosity, and
high-luminosity slope, respectively; both XLF models are truncated above,
$L_c$, the cut-off luminosity.  To make the numbers more intuitive, we take
$L$, $L_b$, and $L_c$ to be in units of $10^{38}$~\lum, when quoting and
describing normalization values.  For a given galaxy, we fit the data to
determine all constants, except for the break and cut-off luminosities, which
we fix at $L_b = 10^{38}$~\lum\ and $L_c = 2 \times 10^{40}$~\lum.  Also, when
the luminosity of the 50\% completeness limit (see below for completeness
description), $L_{50}$, was larger than $0.5 \times L_b$, the fit to $\alpha_1$
was unreliable.  For these cases, $\alpha_1$ was fixed to either 1.2 or 1.6 for
galaxies that are respectively below or above sSFR~=~$10^{-10}$~yr$^{-1}$.
Similarly, in some cases, $L_{50}$ was above the $L_b$ and $\alpha_2$ was
unreliable.  For these cases, $\alpha_2$ was fixed to either 2.2 or 1.6 for
galaxies that are respectively below or above sSFR~=~$10^{-10}$~yr$^{-1}$.

In principle, we can fit for these values for each galaxy, and we have made
attempts to free these parameters; however, in most cases, $L_b$ is not well
constrained, and the best-fit value of $L_c$ often ends up being a lower limit
constraint at the highest luminosity point source for each galaxy.  We
therefore chose to fix these parameters near sample-averaged values, which we
determine in $\S$4.2 below.  There are thus three free parameters, namely,
$K_{\rm BKNPL}$, $\alpha_1$, and $\alpha_2$.

For the CXB contribution, we implemented a fixed form for the number counts,
provided by Kim \etal\ (2007).  The Kim \etal\ (2007) extragalactic
number counts provide estimates of the number of sources per unit area versus
0.5--8~keV flux.  The best-fit function follows a broken power-law distribution
with parameters derived from the combined \chandra\ Multiwavelength Project
(ChaMP) and \chandra\ Deep Field-South (CDF-S) extragalactic survey data sets
(see Table~4 of Kim \etal\ 2007).  For each galaxy, the number counts were
converted to an observed \hbox{0.5--8~keV} XLF contribution by multiplying
the number counts by the areal extent of the galaxy, as defined in Table~1, and 
converting CXB model fluxes to \xray\ luminosities, given the distance to the
galaxy.

A complete model of the observed XLF, $dN/dL({\rm obs})$, consists of the
intrinsic XLF component, $dN/dL({\rm int})$, e.g., from Equation~(4), plus the
fixed CXB curve, $dN/dL({\rm CXB})$, convolved with a galaxy-wide weighted
completeness function, $\xi(L)$, which was constructed using the
radial-dependent completeness functions calculated in $\S$4.  Specifically, $\xi(L)$ was 
calculated by statistically weighting the contributions from the model XLF at
each annulus according to the observed distributions of \xray\ point sources.
Formally, we computed $\xi(L)$ using the following relation:
\begin{equation}
\xi(L) = \sum_i \xi_{i}(L) \times w_i,
\end{equation}
where $\xi_{i}(L)$ is the completeness function for the $i$th
annular bin and $w_i$ is the fraction of total number of galaxy-wide sources
within the $i$th annuluar bin based on the observed point-source distributions.
For all galaxies, $\xi(L)$ is very close to a monotonically increasing function, although some low-level fluctuations exist due to the nature of our simulations.  For points
of reference, we quote and utilize two luminosity limits, $L_{50}$ and
$L_{90}$, which correspond to the point-source luminosity at 50\% and 90\%
completeness (i.e., $\xi(L_{50}) = 0.5$ and $\xi(L_{90}) = 0.9$).  These values
are tabulated in Table~3.

%%%%%%%%%%%%%%%%%%%%%%%%%%%%%%%%%%%%%%%%%%%%%%%%%%%%%%%%%%%%%%%%%%%%%%%%%%%%%%%%%%
% Table 3
%%%%%%%%%%%%%%%%%%%%%%%%%%%%%%%%%%%%%%%%%%%%%%%%%%%%%%%%%%%%%%%%%%%%%%%%%%%%%%%%%%
\begin{longrotatetable}
\begin{deluxetable*}{llcccccccccccccc}
\tablewidth{0pt}
\tabletypesize{\scriptsize}
\tablecaption{X-ray Luminosity Function Fits By Galaxy}
\tablehead{
 \multicolumn{1}{c}{\sc Galaxy} & \colhead{} & \colhead{}  & \colhead{} & \colhead{}  & \multicolumn{4}{c}{\sc Single Power Law$^\dagger$} & \multicolumn{5}{c}{\sc Broken Power Law$^\ddagger$} & \colhead{} & \colhead{}  \\
\vspace{-0.25in} \\
\multicolumn{1}{c}{\sc Name} &  \multicolumn{1}{c}{\sc Alt} & \colhead{} & \colhead{$\log L_{50}$}  &  \colhead{$\log L_{90}$} & \multicolumn{4}{c}{\rule{1.6in}{0.01in}} & \multicolumn{5}{c}{\rule{2.3in}{0.01in}} & \colhead{Model}  & \colhead{$\log L_{\rm X}$} \\
\vspace{-0.25in} \\
\multicolumn{1}{c}{\sc (NGC)} & \multicolumn{1}{c}{\sc Name} & \colhead{$N_{\rm src}$} & \colhead{(\lum)} & \colhead{(\lum)}  & \colhead{$K_{\rm PL}$} & \colhead{$\alpha$} & \colhead{$C$} & \colhead{$P_{\rm Null}$} & \colhead{$K_{\rm BKNPL}$} & \colhead{$\alpha_1$} & \colhead{$\alpha_2$} & \colhead{$C$} & \colhead{$P_{\rm Null}$} & \colhead{(S B)} & \colhead{(ergs~s$^{-1}$)} \\
\vspace{-0.25in} \\
\multicolumn{1}{c}{(1)} & \multicolumn{1}{c}{(2)} & \multicolumn{1}{c}{(3)} & \colhead{(4)} & \colhead{(5)} & \colhead{(6)} & \colhead{(7)} & \colhead{(8)} & \colhead{(9)} & \colhead{(10)} & \colhead{(11)} & \colhead{(12)} & \colhead{(13)}  & \colhead{(14)} & \colhead{(15)}  & \colhead{(16)}
}
\startdata
       337 &            &   6 & 38.5 & 38.7 &      6.0$^{+12.6}_{-4.6}$ &    1.49$^{+0.42}_{-0.31}$ &   22 &      0.944 &    8.08$^{+3.74}_{-3.00}$ &               1.60$^\ast$ &               1.60$^\ast$ &   23 &      0.889 & B &             40.4$\pm$0.2 \\ 
       584 &            &   7 & 38.5 & 38.7 &       193$^{+504}_{-151}$ &    3.60$^{+0.88}_{-0.99}$ &   11 &      0.686 &     21.0$^{+12.4}_{-9.2}$ &               1.20$^\ast$ &               2.20$^\ast$ &    9 &      0.169 & B &             40.0$\pm$0.2 \\ 
       628 &        M74 &  43 & 36.4 & 36.6 &    1.51$^{+0.67}_{-0.55}$ &    1.68$^{+0.16}_{-0.14}$ &   31 &      0.106 &    4.16$^{+2.16}_{-1.58}$ &             1.25$\pm$0.21 &    4.05$^{+0.66}_{-0.95}$ &   26 &      0.109 & B &             38.9$\pm$0.2 \\ 
       925 &            &   7 & 37.5 & 37.7 &    1.47$^{+1.02}_{-0.72}$ &    1.37$^{+0.26}_{-0.24}$ &   26 &      0.785 &    1.03$^{+0.82}_{-0.52}$ &    2.28$^{+1.17}_{-1.30}$ &    1.30$^{+0.27}_{-0.24}$ &   25 &      0.849 & S &             40.1$\pm$0.4 \\ 
      1097 &            &  23 & 38.0 & 38.2 &      17.4$^{+7.4}_{-5.6}$ &    1.75$^{+0.20}_{-0.18}$ &   25 &      0.261 &      28.9$^{+7.2}_{-6.3}$ &               1.20$^\ast$ &               2.20$^\ast$ &   33 &      0.387 & B &             40.1$\pm$0.1 \\ 
\\
      1291 &            &  62 & 37.1 & 37.3 &    9.26$^{+1.77}_{-1.58}$ &    1.69$^{+0.11}_{-0.09}$ &   36 &      0.162 &      21.5$^{+4.7}_{-4.1}$ &    0.90$^{+0.19}_{-0.20}$ &    2.60$^{+0.40}_{-0.31}$ &   25 &      0.173 & B &     39.7$^{+0.2}_{-0.1}$ \\ 
      1316 &            &  81 & 37.9 & 38.1 &    70.6$^{+14.6}_{-12.3}$ &    2.21$^{+0.19}_{-0.16}$ &   28 &      0.383 &    70.4$^{+15.8}_{-13.1}$ &               1.20$^\ast$ &    2.20$^{+0.19}_{-0.17}$ &   28 &      0.383 & S &     40.9$^{+0.2}_{-0.1}$ \\ 
      1404 &            &  61 & 37.6 & 37.9 &      20.7$^{+3.3}_{-3.1}$ &    1.99$^{+0.14}_{-0.12}$ &   27 &      0.061 &      19.6$^{+4.3}_{-3.8}$ &    2.09$^{+0.35}_{-0.34}$ &    1.95$^{+0.20}_{-0.17}$ &   27 &      0.047 & S &             40.4$\pm$0.1 \\ 
      2841 &            &  40 & 37.6 & 37.8 &      12.2$^{+2.9}_{-2.6}$ &    2.05$^{+0.23}_{-0.19}$ &   24 &      0.218 &      13.9$^{+4.2}_{-3.6}$ &    1.70$^{+0.58}_{-0.59}$ &    2.17$^{+0.34}_{-0.26}$ &   24 &      0.283 & S &             40.2$\pm$0.1 \\ 
      3031 &        M81 & 185 & 35.9 & 36.3 &    5.15$^{+1.07}_{-0.93}$ &             1.43$\pm$0.06 &   50 &      0.034 &      10.6$^{+2.7}_{-2.4}$ &             1.18$\pm$0.09 &    2.16$^{+0.31}_{-0.25}$ &   43 &      0.092 & B &     39.7$^{+0.3}_{-0.2}$ \\ 
\\
      3184 &            &  26 & 37.0 & 37.2 &    2.17$^{+0.87}_{-0.74}$ &    1.56$^{+0.19}_{-0.17}$ &   37 &      0.782 &    7.25$^{+3.29}_{-2.66}$ &    0.35$^{+0.35}_{-0.24}$ &    2.73$^{+0.87}_{-0.58}$ &   34 &      0.773 & B &             39.2$\pm$0.3 \\ 
      3198 &            &  11 & 37.1 & 37.3 &    1.51$^{+0.74}_{-0.61}$ &    1.45$^{+0.22}_{-0.19}$ &   30 &      0.748 &    4.01$^{+2.37}_{-1.73}$ &    0.28$^{+0.37}_{-0.20}$ &    2.23$^{+0.92}_{-0.48}$ &   28 &      0.851 & B &     39.2$^{+0.5}_{-0.4}$ \\ 
      3351 &        M95 &  38 & 36.7 & 36.9 &    2.88$^{+0.89}_{-0.76}$ &    1.59$^{+0.14}_{-0.12}$ &   23 &      0.008 &    7.92$^{+3.24}_{-2.53}$ &    0.93$^{+0.25}_{-0.27}$ &    2.78$^{+0.87}_{-0.55}$ &   21 &      0.032 & B &     39.3$^{+0.3}_{-0.2}$ \\ 
      3521 &            &  51 & 37.2 & 37.4 &    9.05$^{+1.79}_{-1.60}$ &             1.55$\pm$0.09 &   45 &      0.545 &      22.1$^{+4.7}_{-4.3}$ &    0.36$^{+0.27}_{-0.22}$ &    2.17$^{+0.23}_{-0.21}$ &   30 &      0.267 & B &             40.0$\pm$0.2 \\ 
      3627 &        M66 &  61 & 37.1 & 37.3 &    8.43$^{+1.60}_{-1.46}$ &             1.55$\pm$0.09 &   45 &      0.554 &      15.5$^{+3.7}_{-3.2}$ &    0.98$^{+0.21}_{-0.22}$ &    1.95$^{+0.20}_{-0.18}$ &   41 &      0.738 & B &             40.1$\pm$0.2 \\ 
\\
      3938 &            &  23 & 37.2 & 37.4 &    4.03$^{+1.24}_{-1.05}$ &    1.65$^{+0.17}_{-0.16}$ &   23 &      0.056 &    8.06$^{+3.24}_{-2.55}$ &             0.76$\pm$0.42 &    2.23$^{+0.54}_{-0.35}$ &   23 &      0.219 & B &             39.5$\pm$0.3 \\ 
      4125 &            &  35 & 37.7 & 37.9 &      15.8$^{+3.6}_{-3.3}$ &    2.26$^{+0.29}_{-0.23}$ &   26 &      0.458 &      20.9$^{+5.6}_{-4.9}$ &               1.20$^\ast$ &    2.44$^{+0.53}_{-0.35}$ &   28 &      0.796 & S &     40.3$^{+0.3}_{-0.1}$ \\ 
      4254 &        M99 &  32 & 37.7 & 37.9 &      14.9$^{+3.8}_{-3.4}$ &    2.02$^{+0.22}_{-0.19}$ &   16 &      0.017 &      16.4$^{+4.7}_{-4.1}$ &               1.60$^\ast$ &    2.08$^{+0.29}_{-0.24}$ &   15 &      0.019 & S &             40.3$\pm$0.1 \\ 
      4321 &       M100 &  60 & 37.1 & 37.3 &    8.18$^{+1.70}_{-1.51}$ &    1.53$^{+0.10}_{-0.09}$ &   44 &      0.363 &      17.8$^{+4.3}_{-3.7}$ &    0.71$^{+0.25}_{-0.26}$ &    2.04$^{+0.24}_{-0.19}$ &   36 &      0.399 & B &             40.1$\pm$0.2 \\ 
      4450 &            &   7 & 38.2 & 38.4 &    45.6$^{+63.0}_{-27.2}$ &    3.47$^{+1.08}_{-0.89}$ &   13 &      0.464 &      14.0$^{+6.5}_{-5.3}$ &               1.20$^\ast$ &               2.20$^\ast$ &   12 &      0.148 & B &             39.8$\pm$0.2 \\ 
\\
      4536 &            &  10 & 38.0 & 38.1 &    6.20$^{+4.41}_{-2.84}$ &    1.76$^{+0.36}_{-0.28}$ &   22 &      0.604 &    6.85$^{+4.94}_{-3.17}$ &               1.60$^\ast$ &    1.83$^{+0.41}_{-0.30}$ &   22 &      0.693 & S &     40.1$^{+0.3}_{-0.2}$ \\ 
      4552 &        M89 & 115 & 37.2 & 37.6 &      23.3$^{+2.7}_{-2.6}$ &    1.76$^{+0.08}_{-0.07}$ &   40 &      0.002 &      34.9$^{+5.1}_{-4.7}$ &             1.28$\pm$0.16 &    2.07$^{+0.15}_{-0.14}$ &   35 &      0.068 & B &             40.3$\pm$0.1 \\ 
      4559 &            &   5 & 37.5 & 37.6 &    0.62$^{+0.66}_{-0.39}$ &    1.17$^{+0.28}_{-0.27}$ &   20 &      0.668 &    0.74$^{+0.80}_{-0.46}$ &    0.90$^{+0.90}_{-0.61}$ &    1.25$^{+0.32}_{-0.29}$ &   20 &      0.768 & S &     40.1$^{+0.4}_{-0.5}$ \\ 
      4569 &            &  26 & 37.7 & 37.8 &    8.85$^{+2.90}_{-2.50}$ &    2.09$^{+0.29}_{-0.26}$ &   20 &      0.132 &      12.1$^{+4.4}_{-3.7}$ &    1.13$^{+0.78}_{-0.69}$ &    2.46$^{+0.60}_{-0.39}$ &   19 &      0.255 & B &     39.7$^{+0.3}_{-0.2}$ \\ 
      4594 &       M104 & 192 & 36.8 & 37.1 &      21.3$^{+2.2}_{-2.1}$ &             1.59$\pm$0.05 &   59 &      0.707 &      48.0$^{+5.5}_{-5.2}$ &    1.06$^{+0.07}_{-0.08}$ &    2.45$^{+0.20}_{-0.17}$ &   22 &      0.010 & B &             40.2$\pm$0.1 \\ 
\\
      4725 &            &  36 & 37.3 & 37.5 &    5.57$^{+1.60}_{-1.38}$ &    1.72$^{+0.17}_{-0.15}$ &   31 &      0.274 &      10.1$^{+3.8}_{-3.1}$ &             1.01$\pm$0.38 &    2.32$^{+0.59}_{-0.36}$ &   30 &      0.694 & B &     39.6$^{+0.3}_{-0.2}$ \\ 
      4736 &        M94 &  71 & 36.5 & 36.8 &    4.97$^{+1.02}_{-0.88}$ &    1.42$^{+0.08}_{-0.07}$ &   55 &      0.554 &    8.82$^{+2.54}_{-2.09}$ &    1.13$^{+0.13}_{-0.14}$ &    1.81$^{+0.21}_{-0.18}$ &   52 &      0.842 & B &             40.1$\pm$0.3 \\ 
      4826 &        M64 &  33 & 37.0 & 37.2 &    2.96$^{+1.05}_{-0.89}$ &    1.51$^{+0.17}_{-0.14}$ &   25 &      0.044 &      10.2$^{+3.6}_{-3.1}$ &    0.29$^{+0.27}_{-0.20}$ &    2.55$^{+0.60}_{-0.43}$ &   17 &      0.023 & B &     39.4$^{+0.3}_{-0.2}$ \\ 
      5033 &            &  24 & 37.7 & 37.9 &      12.5$^{+3.1}_{-2.8}$ &    2.19$^{+0.27}_{-0.24}$ &   30 &      0.884 &      16.3$^{+5.4}_{-4.5}$ &             1.49$\pm$0.74 &    2.60$^{+0.91}_{-0.45}$ &   31 &      0.424 & B &     39.8$^{+0.4}_{-0.2}$ \\ 
      5055 &        M63 &  61 & 37.1 & 37.3 &    7.88$^{+1.60}_{-1.40}$ &             1.59$\pm$0.10 &   34 &      0.101 &      12.6$^{+3.3}_{-2.8}$ &    1.17$^{+0.21}_{-0.22}$ &    1.91$^{+0.21}_{-0.18}$ &   33 &      0.215 & B &             40.1$\pm$0.2 \\ 
\\
      5194 &        M51 & 237 & 36.3 & 36.6 &      10.1$^{+1.5}_{-1.4}$ &             1.59$\pm$0.05 &   49 &      0.034 &      11.4$^{+2.4}_{-2.1}$ &             1.55$\pm$0.07 &    1.71$^{+0.15}_{-0.13}$ &   48 &      0.062 & B &             40.4$\pm$0.2 \\ 
      5236 &        M83 & 363 & 35.9 & 36.2 &    8.94$^{+1.42}_{-1.31}$ &    1.56$^{+0.05}_{-0.04}$ &   57 &      0.073 &      12.0$^{+2.3}_{-2.1}$ &    1.47$^{+0.06}_{-0.05}$ &    1.93$^{+0.22}_{-0.18}$ &   54 &      0.182 & B &             40.1$\pm$0.2 \\ 
      5457 &       M101 & 174 & 36.1 & 36.3 &    3.90$^{+1.07}_{-0.93}$ &             1.62$\pm$0.08 &   38 &      0.019 &    6.14$^{+2.45}_{-1.86}$ &    1.47$^{+0.12}_{-0.13}$ &    2.41$^{+0.78}_{-0.45}$ &   37 &      0.088 & B &     39.4$^{+0.3}_{-0.2}$ \\ 
      5713 &            &  15 & 38.3 & 38.7 &     10.2$^{+10.0}_{-5.7}$ &             1.49$\pm$0.24 &   29 &      0.456 &      13.9$^{+4.4}_{-3.8}$ &               1.60$^\ast$ &               1.60$^\ast$ &   30 &      0.638 & S &             40.7$\pm$0.2 \\ 
      5866 &       M102 &  36 & 37.4 & 37.6 &    8.08$^{+1.95}_{-1.72}$ &    1.95$^{+0.18}_{-0.17}$ &   26 &      0.214 &      19.6$^{+5.1}_{-4.4}$ &    0.64$^{+0.37}_{-0.34}$ &    3.30$^{+0.69}_{-0.53}$ &   17 &      0.188 & B &             39.5$\pm$0.1 \\ 
\\
      6946 &            & 115 & 36.4 & 36.7 &    6.01$^{+1.23}_{-1.07}$ &             1.49$\pm$0.07 &   53 &      0.289 &    8.99$^{+2.49}_{-2.08}$ &    1.30$^{+0.12}_{-0.11}$ &    1.78$^{+0.20}_{-0.18}$ &   52 &      0.552 & B &             40.1$\pm$0.3 \\ 
      7331 &            &  95 & 37.2 & 37.5 &      18.3$^{+2.4}_{-2.3}$ &    1.65$^{+0.08}_{-0.07}$ &   56 &      0.710 &      28.7$^{+4.7}_{-4.3}$ &    1.06$^{+0.19}_{-0.20}$ &    1.92$^{+0.14}_{-0.13}$ &   50 &      0.703 & B &             40.4$\pm$0.2 \\ 
      7552 &            &  14 & 37.4 & 37.6 &    2.95$^{+1.25}_{-0.98}$ &    1.43$^{+0.19}_{-0.16}$ &   28 &      0.190 &    4.72$^{+2.32}_{-1.70}$ &    0.43$^{+0.50}_{-0.31}$ &    1.67$^{+0.28}_{-0.23}$ &   27 &      0.351 & B &     40.0$^{+0.3}_{-0.4}$ \\ 
\enddata
\tablecomments{All fits include the effects of incompleteness and model contributions from the CXB, following Eqn.~(7).  A full description of our model fitting procedure is outlined in $\S$4.1.  Col.(1) and (2): Galaxy NGC and Messier name, as reported in Table~1.  Col.(3): Total number of \xray\ sources detected within the galactic boundaries defined in Table~1.  Col.(4) and (5): Logarithm of the luminosities corresponding to the respective 50\% and 90\% completeness limits. Col.(6) and (7): Median and 1$\sigma$ uncertainty values of the single power-law normalization and slope, respectively (see eqn.~(4)) -- our adopted ``best model'' consists of the median values.  Col.(8): C-statistic, $C$, associated with the best model. Col.(9): Null-hypothesis probability of the best model describing the data.  The null-hypothesis probability is calculated following the prescription in Kaastra~(2017) and is appropriate for the use of the C statistic. Col.(10)--(12):  Median and 1$\sigma$ uncertainty values of the single power-law normalization and slope, respectively (see eqn.~(5)). Col.(13) and (14): Respectively, C-statistic and null-hypothesis probability for the best broken power-law model.  Col.(15): Adopted model, used to calculate integrated \xray\ luminosity.  Here, ``S'' and ``B'' are the single and broken power-law models, respectively. Col.(16): integrated \xray\ luminosity, $L_{\rm X}$, from equation~(9) for the adopted model. \\
$^\ast$Parameter was fixed due to shallow \chandra\ depth.\\
$^\dagger$Single power-law models are derived following Eqn.~(4) with a fixed cut-off luminosity of $L_c = 5 \times 10^{40}$~\lum.\\
$^\ddagger$Broken power-law models are derived following Eqn.~(5) with a fixed break luminosity of $L_b = 10^{38}$~\lum\ and cut-off luminosity of $L_c = 5 \times 10^{40}$~\lum.\\
}
\end{deluxetable*}
\end{longrotatetable}

We thus modeled the observed XLF using a multiplicative model
\begin{equation}
dN/dL({\rm obs}) = \xi(L) [dN/dL({\rm int}) + dN/dL({\rm CXB})].
\end{equation}
Procedurally, for each galaxy, we constructed the observed $dN/dL({\rm obs})$
using luminosity bins of constant $\delta \log L = 0.057$~dex that spanned the range of
\hbox{$L_{\rm min} = L_{50}$} to 
$L_{\rm max} = 5 \times 10^{41}$~\lum.  For most galaxies, the majority of the bins contained zero
sources, with other bins containing small numbers of sources.  As such, we
evaluated the goodness of fit using a modified version of the C-statistic
({\ttfamily cstat}; Cash~1979; Kaastra~2017):
\begin{equation}
C = 2 \sum_{i=1}^{n} M_i - N_i + N_i \ln(N_i/M_i),
\end{equation}
where the summation takes place over the $n=100$ bins of \xray\ luminosity, and
$N_i$ and $M_i$ are the observed and model counts.  We note that when $N_i =
0$, $N_i \ln (N_i/M_i) = 0$, and when $M_i=0$ (e.g., beyond the cut-off
luminosity), the entire $i$th term in the summation is zero.

%
%%%%%%%%%%%%%%%%%%%%%%%%%%%%%%%%%%%%%%%%%%%%%%%%%%%%%%%%%%%%%%%%%%%%%%%%%%%%%%%%%%
% Figure 4
%%%%%%%%%%%%%%%%%%%%%%%%%%%%%%%%%%%%%%%%%%%%%%%%%%%%%%%%%%%%%%%%%%%%%%%%%%%%%%%%%%
%
\begin{figure*}
\figurenum{4}
\centerline{
\includegraphics[width=17cm]{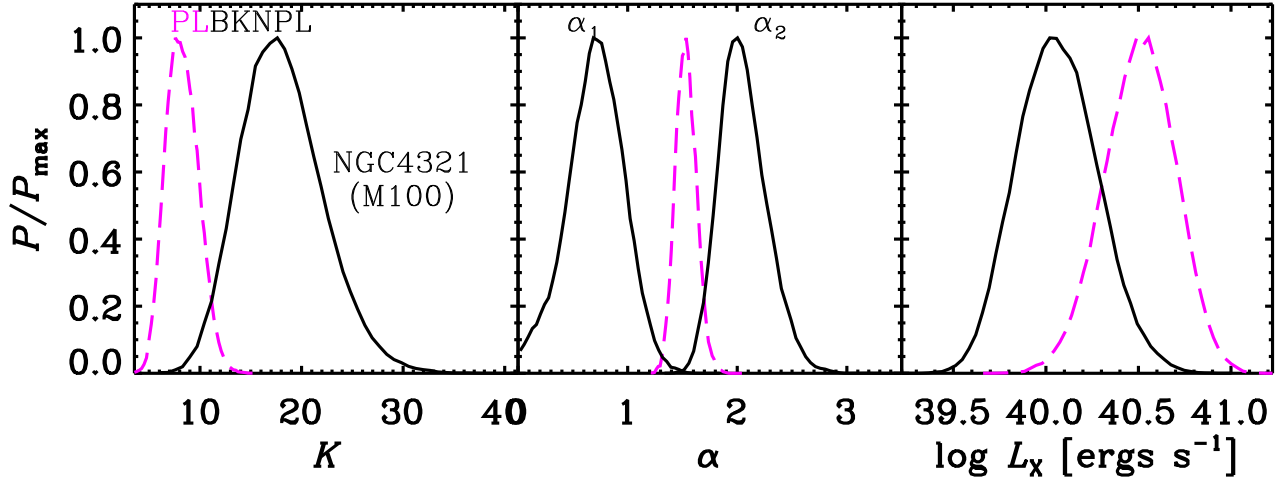}
}
\caption{
%%%
Example probability distribution functions (PDFs) for single ({\it magenta dashed curves\/}) and
broken ({\it black solid curves\/}) power-law parameters, based on fits to the
XLF of NGC~4321 (M100).  The normalization ($K$), XLF slopes, and integrated
point-source luminosity, $L_{\rm X}$, are displayed here.  The data and
best-fit models are shown in Figure~3.  
%%%
}
\end{figure*}
%%%%%%%%%%%%%%%%%%%%%%%%%%%%%%%%%%%%%%%%%%%%%%%%%%%%%%%%%%%%%%%%%%%%%%%%%%%%%%%%%%

When fitting our data and measuring uncertainties on parameters, we made use of
a Markov Chain Monte Carlo (MCMC) procedure that implemented the
Metropolis-Hastings sampling algorithm (Hastings~1970).  In this procedure, the
fitting parameters were first given initial guesses, which we took to be the
same set of values for every galaxy.  The value of {\ttfamily cstat}, $C_{\rm
init}$, was computed for this initial guess, and stored.  Next, the guesses
were perturbed randomly in accordance with a Gaussian distribution with a
user-supplied set of standard deviations for each parameter.  To begin, we
chose the widths of the Gaussians to be large (relative to their likely final
distributions) so as to sample parameter space well.  The {\ttfamily cstat}
value of the model with perturbed parameters was then computed, $C_{\rm pert}$,
and compared with the value obtained from the previous run and the likelihood
ratio, $\mathcal{L}_{\rm rat} = \exp\{-(C_{\rm pert}-C_{\rm init})/2\}$, was
evaluated.  Next, a random number, $A_{\rm random}$, between 0 and 1, was drawn
and compared with $\mathcal{L}_{\rm rat}$.  If $\mathcal{L}_{\rm rat} > A_{\rm
random}$, then the new set of parameters was stored, and if $\mathcal{L}_{\rm
rat} \le A_{\rm random}$, then the old set of parameters was preserved for
subsequent perturbations.  Using the current set of stored parameters, the
above procedure (i.e., perturbation of parameters, evaluation of
$\mathcal{L}_{\rm rat}$, and comparison with $A_{\rm random}$) was then
repeated 100,000 times, with each iteration using only accepted parameters, to
form an initial MCMC chain.  

After the 100,000 iterations, we used the initial MCMC chain to compute updated
standard deviations of the accepted values, and subsequently ran an additional
900,000 final MCMC iterations, using these standard deviations and the final
set of parameters in the initial MCMC chain as a starting point.  The
distributions of parameter values from the final MCMC chain formed our
probability distribution functions (PDFs).  
Furthermore, additional
model-dependent calculated parameter PDFs can be computed by storing their values in MCMC
chains.  For example, for each model in the MCMC chain, we compute the integrated 0.5--8~keV
luminosity, $L_{\rm X}$:
\begin{equation}
L_{\rm X} \equiv \int_{L_{\rm lo}}^{L_c} \frac{dN}{dL} L dL,
\end{equation}
where we adopt a lower integration limit of $L_{\rm lo} = 10^{36}$~\lum.

We note that for a single power-law model, PDFs
can be computed with ease using grid-based sampling of the 2D parameter space
(i.e., normalization and slope of the power law).  
We compared PDFs that were
computed from such grid-based sampling with those obtained from our MCMC
procedure and found essentially identical PDFs.  Since we later incorporate
more complex models, with up to 7 free parameters ($\S$4.2 below), where the
computation time is too large to use a grid-based approach, we chose to use the
MCMC procedure consistently throughout this paper.

%
%%%%%%%%%%%%%%%%%%%%%%%%%%%%%%%%%%%%%%%%%%%%%%%%%%%%%%%%%%%%%%%%%%%%%%%%%%%%%%%%%%
% Figure 5
%%%%%%%%%%%%%%%%%%%%%%%%%%%%%%%%%%%%%%%%%%%%%%%%%%%%%%%%%%%%%%%%%%%%%%%%%%%%%%%%%%
%
\begin{figure}
\figurenum{5}
\centerline{
\includegraphics[width=8cm]{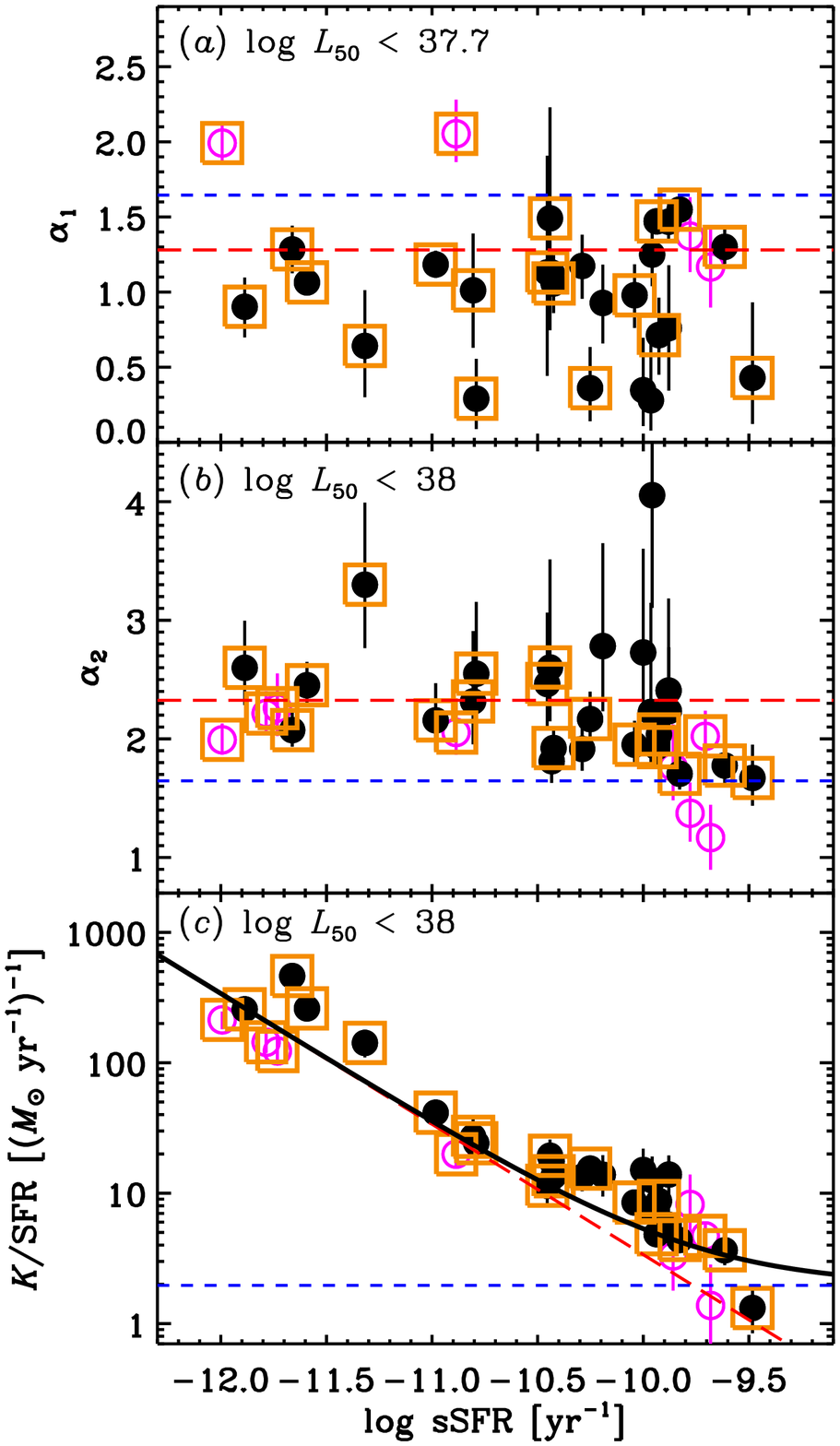}
}
\caption{
%%%
Best-fit XLF parameter values versus sSFR for the galaxy-wide XLF fits.  Solid
and open-magenta symbols indicate parameter values are determined from fits to broken
and single power-law models, respectively.  Orange squares highlight galaxies with
$M_\star > 2 \times 10^{10}$~\msol\ or SFR~$> 2$~\sfr\ to indicate sources that
are least likely to suffer from variance due to poor sampling of the XLF.  For
$\alpha_1$, we plot only objects with faintest sources $L_{\rm 50} < 5 \times
10^{37}$~\lum, where this parameter can be constrained.  Similarly, we only
display $\alpha_2$ and $K$ constraints for galaxies with $L_{\rm 50} <
10^{38}$~\lum. For cases where the single power-law fit was used, values of
$\alpha_1$ and $\alpha_2$ are set to $\alpha$.  In each panel, the trends for
HMXBs ({\it blue short-dashed\/}) and LMXBs ({\it red long-dashed\/}) are
displayed, based on the global model fit presented in $\S$4.2, and their
combined trends are shown with black solid curves.
%%%
}
\end{figure}
%%%%%%%%%%%%%%%%%%%%%%%%%%%%%%%%%%%%%%%%%%%%%%%%%%%%%%%%%%%%%%%%%%%%%%%%%%%%%%%%%%

%
%%%%%%%%%%%%%%%%%%%%%%%%%%%%%%%%%%%%%%%%%%%%%%%%%%%%%%%%%%%%%%%%%%%%%%%%%%%%%%%%%%
% Figure 6
%%%%%%%%%%%%%%%%%%%%%%%%%%%%%%%%%%%%%%%%%%%%%%%%%%%%%%%%%%%%%%%%%%%%%%%%%%%%%%%%%%
%
\begin{figure*}
\figurenum{6}
\centerline{
\includegraphics[width=18cm]{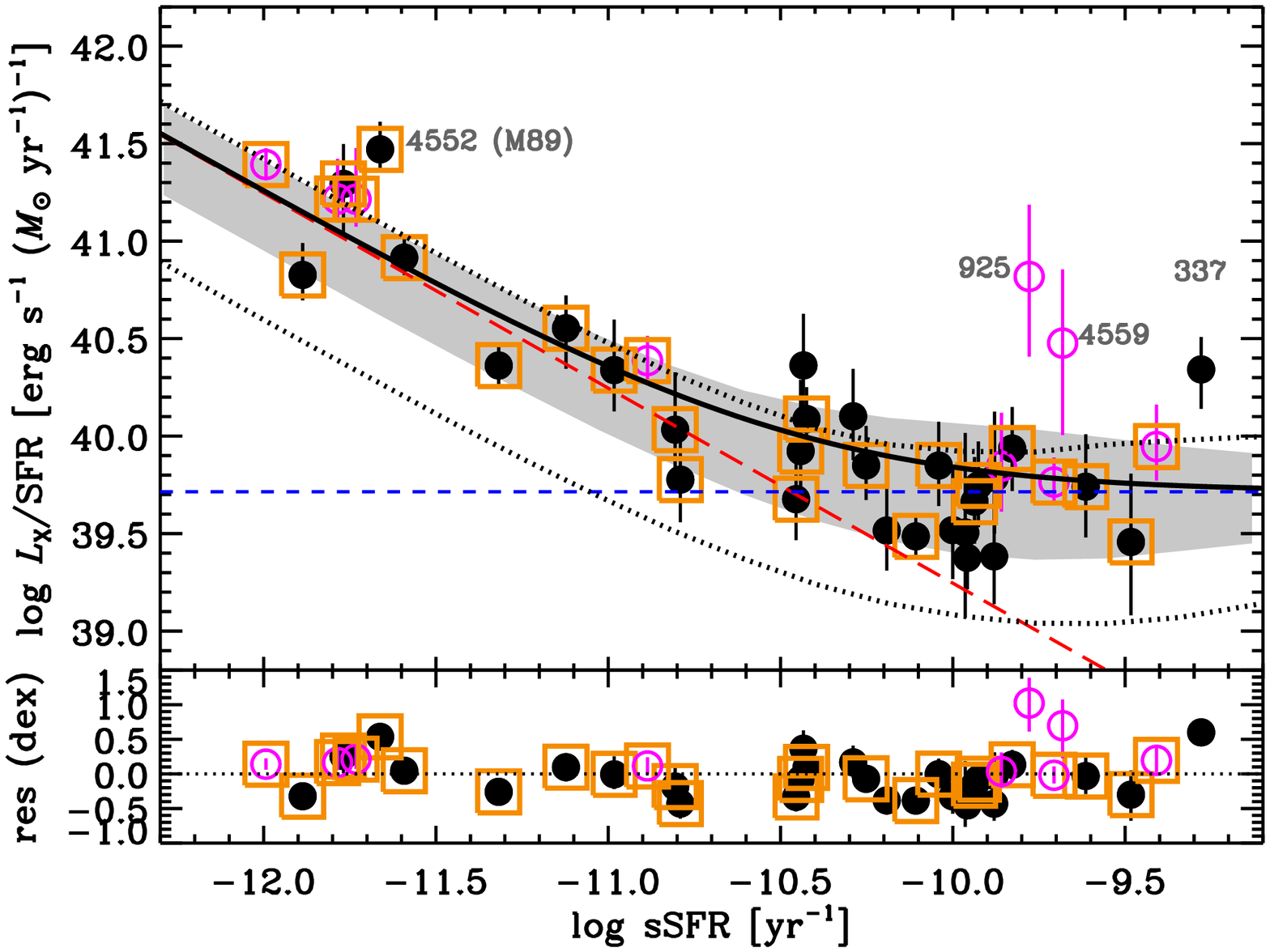}
}
\caption{
%%%
({\it top\/}) Integrated point-source 0.5--8~keV luminosity per unit SFR,
$L_{\rm X}$/SFR, versus sSFR for the fits to the individual galaxies, as
described in $\S$4.1.  Symbols have the same meaning as they did in Figure~5.
Predictions from our best-fit global model ({\it black solid curve\/}; see
$\S$4.2), along with its contributions from LMXBs ({\it red long-dashed
curve\/}) and HMXBs ({\it blue short-dashed curve\/}), are displayed and
residuals between individual galaxy $L_{\rm X}$ and global model prediction is
plotted in the bottom panel.  The gray shaded region shows the expected
1$\sigma$ scatter due to XLF sampling for galaxies with the median mass
$M_\star = 2 \times 10^{10}$~\msol\ (see $\S$5.3 for details); galaxies above this limit are
highlighted with orange squares.  The dotted curves in the top panel show the
expected 1$\sigma$ scatter for galaxies with stellar masses
equal to $3 \times 10^9$~\msol; 95\% of our galaxies are above this limit.  In general,
galaxy-to-galaxy scatter is comparable to that expected from XLF sampling;
however, notable exceptions at low-sSFR (e.g., NGC~4552) and high-sSFR (e.g., NGC~337, 925, and 4559) are observed.
%%%
}
\end{figure*}
%%%%%%%%%%%%%%%%%%%%%%%%%%%%%%%%%%%%%%%%%%%%%%%%%%%%%%%%%%%%%%%%%%%%%%%%%%%%%%%%%%

Hereafter, when quoting best-fit parameter values and uncertainties, we adopt
median values from each PDF with 16\% and 84\% confidence lower and upper
limits.  In Table~3, we tabulate the best-fit parameter values for the single
and broken power-law fits for each each galaxy.  In Figure~3, we show the
best-fit single ({\it magenta dashed curves\/}) and broken ({\it black solid
curves\/}) power-law model cumulative XLFs, which include contributions from
the CXB ({\it green dotted curves\/}) and have incompleteness folded in.
Goodness of fit was evaluated following the methods outlined in Kaastra~(2017),
which provides parameterizations of the expected $C$ statistic and its
variance for a given model and data binning scheme, so that goodness of fit can
be evaluated in an identical way to classical $\chi^2$ fitting.  For each of
our fits, the null hypothesis probability, $P_{\rm null}$, was calculated as
the one minus the probability that the model can be rejected.  The values of
$P_{\rm null}$ are listed in Table~3 for both models.  

For many galaxies, a single power law provides a statistically acceptable fit
to the data (e.g., $P_{\rm null} > 0.01$), with only one of the fits
being rejected at the $>$99.9\% confidence level ($P_{\rm null} < 0.001$).  
The majority of
the poorest fit cases (e.g., $P_{\rm null} < 0.05$) have a large number of sources detected, due to deep observational
data sets.  Visual inspection of the fits suggest that 
some complex structures within the XLFs themselves are not described well with power-laws. 
Not surprisingly, the broken power-law model provides
improvements to the {\ttfamily cstat} values of the XLFs for many cases; however, in very few cases
are the fit improvements statistically significant.  

%
%%%%%%%%%%%%%%%%%%%%%%%%%%%%%%%%%%%%%%%%%%%%%%%%%%%%%%%%%%%%%%%%%%%%%%%%%%%%%%%%%%
% Figure 7
%%%%%%%%%%%%%%%%%%%%%%%%%%%%%%%%%%%%%%%%%%%%%%%%%%%%%%%%%%%%%%%%%%%%%%%%%%%%%%%%%%
%
\begin{figure*}
\figurenum{7}
\centerline{
\includegraphics[width=18cm]{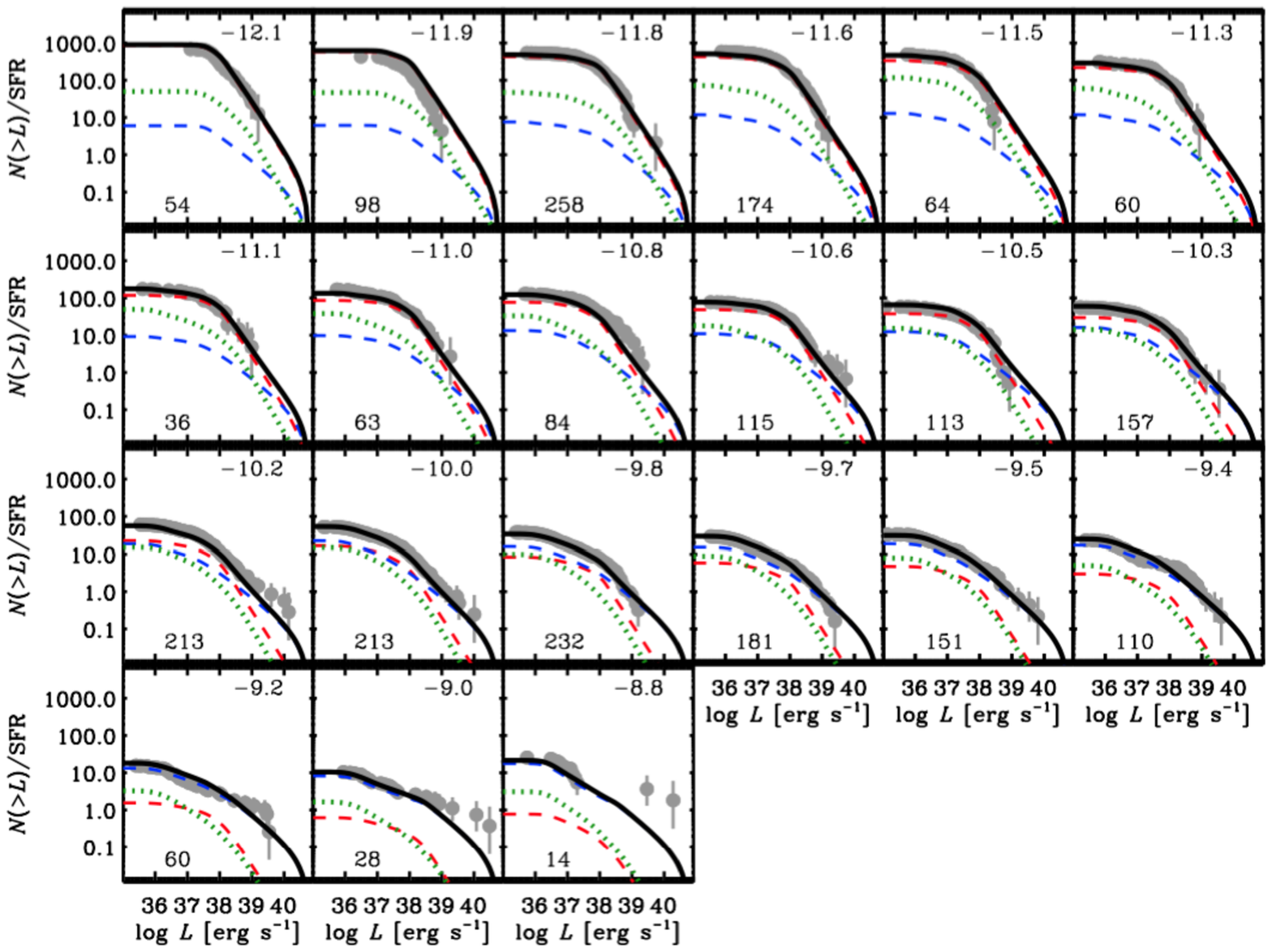}
}
\caption{
%%%
Observed SFR-normalized cumulative XLFs as a function of sSFR.  In each panel, the
observed XLFs ({\it gray circles} with 1$\sigma$ error bars) were generated by
culling \xray\ sources from subgalactic regions within all the galaxies in our
sample that had $\log$~sSFR~(yr$^{-1}$) values annotated in the upper
right-hand corner panel.  The numbers of \xray\ detected sources are annotated in the lower left-hand corners of each panel. The XLFs were normalized by the cumulative SFR from
the sSFR-selected subgalactic regions, as described in $\S$3.2.  Our best-fit
global model is shown with black solid curves, and the contributions from the
CXB ({\it green dotted curves\/}), LMXBs ({\it red long-dashed curves\/}), and
HMXBs ({\it blue short-dashed curves\/}) are included.  From the upper left
panel to the lower-right panel, the SFR-normalized XLFs both decline in
normalization and become shallower in slope, as the population shifts from LMXB
to HMXB dominated.
%%%
}
\end{figure*}
%%%%%%%%%%%%%%%%%%%%%%%%%%%%%%%%%%%%%%%%%%%%%%%%%%%%%%%%%%%%%%%%%%%%%%%%%%%%%%%%%%

%%%%%%%%%%%%%%%%%%%%%%%%%%%%%%%%%%%%%%%%%%%%%%%%%%%%%%%%%%%%%%%%%%%%%%%%%%%%%%%%%%
% Table 4
%%%%%%%%%%%%%%%%%%%%%%%%%%%%%%%%%%%%%%%%%%%%%%%%%%%%%%%%%%%%%%%%%%%%%%%%%%%%%%%%%%
\begin{table*}
{\footnotesize
\begin{center}
\caption{Best Fit Parameters for Global Fits}
\begin{tabular}{lcccccc}
\hline\hline
\multicolumn{1}{c}{\sc Parameter} &  & {\sc First} & {\sc Second} & {\sc Cleaned} & {\sc Full}  & {\sc M12/Z12} \\
\multicolumn{1}{c}{\sc Name} & {\sc Units} & {\sc Subsample} & {\sc Subsample} & {\sc Sample} & {\sc Sample} & {\sc Value} \\
\multicolumn{1}{c}{(1)} & (2) & (3) & (4) & (5) & (6) & (7) \\
\hline\hline
SFR  & \sfr\ & 18.5 & 26.9 & 40.9 & 45.4 & \\ 
$M_\star$  & $10^{11}$~\msol\ & 7.97 & 4.50 & 10.21 & 12.47 & \\ 
$\log$~sSFR  & $\log$~yr$^{-1}$ & $-$10.63 & $-$10.22 & $-$10.40 & $-$10.44 & \\ 
$N_{\rm det}$  &  &852 & 1626 & 2071 & 2478  & \\
\hline
\multicolumn{7}{c}{Parameter Fit Values} \\
\hline
$K_{\rm LMXB}$ & ($10^{11}$~\msol)$^{-1}$ & 32.3$^{+5.7}_{-5.7}$ & 39.6$^{+2.1}_{-2.0}$ & 26.0$^{+3.4}_{-2.4}$ & 33.8$^{+7.3}_{-3.6}$ & $41.5 \pm 11.5$ \\
$\alpha_1$ &  & 1.21$^{+0.08}_{-0.08}$ & 1.31$^{+0.03}_{-0.04}$ & 1.31$^{+0.05}_{-0.07}$ & 1.28$^{+0.06}_{-0.09}$ & 1.02$^{+0.07}_{-0.08}$ \\
$L_b$ & $10^{38}$~\lum  & 0.77$^{+0.39}_{-0.16}$ & 3.27$^{+0.51}_{-0.55}$ &2.16$^{+1.39}_{-0.71}$ &1.48$^{+0.70}_{-0.66}$ & $0.546^{+0.043}_{-0.037}$ \\
$\alpha_2$ &  &  2.15$^{+0.15}_{-0.11}$ & 3.15$^{+0.56}_{-0.42}$ & 2.57$^{+0.54}_{-0.28}$ & 2.33$^{+0.27}_{-0.21}$ & 2.06$^{+0.06}_{-0.05}$ \\
$L_{b,2}^\dagger$ & $10^{38}$~\lum  &  & & & & $5.99^{+0.95}_{-0.67}$ \\
$\alpha_3^\dagger$ &  &  & & & & $3.63^{+0.67}_{-0.49}$ \\
$K_{\rm HMXB}$ & (\sfr)$^{-1}$ & 2.43$^{+0.27}_{-0.27}$ & 1.48$^{+0.14}_{-0.14}$ & 2.06$^{+0.16}_{-0.15}$ & 1.96$^{+0.14}_{-0.14}$ & $2.68 \pm 0.13$ \\
$\gamma$ &  & 1.53$^{+0.05}_{-0.05}$ & 1.71$^{+0.03}_{-0.03}$ & 1.66$^{+0.02}_{-0.02}$ & 1.65$^{+0.03}_{-0.02}$ & $1.58 \pm 0.02$ \\
$\log Lc$ & $\log$ \lum\  & 40.5$^{+0.4}_{-0.1}$ & 41.0$^{+0.5}_{-0.3}$ & 40.8$^{+0.5}_{-0.2}$ & 40.7$^{+0.4}_{-0.2}$ & $40.04^{+0.18}_{-0.16}$\\
$C$ &  & 1014 & 1185 & 1331 & 1410 & \ldots \\
$P_{\rm null}$ &  & 0.705 & 0.017 & 0.177 & 0.145 & \ldots \\
\hline
\multicolumn{7}{c}{Calculated Parameters} \\
\hline
$\log \alpha_{\rm LMXB}$ & $\log$ \lum~$M_\odot^{-1}$  & 29.14$^{+0.07}_{-0.06}$ & 29.31$^{+0.05}_{-0.04}$ & 29.15$^{+0.07}_{-0.05}$ & 29.25$^{+0.07}_{-0.06}$ & $29.2 \pm 0.1$ \\
$\log \beta_{\rm HMXB}$ & $\log$ \lum~(\sfr)$^{-1}$  & 39.89$^{+0.15}_{-0.11}$ & 39.56$^{+0.15}_{-0.13}$ & 39.73$^{+0.15}_{-0.10}$ & 39.71$^{+0.14}_{-0.09}$ & $39.67 \pm 0.06$ \\
\hline
\end{tabular}
\end{center}
Note.---Col.(1) and (2): Parameter and units. Col.(3)--(6): Value of each parameter for the first subsample, second subsample, ``cleaned' sample, and full sample of sources.  The two subsamples represent fits based on simply dividing the full sample in half, when ordered by NGC name.  The first and second subsamples include NGC~337--4321 and NGC~4450--7552, respectively.  The cleaned sample excludes galaxies with low metallicity (NGC~337, 925, 3198, 4536, and 4559) and galaxies with relatively large GC $S_N$ (NGC~1404, 4552, and 4594). Col.(7): Comparison values of HMXB and LMXB scaling relations from M12 and Z12, respectively.\\
$^\dagger$Parameter was used in Z12, but not in our study.\\
}
\end{table*}

Despite the lack of statistical improvement, we expect that in most cases, the
broken power-law fits provide more realistic estimates of the integrated total
luminosity, $L_{\rm X}$, than the single power-law fits.  One clear example where the solutions are notably different 
is illustrated in Figure~4 for NGC~4321 (M100).  While statistically, the
single and broken power-law fits have very close $P_{\rm null}$ values to each other,
the overall $C$ is notably improved by the broken power-law fit and the
calculated $L_{\rm X}$ values are substantially different between models.  We
note that this is an extreme case, and that most galaxies have better agreement
between $L_{\rm X}$ values when both models are statistically acceptable.
We therefore chose to adopt parameters derived using the broken power-law
model, unless either (1) the $C$ value for the broken power law provided no
improvement over the single power-law value or (2) the two slopes implied by
the broken power-law (i.e., $\alpha_1$ and $\alpha_2$) were within 1$\sigma$ of
each other.  In Table~3, we indicate our adopted model and list $L_{\rm X}$
based on that model.

In Figure~5, we show the best-fit XLF parameter values versus sSFR for all
galaxies in our sample.  In terms of trends, $\alpha_1$ is consistent with
being constant across all sSFR values, suggesting little variation in the
low-luminosity slope of the XLF for young versus old populations.  $\alpha_2$,
on the other hand, exhibits an average decline with increasing sSFR (Spearman's
$\rho$ correlation significance $>$99.95\% confidence level), presumably
indicating that as the XRB population transitions from LMXBs to HMXBs.  If we
restrict the sample to massive galaxies ($\simgt$$2 \times 10^{10}$~\msol) or
galaxies with substantial SFRs ($\simgt$2~\sfr), so that the respective LMXB
and HMXB population statistics allow for less galaxy-to-galaxy sampling
stochasticity (e.g., Gilfanov~2004; Justham \& Schawinski~2012; see $\S$5.3 below), we get a clearer sense
of this trend (see orange boxes in Fig.~5).  Finally, we find that the
normalization per unit SFR declines with increasing sSFR, as would be expected
as the population shifts from being LMXB dominated at low-sSFR to more HMXB
dominated at high-sSFR.

In Figure~6, we show $L_{\rm X}$/SFR versus sSFR for the sample.  As reported
by previous authors, this curve shows a clear decline of $L_{\rm X}$/SFR with
increasing sSFR, due to the transition from LMXBs to
HMXBs (e.g., Colbert \etal\ 2004; Lehmer \etal\ 2010).    From
Figure~5, it can be inferred that this trend is largely driven by the decline
in normalization per unit SFR of the XLF.  However, for galaxies where the XLFs
are expected to be well sampled (i.e., the orange squares in Figs.~5 and 6), we
find a larger range in $K$/SFR than $L_{\rm X}$/SFR, due to the fact
that the high-luminosity-end XLF slope ($\alpha_2$) becomes shallower for galaxies with high-sSFR (Fig.~5$b$),
due to the relatively shallow-sloped HMXB XLF becoming more dominant
(e.g., Grimm \etal\ 2003; M12).

\subsection{Global Fit to Specific-SFR Binned Regions}

As discussed above, it is expected that the decline in $L_{\rm X}$/SFR with
sSFR is driven by a transition from LMXB to HMXB dominance, and the rate of
decline is affected by changes in {\it both} XLF normalizations and slopes.
Here we examine XLFs in subgalactic regions, selected from the SFR and
$M_\star$ maps discussed in $\S$3.1, to better isolate XRB populations as a
function of sSFR, and decompose the XLFs into the SFR-scaled HMXB and
$M_\star$-scaled LMXB components.  Hereafter, we make the assumption that the
\xray\ point source population that is not part of the CXB is dominated by
XRBs; however, we note that there will be some contribution from other sources,
in particular supernova remnants (SNR) and Galactic stars.  Unfortunately, a
clean identification of the nature of every point source in our catalog is
beyond the scope of this work.  However, we expect that the contributions of
these sources to the XLFs will be smaller than CXB sources (see, e.g.,
Fig.~10 of Long \etal\ 2014 for M83), and will therefore not have a major
impact on our conclusions.

To address the above goal, we began by generating {\it local sSFR maps} on the
pixel scale of our SFR and $M_\star$ maps.  For each pixel, we computed the
total SFR and $M_\star$ within a square $500 \times 500$~pc$^2$ region,
centered on the pixel.  Such pixels have sizes of $3.5 \times
3.5$~arcsec$^2$~pixel$^{-1}$ for the most distant galaxy in the sample,
NGC~5713, to $29.5 \times 29.5$~arcsec$^2$~pixel$^{-1}$ for the nearest galaxy,
M81.  Thus each pixel can be used to signify the ``local'' conditions
surrounding a given location, all on the same physical scale.  Using these
maps, we sorted all pixels for all galaxies into bins of sSFR with bin width, or ``resolution,'' of $\Delta
\log$~sSFR~=~0.16~dex, which is the root-mean-square error on the SFR and
$M_\star$ calibration uncertainties (see $\S$2 for details).  For the lowest
and highest sSFR bins, we required at least one \xray\ source be detected
within and placed no limits on the respective lower and upper bounds for the
inclusion of sSFR pixels in those bins.  In total, we identified \nssfr\ sSFR bins,
continuously covering the sSFR range from $\approx$$2.5 \times
10^{-13}$~yr$^{-1}$ to $\approx$$1.6 \times 10^{-9}$~yr$^{-1}$.  The bins
contain between 14 and 260 \xray\ sources per bin.
For
each of the sSFR bins, we selected all pixels within the galactic regions
(defined in Table~1) that were within the sSFR range of that bin, and calculated
the total SFR and $M_\star$ corresponding to those pixels.

%
%%%%%%%%%%%%%%%%%%%%%%%%%%%%%%%%%%%%%%%%%%%%%%%%%%%%%%%%%%%%%%%%%%%%%%%%%%%%%%%%%%
% Figure 8
%%%%%%%%%%%%%%%%%%%%%%%%%%%%%%%%%%%%%%%%%%%%%%%%%%%%%%%%%%%%%%%%%%%%%%%%%%%%%%%%%%
%
\begin{figure*}
\figurenum{8}
\centerline{
\includegraphics[width=17cm]{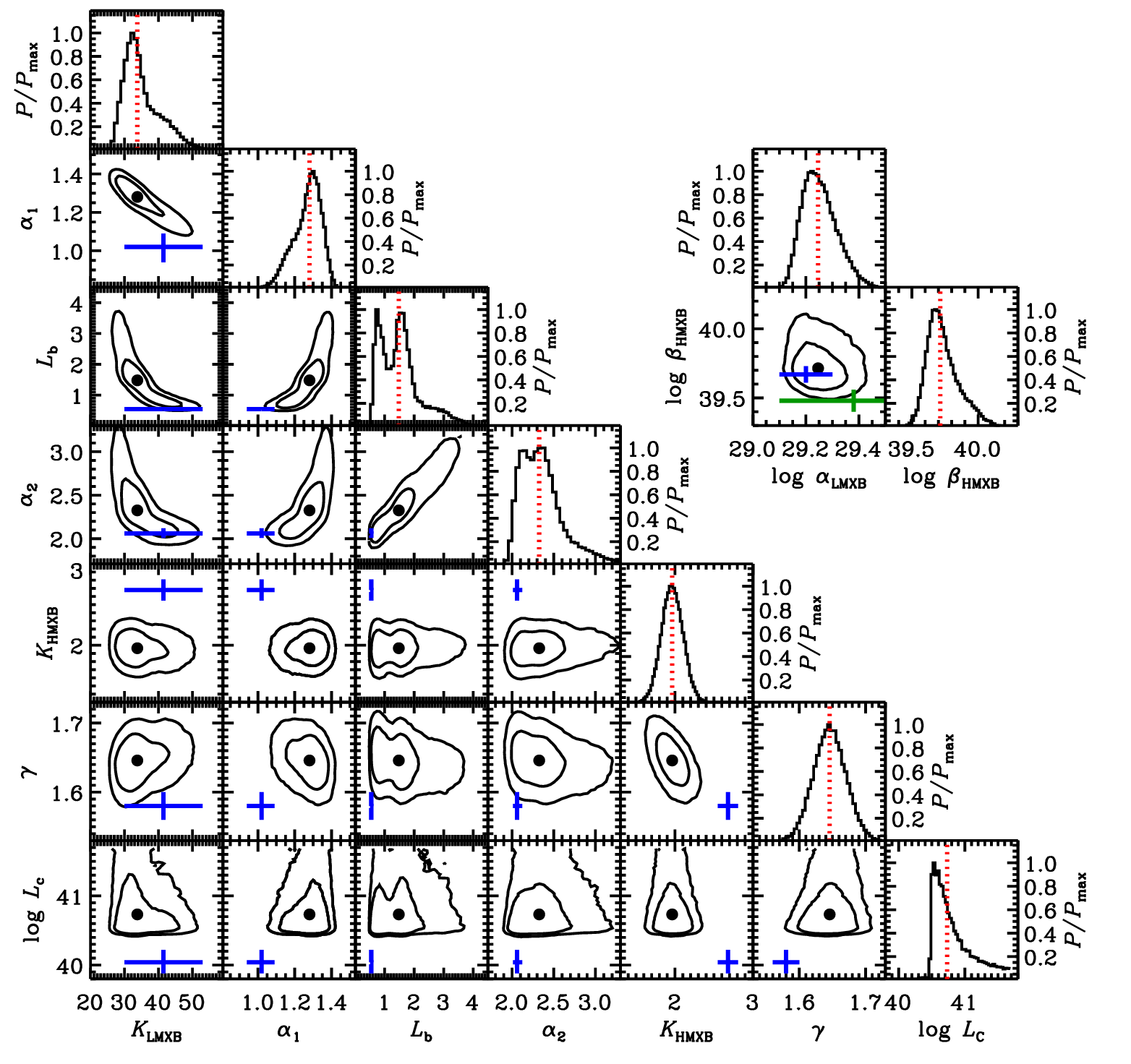}
}
\caption{
%%%
Probability distribution functions ($P/P_{\rm max}$) and confidence contours
for parameter pairs (showing 68\% and 95\% confidence contours drawn) for our
best-fit global model, which is based on fits to \nssfr\ sSFR-selected subgalactic regions (see $\S$4.2 for complete details).  The vertical red dotted lines
and solid black points indicate the median values of each parameter, which are
adopted as our best global model.  The 2-D parameter correlation distributions
include the seven free parameters ($K_{\rm LMXB}$, $\alpha_1$, $L_b$,
$\alpha_2$, $K_{\rm HMXB}$, $\gamma$, and $L_c$) that were fit with our global
model.  The distribution functions for the integrated $L_{\rm
X}$(LMXB)/$M_\star$ ($\alpha_{\rm LMXB}$) and $L_{\rm X}$(HMXB)/SFR
($\beta_{\rm HMXB}$), implied by our model, are shown in the upper-right panels.
Comparison values and 1$\sigma$ errors from M12 and Z12 for HMXB and LMXB
parameters are indicated with blue crosses, and the \chandra\ Deep Field-South
independent estimates of $\alpha_{\rm LMXB}$ and $\beta_{\rm HMXB}$ from Lehmer \etal\ (2016) are
shown with a green cross representing the 1$\sigma$ range.
%%%
}
\end{figure*}
%%%%%%%%%%%%%%%%%%%%%%%%%%%%%%%%%%%%%%%%%%%%%%%%%%%%%%%%%%%%%%%%%%%%%%%%%%%%%%%%%%

In Figure~7, we show an array of observed SFR-normalized cumulative XLFs for the
\nssfr\ sSFR bins.
From this representation, it is clear that the XRB
XLF both declines in normalization per unit SFR and becomes shallower in
overall slope with increasing sSFR, as described in $\S$4.1.  

Assuming that these trends are driven by changes in the relative LMXB to HMXB
populations, we chose to fit all \nssfr\ sSFR-binned XLFs globally using a
single XLF model that self-consistently describes the contributions from each
XRB population.  For a given bin of sSFR, the XLF is modeled using the
following set of equations:
\begin{equation}
\frac{dN}{dL} = \xi(L) \left[\frac{dN_{\rm LMXB}}{dL} + \frac{dN_{\rm
HMXB}}{dL} + {\rm CXB} \right]
\end{equation}
\begin{equation}
\frac{dN_{\rm HMXB}}{dL} = {\rm SFR} \; K_{\rm HMXB} \left \{ \begin{array}{lr}
L^{-\gamma}  & \;\;\;\;\;\;\;\;(L < L_c) \\ 
0,  & (L \ge L_c) \\ 
\end{array}
  \right.
\end{equation}
\begin{equation}
\frac{dN_{\rm LMXB}}{dL} = M_\star \; K_{\rm LMXB}  \left \{ \begin{array}{lr}
L^{-\alpha_1}  & \;\;\;\;\;\;\;\;(L < L_b) \\ 
L_b^{\alpha_2-\alpha_1} L^{-\alpha_2},  & (L_b \le L < L_c) \\ 
0,  & (L \ge L_c) \\ 
\end{array}
  \right.
\end{equation}
where Eqn.~(11) and (12) mirror Eqn.~(3) and (4), respectively.  In this case,
$K_{\rm HMXB}$ and $K_{\rm LMXB}$ are, respectively, normalizations per unit
SFR ([\sfr]$^{-1}$) and $M_\star$ ([$10^{11}$~\msol]$^{-1}$) at $L =
10^{38}$~\lum.  Here, since our data set is much more expansive than for
individual galaxies, we are able to perform fitting for seven parameters:
$K_{\rm LMXB}$, $\alpha_1$, $L_b$, $\alpha_2$, $K_{\rm HMXB}$, $\gamma$, and
$L_c$.  We utilize the same statistical methodology for determining the best
fit solution and parameter uncertainties, and minimize $C$ following:
\begin{equation}
C = 2 \sum_{i = 1}^{n_{\rm sSFR}} \left(\sum_{j=1}^{n_{\rm X}} M_{i,j} -
N_{i,j} + N_{i,j} \ln(N_{i,j}/M_{i,j}) \right),
\end{equation}
where $C$ is now determined ``globally'' through the double summation over
all $n_{\rm sSFR} =$~\nssfr\ sSFR bins ($i$th index) and $n_{\rm X}=100$ \xray\
luminosity bins ($j$th index; see $\S$4.1 for details related to luminosity
binning).

%
%%%%%%%%%%%%%%%%%%%%%%%%%%%%%%%%%%%%%%%%%%%%%%%%%%%%%%%%%%%%%%%%%%%%%%%%%%%%%%%%%%
% Figure 9
%%%%%%%%%%%%%%%%%%%%%%%%%%%%%%%%%%%%%%%%%%%%%%%%%%%%%%%%%%%%%%%%%%%%%%%%%%%%%%%%%%
%
\begin{figure}
\figurenum{9}
\centerline{
\includegraphics[width=8.5cm]{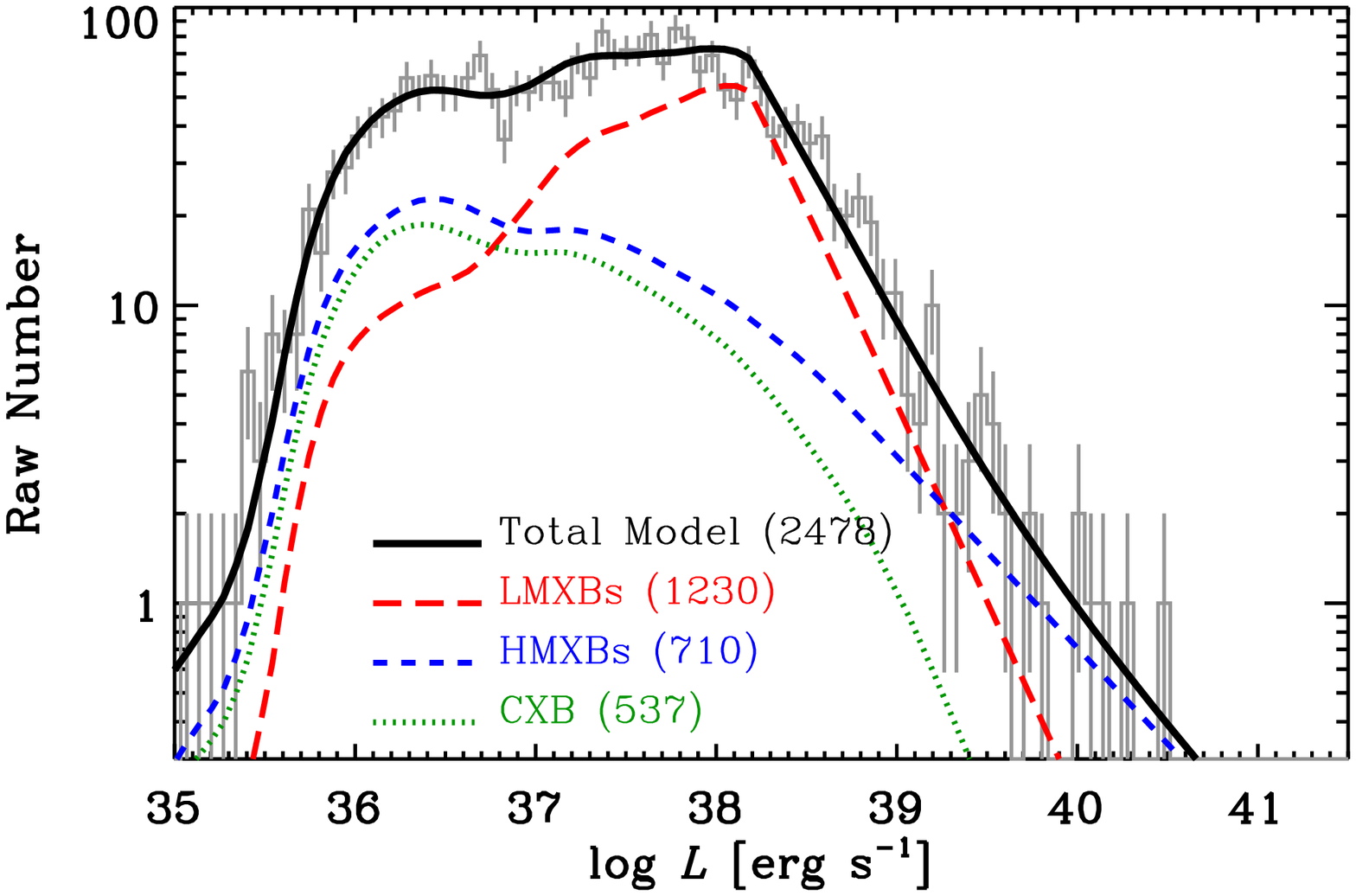}
}
\caption{
%%%
Raw number of sources detected as a function of \xray\ luminosity $L$, in bins
of $\Delta \log L = 0.057$~dex.  The shape of this curve is dependent on the XRB
XLFs, contributions from the CXB, and varying depths of \chandra\ observations
across the galaxy sample.  The cumulative model, based on summing contributions
from all sSFR bins, is shown as a solid curve, with HMXB ({\it blue
short-dashed\/}), LMXB ({\it red long-dashed\/}), and CXB ({\it green dotted})
components indicated.  The total number of sources predicted by each
model component are annotated in the key.  For comparison, the total number of
sources detected in the sample above
$10^{35}$~\lum\ is \nxrb, which is very close to that predicted (see annotation).
}
%%%
\end{figure}
%%%%%%%%%%%%%%%%%%%%%%%%%%%%%%%%%%%%%%%%%%%%%%%%%%%%%%%%%%%%%%%%%%%%%%%%%%%%%%%%%%

In Figure~8, we show the best-fit values, PDFs, and parameter correlations for
the above model, and in Table~4, we tabulate parameter values from this model.
Figure~9 shows the culled differential raw numbers of sources in luminosity
bins of $\Delta \log L = 0.057$~dex, with Poisson errors plotted (derived
following Gehrels~1986).  This distribution is compiled from all galaxies in
our sample, which have varying \chandra\ exposures, completeness functions, and
properties (e.g., sSFR).  In total, our data set contains \nxrb\ \xray\
detected point sources.  Our model suggests that \nlmxb, \nhmxb, and \nmod\ of
the sources are LMXBs, HMXBs, and CXB sources, respectively.  In a cumulative
sense, our overall model ({\it black curve\/}) reproduces very well the raw
distribution of source counts, including the complex contours associated with
incompleteness.  However, our fits are based on minimizing $C$ from Eqn.~(13),
which requires fitting a decomposition of these data into \nssfr\ such curves,
binned by sSFR.  Using the Kaastra~(2017) prescription for evaluating goodness
of fit, based on {\ttfamily cstat}, we find that the best-fit for
the \nssfr\ sSFR and 100 $L_{\rm X}$ bins is an acceptable model to the ensemble
data set, with $P_{\rm null} = 0.145$.

%
%%%%%%%%%%%%%%%%%%%%%%%%%%%%%%%%%%%%%%%%%%%%%%%%%%%%%%%%%%%%%%%%%%%%%%%%%%%%%%%%%%
% Figure 10
%%%%%%%%%%%%%%%%%%%%%%%%%%%%%%%%%%%%%%%%%%%%%%%%%%%%%%%%%%%%%%%%%%%%%%%%%%%%%%%%%%
%
\begin{figure*}
\figurenum{10}
\centerline{
\includegraphics[width=18cm]{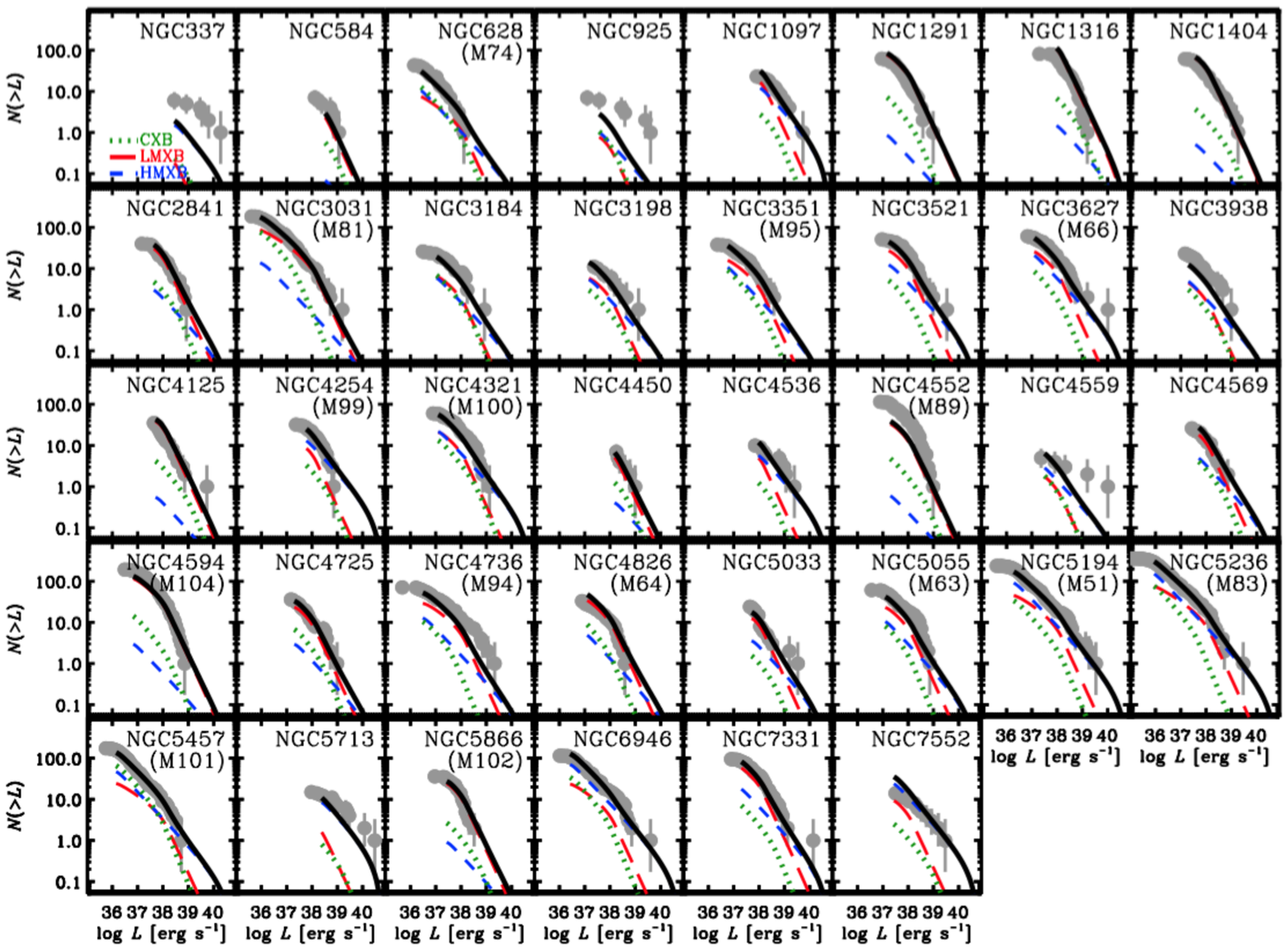}
}
\caption{
%%%
Observed galaxy-wide cumulative XLFs ({\it gray points} with 1$\sigma$ Poisson
error bars), displayed the same as in Fig.~3, but with our global decomposition
model predictions plotted ({\it black solid curves\/}).  Our model is detailed
in $\S$4.2, and consists of contributions from the CXB ({\it green dotted
curves\/}), LMXBs ({\it red long-dashed curves\/}), and HMXBs ({\it blue
short-dashed curves\/}), the normalizations of which scale with galaxy area,
$M_\star$, and SFR, respectively.  For all cases, the global XLF model provides
a good description of the overall XLF shapes, with the exceptions where the XLF
data are elevated over the model, which we suspect is due to anomalously low
metallicity (e.g., NGC~4559) or a relatively
large population of GC LMXBs (e.g., NGC~4552).
%%%
}
\end{figure*}
%%%%%%%%%%%%%%%%%%%%%%%%%%%%%%%%%%%%%%%%%%%%%%%%%%%%%%%%%%%%%%%%%%%%%%%%%%%%%%%%%%

We further present the calculated parameters, 
$$\beta_{\rm HMXB} \equiv \frac{1}{{\rm SFR}} \int_{L_{\rm lo}}^{L_{\rm c}}
\frac{dN_{\rm HMXB}}{dL}\, L \, dL = L_{\rm X}({\rm HMXB})/{\rm SFR}$$
and 
\begin{equation}
\alpha_{\rm LMXB} \equiv \frac{1}{M_\star} \int_{L_{\rm lo}}^{L_{\rm c}}
\frac{dN_{\rm LMXB}}{dL} \, L \, dL = L_{\rm X}({\rm LMXB})/M_\star,
\end{equation}
two widely used scaling relations, in Figure~8 and Table~4.  In
Figure~5, we show the model-implied XLF slopes and SFR-normalized XLF
normalizations for HMXB and LMXB populations, and in Figure~6, we display the
implied $L_{\rm X}$/SFR vs.~sSFR relation based on the $\alpha_{\rm LMXB}$ and
$\beta_{\rm HMXB}$ model values.  For the galaxies where we expect the XLFs to
be well sampled (i.e., those with $M_\star > 2 \times 10^{10}$~\msol\ or SFR~$> 2$~\sfr; {\it
orange boxes} in Figs.~5 and 6), we find that the galaxy-by-galaxy XLF
parameters follow the global model expectation, in which the high-luminosity
slopes ($\alpha_2$), SFR-normalized XLF normalizations ($K$/SFR), and $L_{\rm
X}$/SFR transition from LMXB-like at low-sSFR to HMXB-like at high-sSFR.
Galaxies with lower $M_\star$ or SFR show more significant scatter away from the
average trend, and in $\S$5.3 below, we examine closely the significance of
this scatter.

In Figure~7, we display the sSFR-dependent best-fit cumulative XLF model fits
to the data, including contributions from LMXB, HMXB, and CXB components.  Our
model reproduces the trends and basic shapes of these curves well, going from a
low-sSFR XLF with relatively high normalization per SFR and broken power-law
shape to a high-sSFR XLF with low normalization per SFR and single-sloped
power-law shape.  

In Figure~10, we show the cumulative XLFs for all \ngal\ galaxies in our sample (same as
Fig.~3) with the predicted XLFs from our global model overlaid.  That is, the
modeled XLF for a given galaxy is generated using our best global solution, which is based on simultaneous fitting to the \nssfr\ sSFR-selected subgalactic regions,
along with the galaxy-wide completeness function, SFR, $M_\star$, and sky area.
As such, the \xray\ data for a given galaxy is not used in these models, aside
from its minor influence on the global model solution itself (see below).  In
Table~5, we provide the {\ttfamily cstat} value and null-hypothesis
probability, $P_{\rm Null}^{\rm global}$, for the \xray\ data for each galaxy, and for convenience of comparison, we re-tabulate the $P_{\rm Null}$ values from the best-fit single and broken power-law models (Col.(12) and (14), respectively).
With a few notable exceptions, which we will discuss in $\S$5.2 below, the
global XLF model predicts very well the XLFs of several galaxies (considering
the model is not tuned to any one galaxy individually).  In fact, for several
cases (24 out of the 38), the global model produces an equivalent or better
statistical characterization (in terms of $P_{\rm Null}$; compare Col.(4) with
Cols.(12) and (14) in Table~5) of the \xray\ data than the best-fit power-law
models in $\S$3.2!  Some notable cases include NGC3031 (M81),
NGC~5194 (M51), NGC~5236 (M83), and NGC~5457 (M101), all of which include more
than 100 \xray\ sources detected and are better characterized by our global model due to
the somewhat complex contours that naturally result from the varying
contributions from HMXBs and LMXBs.  

To test the level of agreement between our global model and the observed XLFs
of each galaxy, we fit a ``scaled'' version of the global model to each of our
galaxies.  In this model, we fixed the shape of the model XLF, implied by the
global model and the SFR and $M_\star$ of the galaxy, but varied the
normalization of the XLF by a constant factor, $\omega$, such that 
\begin{equation}
\frac{dN_{\rm XRB}}{dL}\biggr\rvert_{\rm scaled} = \omega \left(\frac{dN_{\rm
LMXB}}{dL} + \frac{dN_{\rm HMXB}}{dL} \right).
\end{equation}
An $\omega = 1$, implies no additional scaling of the global model is needed.
Using this form of the XRB XLF in the overall model provided in eqn.~(10), we
fit for only $\omega$ following the procedures defined above.  In Figure~11, we
display the value of the scaling constant versus NGC name.  We find that all
but three galaxies (NGC~337, 925, and 4552) have $\omega$ consistent with unity
to within a factor of two.  For the rest of the galaxies, there is some scatter
in $\omega$ around unity (as required by the global moidel fit itself) of
$\approx$0.14~dex, which is consistent with the SFR and $M_\star$ calibration
uncertainty (i.e., $\approx$0.16~dex; see gray band in Figure~11).  The three
galaxies with substantial deviations will be analyzed in more detail in
$\S$5.2.

Since the global model describes well the majority of the galaxy XLFs in our
sample, it is unlikely that our average XLF scalings suffer from major
galaxy-sample variance.  However, to test for any notable variations between subsets, we
divided our sample into two subsets, retaining the NGC ordering in Table~1,
and re-ran our global XLF calculations.  In Table~4, we present the results
from this run (see ``First Subsample'' and ``Second Subsample'' parameters).
Although some minor differences are found, the parameters and computed
properties ($\alpha_{\rm LMXB}$ and $\beta_{\rm HMXB}$) are consistent between
subsamples at the 1$\sigma$ level.  

%
%%%%%%%%%%%%%%%%%%%%%%%%%%%%%%%%%%%%%%%%%%%%%%%%%%%%%%%%%%%%%%%%%%%%%%%%
\section{Discussion}
%%%%%%%%%%%%%%%%%%%%%%%%%%%%%%%%%%%%%%%%%%%%%%%%%%%%%%%%%%%%%%%%%%%%%%%%
%

\subsection{Comparison with Previous Results and Population Synthesis Models}

Our constraints on the HMXB and LMXB XLFs are similar in form to those
presented in past works (see, e.g., $\S$1 and references therein).  However, as
mentioned in $\S$1, this is the first systematic attempt to decompose the XLF
into LMXB and HMXB components for a sample of mainly late-type galaxies,
regardless of their galaxy-wide sSFR.  Furthermore, our XLF analyses contain a
somewhat larger sample of galaxies, and include ultradeep data from several
galaxies that were not available in past studies.  Notably, this provides (1) a
unique characterization of the LMXB XLF appropriate for late-type galaxies,
which may not necessarily be consistent with the LMXB XLF derived from
elliptical galaxies (see $\S$1) and (2) a cleaner characterization of the HMXB
XLF shape, down to faint limits.  Here, we examine the differences between our
XLFs and those reported in the literature.

%
%%%%%%%%%%%%%%%%%%%%%%%%%%%%%%%%%%%%%%%%%%%%%%%%%%%%%%%%%%%%%%%%%%%%%%%%%%%%%%%%%%
% Figure 11
%%%%%%%%%%%%%%%%%%%%%%%%%%%%%%%%%%%%%%%%%%%%%%%%%%%%%%%%%%%%%%%%%%%%%%%%%%%%%%%%%%
%
\begin{figure*}
\figurenum{11}
\centerline{
\includegraphics[width=18cm]{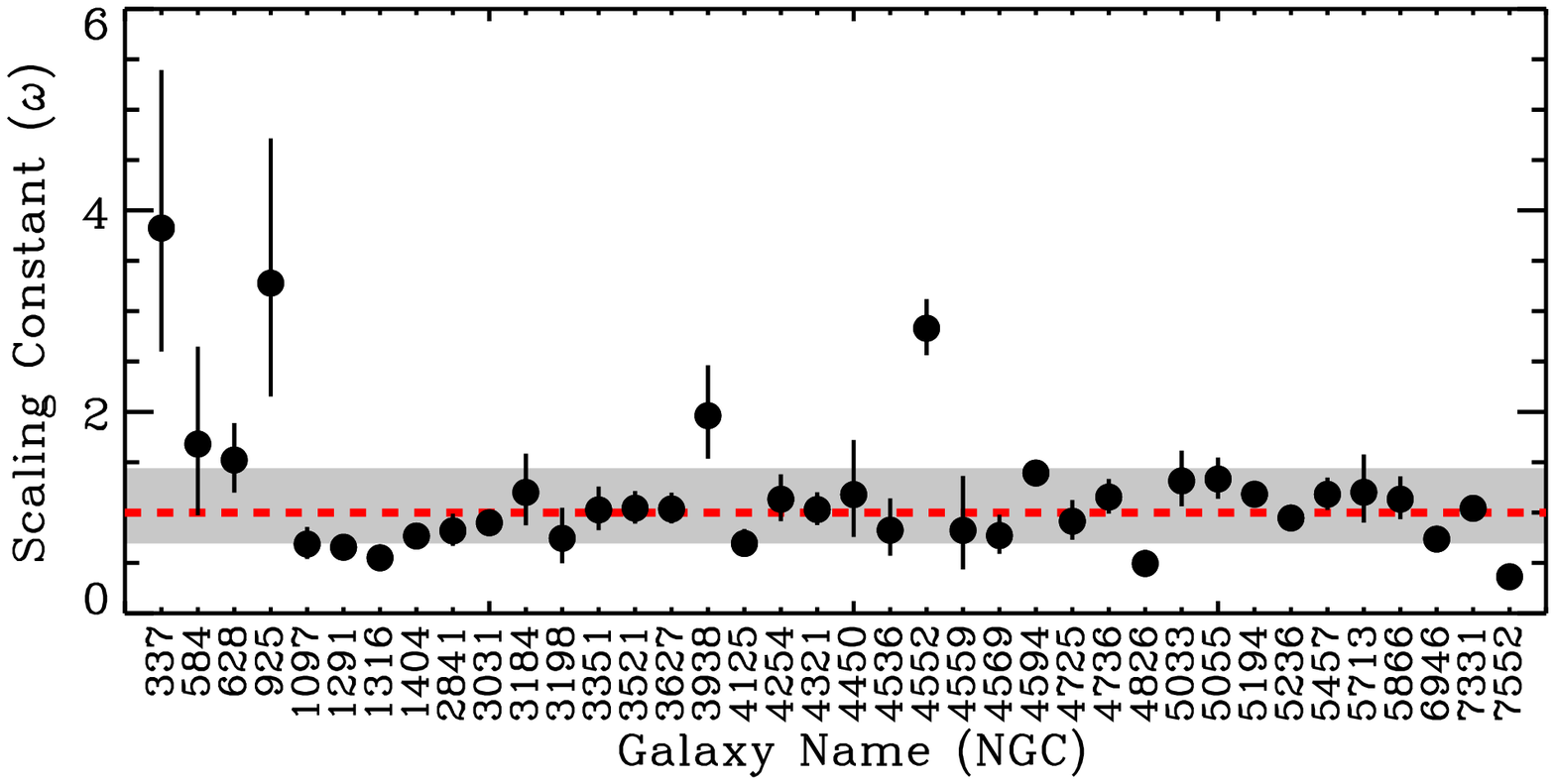}
}
\caption{
%%%
Best-fitting global-model scaling constant, $\omega$, versus
NGC-designated galaxy name.  The red dashed line at $\omega = 1$ and gray band
of width 0.16~dex respectively indicate the expected value from the global model and the
combined calibration uncertainty of the SFR and $M_\star$.  Only a few
galaxies, NGC~337, 925, and 4552 are clear outliers with $\omega > 2$ (see
discussion of these sources in $\S$5.2).
%%%
}
\end{figure*}
%%%%%%%%%%%%%%%%%%%%%%%%%%%%%%%%%%%%%%%%%%%%%%%%%%%%%%%%%%%%%%%%%%%%%%%%%%%%%%%%%%

%
%%%%%%%%%%%%%%%%%%%%%%%%%%%%%%%%%%%%%%%%%%%%%%%%%%%%%%%%%%%%%%%%%%%%%%%%%%%%%%%%%%
% Figure 12
%%%%%%%%%%%%%%%%%%%%%%%%%%%%%%%%%%%%%%%%%%%%%%%%%%%%%%%%%%%%%%%%%%%%%%%%%%%%%%%%%%
%
\begin{figure*}
\figurenum{12}
\centerline{
\includegraphics[width=9cm]{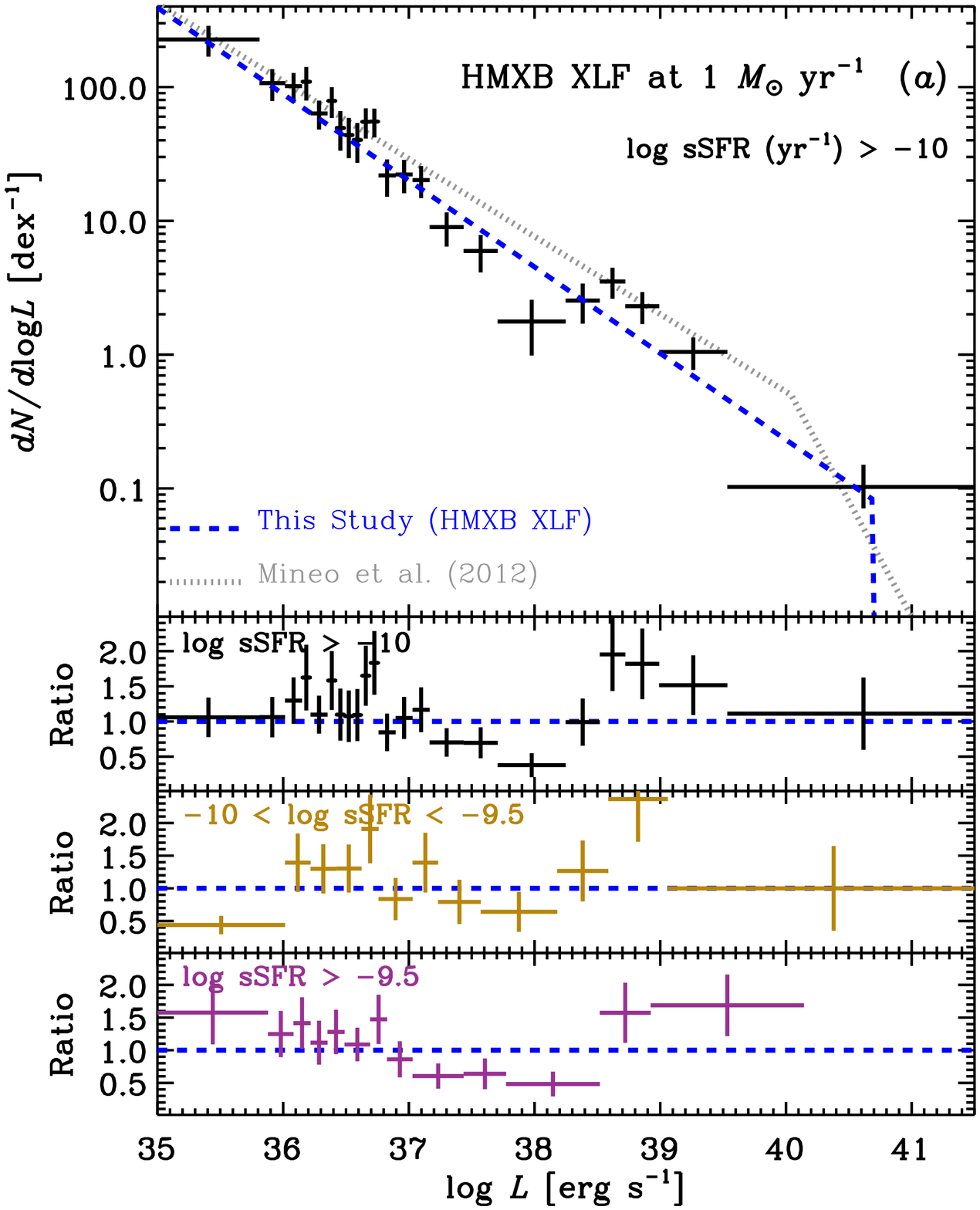}
\hfill
\includegraphics[width=9cm]{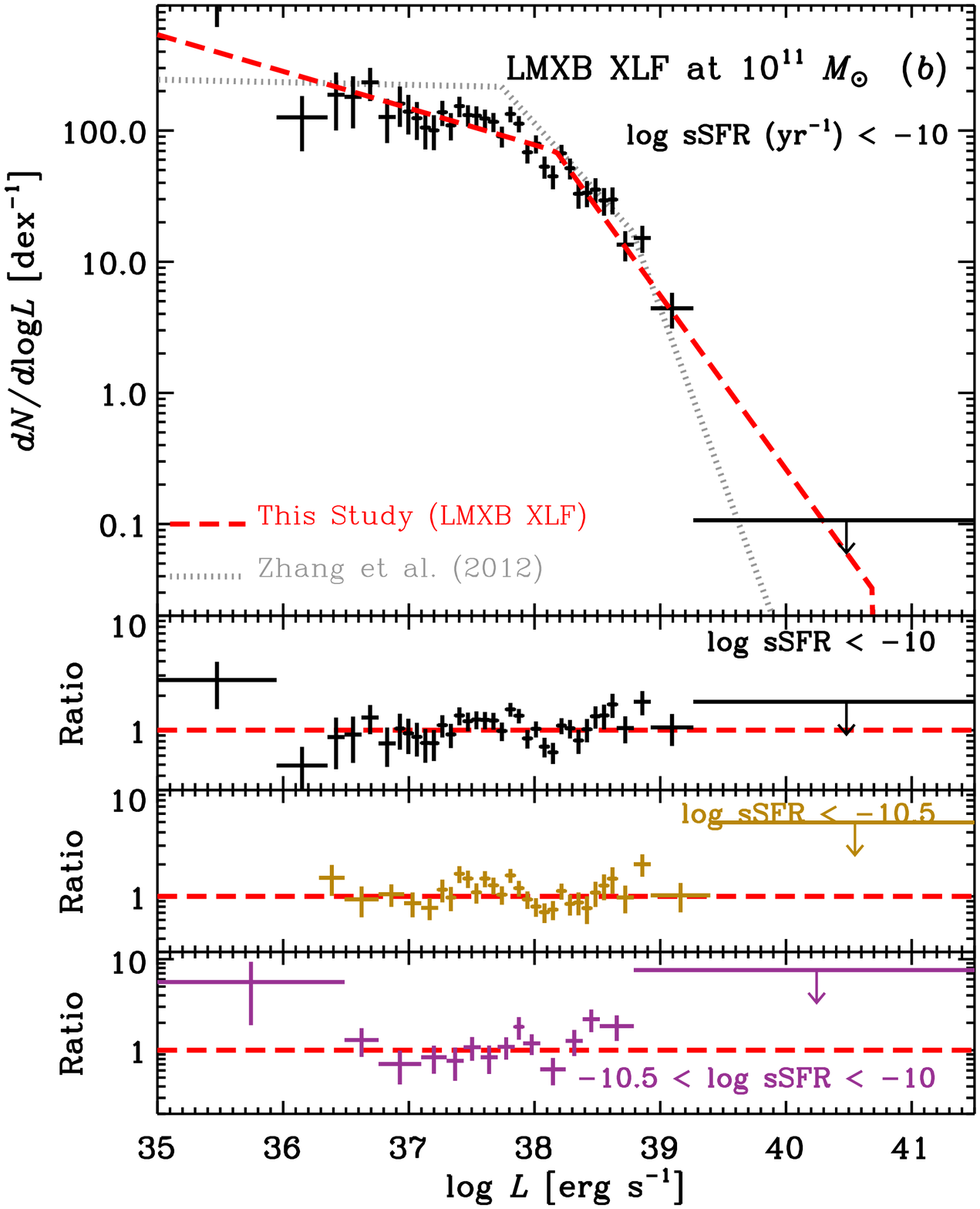}
}
\caption{
%%%
({\it a, top\/}) Constraints on the HMXB XLF, based on subgalactic regions with
sSFR~$> 10^{-10}$~yr$^{-1}$.  The data points and 1$\sigma$ Poisson error bars
represent completeness corrected and CXB-and-LMXB model subtracted constraints
on the HMXB XLF, normalized to a SFR~=~1~\sfr.  The blue short-dashed curve
shows our best-fit model and the dotted curve shows the Mineo \etal\ (2012)
constraint. {\it Bottom panels} show the data-to-model ratio, based on different
sSFR ranges (see annotations).
%%%
({\it b, top\/}) Constraints on the LMXB XLF, based on subgalactic regions with
sSFR~$< 10^{-10}$~yr$^{-1}$.  The data points and 1$\sigma$ Poisson
error bars represent completeness corrected and CXB-and-HMXB model subtracted
constraints on the LMXB XLF, normalized to $M_\star = 10^{11}$~\msol.  The red
long-dashed curve shows our best-fit model and the dotted curve shows the Zhang
\etal\ (2012) constraint from elliptical galaxies.  {\it Bottom panels} show
the data-to-model ratio, based on different sSFR ranges (see annotations).
%%%
}
\end{figure*}
%%%%%%%%%%%%%%%%%%%%%%%%%%%%%%%%%%%%%%%%%%%%%%%%%%%%%%%%%%%%%%%%%%%%%%%%%%%%%%%%%%

For the HMXB XLFs, we chose to compare with M12, who derive HMXB parameters
based on a 1055 \xray\ sources (including $\approx$700 XRBs) in a sample of 29
nearby galaxies with sSFR$~> 10^{-10}$~yr$^{-1}$ in an attempt to avoid LMXB
contributions.  For the LMXB XLF, we compare with the Z12 study of 20
elliptical galaxies, including a total of 1626 \xray\ sources (including
$\approx$1580 XRBs).\footnote{We note that the M12 and Z12 XLFs were derived
using a Salpeter~(1955) IMF, which produces SFR and $M_\star$ values that
differ from our Kroupa~(2001) IMF by factors of 1.56 and 1.24, respectively.
When making comparisons, we have corrected published values by these factors.
We also note that the assumed conversion factors that we use here to
compute physical properties (e.g., UV plus IR tracer of SFR) differ somewhat from
those used by M12 and Z12.  M12 make use of Bell~(2003) when determining
SFR and Z12 utilize Bell \& de~Jong~(2001) for $M_\star$, while we use Hao
\etal\ (2011) and Zibetti \etal\ (2009) for SFR and $M_\star$, respectively.
The only non-negligible differences come from the $M_\star$ conversion factors
for the bluest regions, where the Bell \& de~Jong~(2001) $M/L_K$ is up to a
factor of $\sim$10 times higher (although typically much less discrepant) than
that used by Zibetti \etal\ (2009).  We have chosen to not make adjustments based 
on these conversion factors, when comparing XLF properties, due to the complex form and non-trivial influence on
the results; however, we point out that some discrepancies between results may
in part be due to these assumptions.}  We note that the Z12 LMXB XLF uses a
broken power-law model with two breaks at $L_{\rm b,1} \approx 5 \times
10^{37}$~\lum\ and $L_{\rm b,2} \approx 6 \times 10^{38}$~\lum, instead of the
one break at $\approx$$5 \times 10^{37}$~\lum\ that is used in our model.  We
experimented with an LMXB XLF that involved two breaks, but found poor
constraints on the two separate break locations, and no improvement to the
overall quality of the fits to our data.  As such, we compare our LMXB XLF
parameters $\alpha_1$, $\alpha_2$ and $L_b$ with the Z12 parameters derived
below their $L_{b,2}$ (e.g., our $L_b$ is compared with their $L_{b,1}$).  

In Figure~8, we highlight comparison parameter values from the literature with
blue crosses, representing 1$\sigma$ error bars, as reported in the literature;
these comparisons are tabulated in Table~4.  We find that the parameters of our
LMXB XLF are similar to those of Z12, except that we favor a somewhat higher
normalization and steeper faint-end slope ($\alpha_1$).  These differences,
combined with our lack of a third steep power-law component at high $L$ yields
a somewhat larger estimate for the integrated LMXB \xray\ luminosity per unit mass,
$\alpha_{\rm LMXB}$; however, our estimates are consistent with those of Z12 within the
uncertainties (see upper right panel in Fig.~8).
For the HMXBs, our fit parameters significantly differ from those reported by
M12, due primarily to a preference for a steeper slope ($\gamma$) and lower
normalization ($K_{\rm HMXB}$) for our sample.  These parameters are
anticorrelated in such a way that the integrated \xray\ luminosity per unit SFR
$\beta_{\rm SFR}$ is in good agreement with that of M12.

%
%%%%%%%%%%%%%%%%%%%%%%%%%%%%%%%%%%%%%%%%%%%%%%%%%%%%%%%%%%%%%%%%%%%%%%%%%%%%%%%%%%
% Figure 13
%%%%%%%%%%%%%%%%%%%%%%%%%%%%%%%%%%%%%%%%%%%%%%%%%%%%%%%%%%%%%%%%%%%%%%%%%%%%%%%%%%
%
\begin{figure*}
\figurenum{13}
\centerline{
\includegraphics[width=16cm]{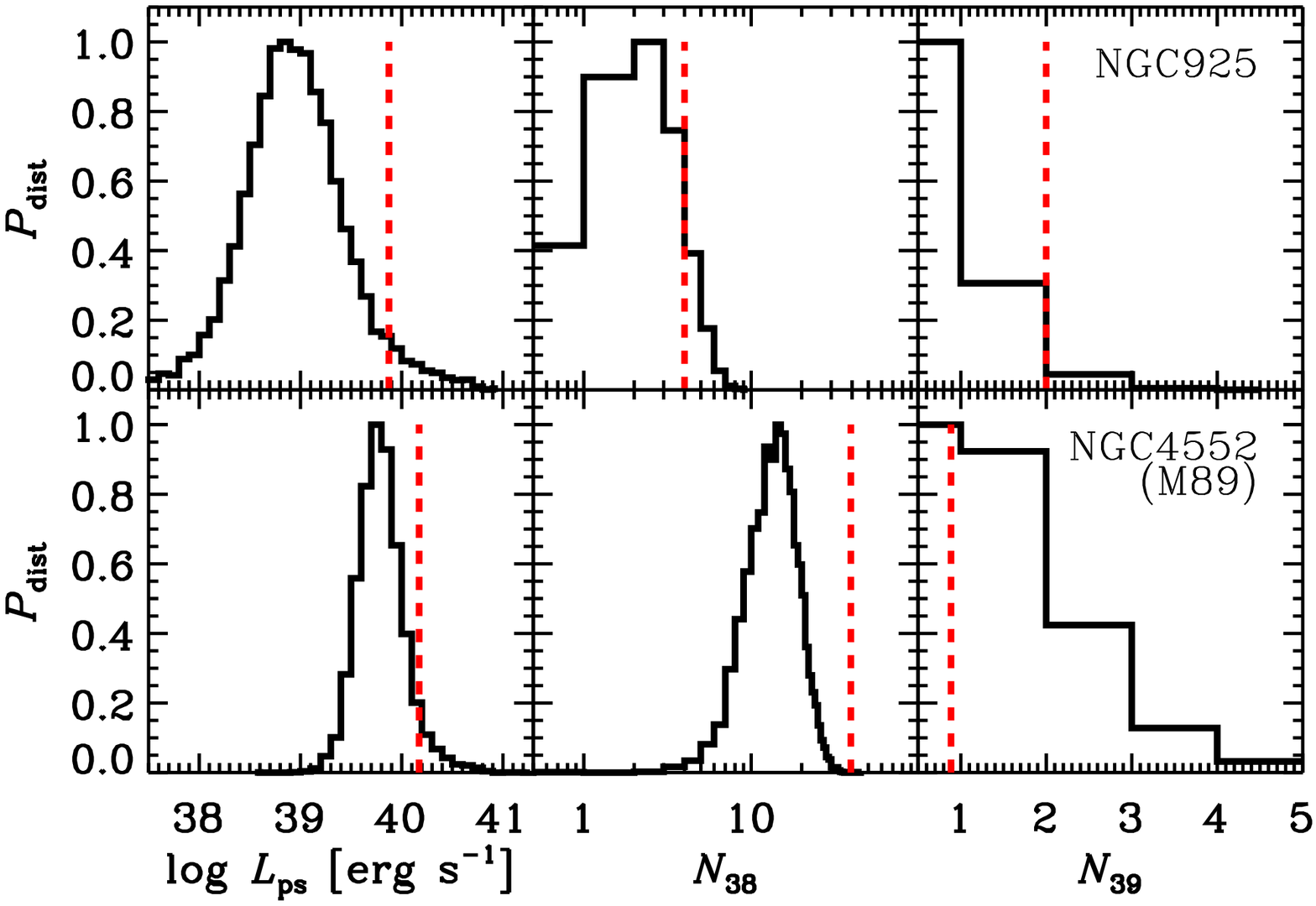}
}
\caption{
%%%
Sample of Monte Carlo {\it predicted} probability distributions of 
point-source luminosities we would expect to detect ($L_{\rm ps}$; {\it left panels\/}), and numbers of
such sources above $10^{38}$~\lum\ ($N_{\rm 38}$; {\it middle panels\/}) and
$10^{39}$~\lum\ ($N_{\rm 39}$; {\it right panels}), based on our global
best-fit models of NGC~925 ({\it top\/}) and NGC~4552 ({\it bottom\/}).  The
observed values of each parameter are shown as vertical red dashed lines.
These galaxies exhibit the most statistically significant deviations away from
the global model, beyond statistical scatter (see Table~5 for specific probability values).  For NGC~925 an excess of $L >
10^{39}$~\lum\ sources are observed, potentially due to the galaxy's relatively
low-metallicity (see $\S$5.2.1).  For NGC~4552, an excess of $L >
10^{38}$~\lum\ sources are observed, potentially due to a correspondingly large
populations of GC LMXBs
(see $\S$5.2.2).
%%%
}
\end{figure*}
%%%%%%%%%%%%%%%%%%%%%%%%%%%%%%%%%%%%%%%%%%%%%%%%%%%%%%%%%%%%%%%%%%%%%%%%%%%%%%%%%%

To reveal any unmodeled complex features in the shapes of the XLFs, and more
clearly compare differences with those from M12 and Z12, we created Figure~12,
which shows our HMXB and LMXB XLFs in differential form.  These ``clean'' HMXB
and LMXB XLFs were created by (1) extracting the observed XLFs from regions
with sSFR~$> 10^{-10}$~yr$^{-1}$ and sSFR~$< 3 \times 10^{-11}$~yr$^{-1}$,
respectively; (2) subtracting the low-level model components related to LMXB and HMXB
populations, respectively, as well as the CXB model components; and (3)
unfolding our data using the completeness functions generated in $\S$3.3.  The
data points in Figure~12, represent the unfolded data and 1$\sigma$ Poisson
errors in the top panels, and the ratio of the data to our best-fit models in
the bottom panels.  We further display the M12 and Z12 models for comparisons.  

Clearly, the sSFR~$> 10^{-10}$~yr$^{-1}$ HMXB data (Fig.~12$a$) shows a complex
shape beyond that described by a simple power-law model.  The HMXB XLF can be
better described as rapidly declining ($\gamma > 1.6$) between $L =
10^{36}$--$10^{38}$~\lum, and following a more exponential-like decline above
$L = 10^{38}$~\lum.  We found this shape was preserved when changing our sSFR
selection limits.  For example, the HMXB XLF for regions with $\log$~sSFR
=~$-10$ to $-9.5$ and $\log$~sSFR~$> -9.5$ both show the same basic shapes (see bottom panels of Fig.~12$a$).
Such a change in slope of the HMXB XLF has been predicted by previous
population synthesis models (e.g., Tzanavaris \etal\ 2013; Zuo \etal\ 2014;
Artale \etal\ 2018), and is potentially due to a dominance in wind-fed, young
($\simlt$20~Myr) BH-HMXBs.

The sSFR~$< 10^{-10}$~yr$^{-1}$ LMXB data (Fig.~12$b$) appear to be generally
consistent with the model across the full luminosity range.  However, when we
examine the data over different sSFR intervals, we see that the residuals are
somewhat more complex and indicate that the high-luminosity ($L \simgt 3 \times
10^{37}$~\lum) LMXB XLF slope gets shallower with increasing sSFR (see bottom
panels of Fig.~12$b$).  This is consistent with a scenario where higher sSFR
regions harbor younger populations of LMXBs that reach higher luminosities than
older LMXB populations (e.g., Fragos \etal\ 2008; Kim \& Fabbiano~2010; Lehmer
\etal\ 2014, 2017).

\subsection{Variations in the Galaxy-Wide XLFs}

%
%%%%%%%%%%%%%%%%%%%%%%%%%%%%%%%%%%%%%%%%%%%%%%%%%%%%%%%%%%%%%%%%%%%%%%%%%%%%%%%%%%
% Figure 14
%%%%%%%%%%%%%%%%%%%%%%%%%%%%%%%%%%%%%%%%%%%%%%%%%%%%%%%%%%%%%%%%%%%%%%%%%%%%%%%%%%
%
\begin{figure}
\figurenum{14}
\centerline{
\includegraphics[width=9cm]{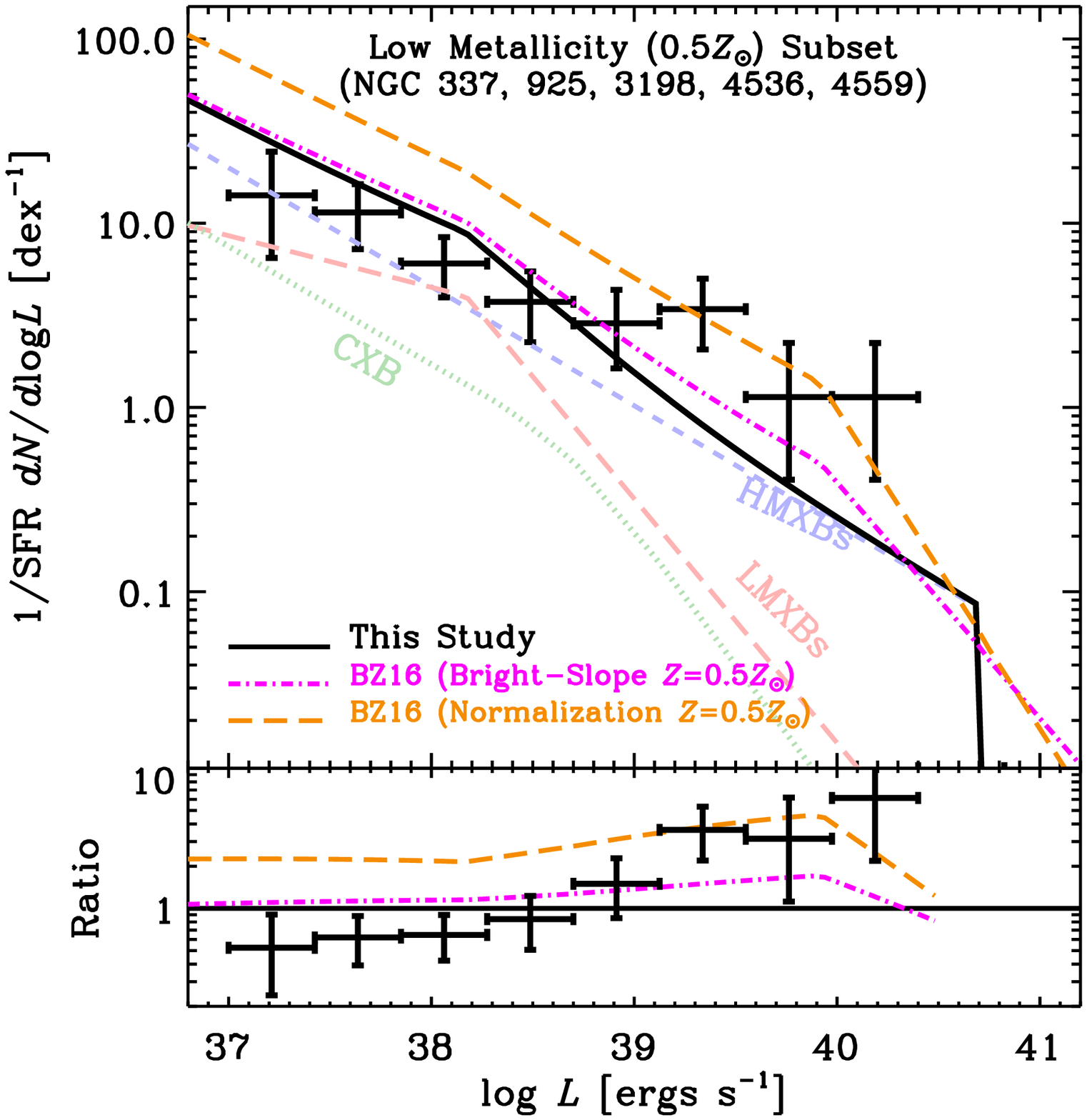}
}
\caption{
%%%
({\it Top\/}) SFR-normalized total XLF for the five lowest metallicity galaxies
in our sample (NGC~337, 925, 3198, 4536, and 4559), which have metallicities of
$\approx$0.5~$Z_\odot$.  The data points and 1$\sigma$ error bars are corrected for completeness, but
include contributions from HMXBs, LMXBs, and CXB sources.  The black curve
shows our global model prediction for this population, including HMXB, LMXB,
and CXB contributions in {\it faded blue, red, and green}, respectively.
Enhancements in the $L \simgt 10^{39}$~\lum\ source population are clearly
observed.  The BZ16 model predictions for enhancements of 0.5~$Z_\odot$ HMXB
populations are overlaid for scenarios where the canonical M12 HMXB XLF
normalization increases ({\it orange dashed curve\/}) or bright-end slope
flattens ({\it magenta dot-dashed curve\/}) with decreasing metallicity (see
$\S$5.2.1 for details).
({\it Bottom\/}) Ratio of data and BZ16 models with respect to our best-fit
global model prediction.  
%%%
}
\end{figure}
%%%%%%%%%%%%%%%%%%%%%%%%%%%%%%%%%%%%%%%%%%%%%%%%%%%%%%%%%%%%%%%%%%%%%%%%%%%%%%%%%%

As described in $\S$4.2, there are a few galaxies, for which the global model
does not provide a good description of the data (see $P_{\rm Null}$ in Col.(4)
of Table~5).  For many of these galaxies, the differences between the model and
data are within the uncertainties of the SFR or stellar mass calibrations (see
Fig.~11), but there are three examples (NGC~337, 925, and 4552) where the XLFs
are dramatically discrepant with the model, $P_{\rm Null} < 0.01$, resulting in galaxy luminosities
that are dramatically offset from the average relation shown in Figure~6.  As
detailed by Gilfanov \etal\ (2004) and Justham \& Schawinski (2012), a shallow-sloped XLF can produce large
variations in the distributions of bright XRBs, and thus $L_{\rm X}$, if the
XLF is poorly sampled.  Such poor XLF sampling is likely to be prevalent in
low-SFR galaxies, where the shallow-sloped HMXB XLF will be poorly sampled at
the high-$L$ end.  To a less important degree, low-$M_\star$ galaxies, that are
dominated by LMXBs (i.e., with low sSFRs), may also suffer from poor XLF
sampling, but this is less important than it is for HMXBs, due to the steep XLF
slope at high-$L$.  Nonetheless, it is instructive to quantify to what degree
the XRB XLFs, and implied integrated $L_{\rm X}$ of our galaxies can be
influenced by simple statistical sampling scatter of the HMXB and LMXB XLFs, so
that we can identify objects that are clear outliers.

For each galaxy in our sample, we performed a 1000-trial Monte Carlo analysis
to construct probability distributions of the summed point-source \xray\
luminosity, $L_{\rm ps}$, as well as the cumulative number of sources detected
above $10^{38}$~\lum\ and $10^{39}$~\lum, $N_{\rm 38}$ and $N_{\rm 39}$,
respectively, assuming that the XRBs in the galaxy follow our global-model XLF
(e.g., the black curves in Fig.~10).  For a given Monte Carlo trial, we first
perturbed the SFR and $M_\star$ values of a given galaxy (starting with the
values in columns~11 and 12 in Table~1) in accordance with a Gaussian
distribution of fractional 1$\sigma$ uncertainties of 0.1 and 0.13~dex,
respectively, corresponding to the uncertainties on the calibrations (see
$\S$3.1).  We note that distance-related uncertainties could affect our
calculations of SFR, $M_\star$ and $L$.  The median distance-related
uncertainty on these quantities is $\approx$0.06~dex (with a range of
0.004--0.2~dex), which is the size of our \xray\ luminosity bins and
significantly smaller than the calibration uncertainties on SFR and $M_\star$.
Furthermore, since distance-related errors affect SFR and $M_\star$ in the same
way that they affect $L$ (and integrated $L_{\rm X}$), the impact of the distance-related
uncertainties are substantially reduced.  We therefore ignore these
uncertainties in our simulations. Using the perturbed values of SFR and
$M_\star$, along with our best-fit global model, CXB estimates, and
completeness functions, we calculated the numbers of HMXBs, LMXBs, and CXB
sources with $L > 10^{36}$~\lum\ that we would expect to detect.  

We perturbed these numbers
using Poisson statistics, and calculated numbers of HMXBs, LMXBs, and CXB
sources ($N_{\rm HMXB}^{\rm MC}$, $N_{\rm LMXB}^{\rm MC}$, and $N_{\rm
CXB}^{\rm MC}$) for the Monte Carlo trial.  Using the integrated HMXB, LMXB,
and CXB XLF components as probability distributions, we assigned each of the
$N_{\rm HMXB}^{\rm MC}$, $N_{\rm LMXB}^{\rm MC}$, and $N_{\rm CXB}^{\rm MC}$
sources luminosity values to construct a simulated list of \xray\ point-sources
for the trial.  The simulated list provides a simulation of the observed XLF, $N_{\rm ps}(L)$
(e.g., equivalent to the gray data points in Fig.~10), and the source list
luminosities can be summed to yield expected total point-source luminosities:
$L_{\rm ps}^{\rm MC} = L_{\rm LMXB}^{\rm MC} + L_{\rm  HMXB}^{\rm MC} + L_{\rm
CXB}^{\rm MC}$

%%%%%%%%%%%%%%%%%%%%%%%%%%%%%%%%%%%%%%%%%%%%%%%%%%%%%%%%%%%%%%%%%%%%%%%%%%%%%%%%%%
% Table 5
%%%%%%%%%%%%%%%%%%%%%%%%%%%%%%%%%%%%%%%%%%%%%%%%%%%%%%%%%%%%%%%%%%%%%%%%%%%%%%%%%%
\begin{deluxetable*}{llcccccccccccc}
\tablewidth{0pt}
\tabletypesize{\scriptsize}
\tablecaption{Global X-ray Luminosity Function Fits By Galaxy}
\tablehead{
 \multicolumn{1}{c}{\sc Galaxy} & \colhead{} &  \multicolumn{5}{c}{\sc Global Model} & \multicolumn{3}{c}{\sc Scaled Global Model} & \multicolumn{4}{c}{\sc Power-Law}\\
\vspace{-0.25in} \\
 \multicolumn{1}{c}{\sc Name} &  \multicolumn{1}{c}{\sc Alt} &  \multicolumn{5}{c}{\rule{2in}{0.01in}} & \multicolumn{3}{c}{\rule{1.3in}{0.01in}} & \multicolumn{4}{c}{\rule{1.5in}{0.01in}} \\
\vspace{-0.25in} \\
\multicolumn{1}{c}{\sc (NGC)} & \multicolumn{1}{c}{\sc Name} &  \colhead{$C$} & \colhead{$P_{\rm Null}$} & \colhead{$P(L_{\rm ps})$} & \colhead{$P(N_{\rm 38})$} & \colhead{$P(N_{\rm 39})$} & \colhead{$\omega$} & \colhead{$C$} & \colhead{$P_{\rm Null}$} & \colhead{$C$} & \colhead{$P_{\rm Null}^{\rm PL}$} & \colhead{$C$} & \colhead{$P_{\rm Null}^{\rm BKNPL}$} \\
\vspace{-0.25in} \\
\multicolumn{1}{c}{(1)} & \multicolumn{1}{c}{(2)} & \multicolumn{1}{c}{(3)} & \colhead{(4)} & \colhead{(5)} & \colhead{(6)} & \colhead{(7)} & \colhead{(8)} & \colhead{(9)} & \colhead{(10)} & \colhead{(11)} & \colhead{(12)} & \colhead{(13)} & \colhead{(14)}
}
\startdata
       337 &            &   29 &   $<$0.001 &      0.034 &     \ldots &      0.005 &  3.82$^{+1.57}_{-1.22}$ &   23 &      0.440 &   22 &      0.944 &   23 &      0.889 \\ 
       584 &            &    9 &      0.899 &      0.719 &     \ldots &      0.189 &  1.68$^{+0.97}_{-0.71}$ &    9 &      0.478 &   11 &      0.686 &    9 &      0.169 \\ 
       628 &        M74 &   32 &      0.543 &      0.720 &      0.852 &      0.360 &  1.52$^{+0.37}_{-0.32}$ &   30 &      0.239 &   31 &      0.106 &   26 &      0.109 \\ 
       925 &            &   33 &   $<$0.001 &      0.045 &      0.067 &      0.004 &  3.28$^{+1.44}_{-1.12}$ &   30 &      0.067 &   26 &      0.785 &   25 &      0.849 \\ 
      1097 &            &   32 &      0.730 &      0.742 &     \ldots &      0.308 &  0.69$^{+0.17}_{-0.15}$ &   29 &      0.814 &   25 &      0.261 &   33 &      0.387 \\ 
\\
      1291 &            &   37 &      0.853 &      0.924 &      0.942 &      0.776 &  0.66$^{+0.10}_{-0.09}$ &   27 &      0.389 &   36 &      0.162 &   25 &      0.173 \\ 
      1316 &            &   54 &      0.001 &      0.995 &     \ldots &      0.938 &           0.55$\pm$0.07 &   29 &      0.852 &   28 &      0.383 &   28 &      0.383 \\ 
      1404 &            &   36 &      0.517 &      0.715 &      0.930 &      0.385 &  0.77$^{+0.11}_{-0.10}$ &   32 &      0.781 &   27 &      0.061 &   27 &      0.047 \\ 
      2841 &            &   25 &      0.630 &      0.774 &      0.724 &      0.781 &  0.82$^{+0.17}_{-0.15}$ &   24 &      0.585 &   24 &      0.218 &   24 &      0.283 \\ 
      3031 &        M81 &   44 &      0.503 &      0.362 &      0.567 &      0.206 &           0.90$\pm$0.12 &   43 &      0.486 &   50 &      0.034 &   43 &      0.092 \\ 
\\
      3184 &            &   35 &      0.527 &      0.428 &      0.049 &      0.440 &  1.20$^{+0.38}_{-0.33}$ &   35 &      0.670 &   37 &      0.782 &   34 &      0.773 \\ 
      3198 &            &   31 &      0.639 &      0.392 &      0.230 &      0.139 &  0.74$^{+0.30}_{-0.25}$ &   30 &      0.477 &   30 &      0.748 &   28 &      0.851 \\ 
      3351 &        M95 &   20 &      0.035 &      0.609 &      0.471 &      0.507 &  1.03$^{+0.23}_{-0.20}$ &   20 &      0.034 &   23 &      0.008 &   21 &      0.032 \\ 
      3521 &            &   39 &      0.642 &      0.386 &      0.091 &      0.494 &  1.04$^{+0.17}_{-0.15}$ &   39 &      0.666 &   45 &      0.545 &   30 &      0.267 \\ 
      3627 &        M66 &   44 &      0.498 &      0.221 &      0.215 &      0.282 &  1.04$^{+0.16}_{-0.14}$ &   44 &      0.516 &   45 &      0.554 &   41 &      0.738 \\ 
\\
      3938 &            &   27 &      0.790 &      0.294 &      0.118 &      0.470 &  1.96$^{+0.50}_{-0.43}$ &   23 &      0.244 &   23 &      0.056 &   23 &      0.219 \\ 
      4125 &            &   33 &      0.511 &      0.506 &      0.949 &      0.498 &  0.69$^{+0.14}_{-0.12}$ &   29 &      0.794 &   26 &      0.458 &   28 &      0.796 \\ 
      4254 &        M99 &   16 &      0.036 &      0.823 &      0.287 &      0.917 &  1.13$^{+0.25}_{-0.22}$ &   16 &      0.027 &   16 &      0.017 &   15 &      0.019 \\ 
      4321 &       M100 &   42 &      0.697 &      0.497 &      0.073 &      0.563 &  1.03$^{+0.17}_{-0.15}$ &   42 &      0.712 &   44 &      0.363 &   36 &      0.399 \\ 
      4450 &            &   12 &      0.485 &      0.823 &     \ldots &      0.548 &  1.18$^{+0.54}_{-0.42}$ &   12 &      0.399 &   13 &      0.464 &   12 &      0.148 \\ 
\\
      4536 &            &   23 &      0.953 &      0.568 &     \ldots &      0.254 &  0.83$^{+0.32}_{-0.25}$ &   23 &      0.819 &   22 &      0.604 &   22 &      0.693 \\ 
      4552 &        M89 &   99 &   $<$0.001 &      0.062 &   $<$0.001 &      0.603 &  2.83$^{+0.29}_{-0.27}$ &   32 &      0.675 &   40 &      0.002 &   35 &      0.068 \\ 
      4559 &            &   26 &      0.203 &      0.053 &      0.455 &      0.017 &  0.82$^{+0.54}_{-0.39}$ &   25 &      0.142 &   20 &      0.668 &   20 &      0.768 \\ 
      4569 &            &   20 &      0.280 &      0.860 &      0.703 &      0.771 &  0.77$^{+0.21}_{-0.18}$ &   19 &      0.259 &   20 &      0.132 &   19 &      0.255 \\ 
      4594 &       M104 &   40 &      0.944 &      0.401 &      0.234 &      0.814 &  1.39$^{+0.11}_{-0.10}$ &   26 &      0.071 &   59 &      0.707 &   22 &      0.010 \\ 
\\
      4725 &            &   30 &      0.938 &      0.495 &      0.687 &      0.591 &  0.91$^{+0.21}_{-0.18}$ &   30 &      0.952 &   31 &      0.274 &   30 &      0.694 \\ 
      4736 &        M94 &   57 &      0.077 &      0.165 &      0.089 &      0.047 &  1.15$^{+0.18}_{-0.16}$ &   57 &      0.115 &   55 &      0.554 &   52 &      0.842 \\ 
      4826 &        M64 &   31 &      0.672 &      0.861 &      0.672 &      0.613 &  0.49$^{+0.13}_{-0.12}$ &   20 &      0.143 &   25 &      0.044 &   17 &      0.023 \\ 
      5033 &            &   33 &      0.226 &      0.339 &      0.513 &      0.124 &  1.31$^{+0.30}_{-0.25}$ &   32 &      0.453 &   30 &      0.884 &   31 &      0.424 \\ 
      5055 &        M63 &   35 &      0.988 &      0.376 &      0.043 &      0.682 &  1.33$^{+0.21}_{-0.19}$ &   33 &      0.637 &   34 &      0.101 &   33 &      0.215 \\ 
\\
      5194 &        M51 &   53 &      0.895 &      0.427 &      0.319 &      0.215 &  1.18$^{+0.10}_{-0.09}$ &   50 &      0.751 &   49 &      0.034 &   48 &      0.062 \\ 
      5236 &        M83 &   52 &      0.650 &      0.597 &      0.274 &      0.652 &           0.95$\pm$0.07 &   51 &      0.635 &   57 &      0.073 &   54 &      0.182 \\ 
      5457 &       M101 &   38 &      0.288 &      0.553 &      0.373 &      0.649 &  1.18$^{+0.17}_{-0.15}$ &   37 &      0.207 &   38 &      0.019 &   37 &      0.088 \\ 
      5713 &            &   32 &      0.142 &      0.135 &     \ldots &      0.167 &  1.20$^{+0.37}_{-0.30}$ &   32 &      0.259 &   29 &      0.456 &   30 &      0.638 \\ 
      5866 &       M102 &   21 &      0.322 &      0.810 &      0.492 &      0.562 &  1.13$^{+0.23}_{-0.20}$ &   21 &      0.282 &   26 &      0.214 &   17 &      0.188 \\ 
\\
      6946 &            &   60 &      0.237 &      0.481 &      0.271 &      0.602 &           0.74$\pm$0.10 &   54 &      0.415 &   53 &      0.289 &   52 &      0.552 \\ 
      7331 &            &   50 &      0.117 &      0.216 &      0.137 &      0.518 &  1.04$^{+0.12}_{-0.11}$ &   50 &      0.126 &   56 &      0.710 &   50 &      0.703 \\ 
      7552 &            &   47 &      0.148 &      0.803 &      0.934 &      0.519 &  0.36$^{+0.12}_{-0.10}$ &   30 &      0.803 &   28 &      0.190 &   27 &      0.351 \\ 
\enddata
\tablecomments{Goodness of fit assessments for all galaxies, based on our global model, scaled global model, and power-law fits.  Col.(1) and (2): Galaxy NGC and Messier name, as reported in Table~1.  Col.(3) and (4): Respectively, C-statistic and null-hypothesis probability for the best global model (see $\S$4.2 for details), which is based on only the SFR and $M_\star$ of the galaxy.  Col.(5)--(7): Probabilities of observing the total detected point-source luminosity $L_{\rm ps}$, total number of sources brighter than $L = 10^{38}$~\lum, and total number of sources brighter than $L = 10^{39}$~\lum, respectively, if the data are drawn from the global model.  The probabilities are based on Monte Carlo simulations, which include the effects of statistical variance and uncertainty in SFR and $M_\star$ calibrations (see $\S$5.2 for detailed description). Col.(8): Constant scaling factor $\omega$ and its 1$\sigma$ error.  The constant scaling factor for a given galaxy multiplies by the XLF predicted by the global model, following equation~(15).  A value of $\omega = 1$ indicates consistency with the global model.  Col.(9) and (10): Respectively, C-statistic and null-hypothesis probability for the scaled global model.  Col.(11)--(14): C-statistic and null-hypothesis probability pairs for power-law and broken power-law models.  These columns are re-tabulations of Col.(8)--(9) and Col.(13)--(14) from Table~3.\\
}
\end{deluxetable*}

Our Monte Carlo procedure, run 1000 times per galaxy, thus provides
probability distributions of $N_{\rm ps}(L)$ and $L_{\rm ps}$.
To identify
potential outliers, we computed three quantities: $P(L_{\rm ps})$,
$P(N_{\rm 38})$, and $P(N_{\rm 39})$, which are the
probabilities of observing a population of sources above the measured $L_{\rm
ps}$, $N_{\rm 38}$, and $N_{\rm 39}$, respectively, given
the model.  The values of these probabilities are provided for each galaxy in Col.(5)--(7) of Table~5.  

Given that there are \ngal\ galaxies in our full sample, we expect that these
probability values may span $0.03 \simlt P \simlt 0.97$ due to random scatter.  Sources outside of
this range are good candidates for outliers that do not follow the relation due
to some inherently different physical property beyond just statistical
variance.  For our sample, we find four cases where $P < 0.03$: NGC~337, 925,
4552, and 4559.  NGC~337, 925, and 4559 are high-sSFR galaxies that show an excess of $L >
10^{39}$~\lum\ point sources, while NGC~4552 is a low-sSFR elliptical galaxy
that shows a significant excess of $L > 10^{38}$~\lum\ point sources.
Figure~13 shows example probability distributions for the three quantities for
NGC~925 and NGC~4552, along with their observed values.
Comparisons of the properties of these galaxies with the rest of the sample
reveal two compelling physical reasons why these galaxies would be offset from
the global model distribution: the effects of low-metallicity on HMXB formation or large contributions from GC LMXB populations.  Below, we discuss each of these scenarios in turn.

\subsubsection{Enhanced HMXBs in Low Metallicity Galaxies}

In terms of metallicity, NGC~337, 925, and 4559 are among the five galaxies
with the lowest metallicities in our sample, together with NGC~3198 and 4536.
These five galaxies have metallicities that are around $\approx$1/2~$Z_\odot$,
factors of 0.4--0.5 times the median metallicity of our sample, and all have
relatively small values of $P(N_{\rm 39})$, indicating a likely excess of
luminous sources within the subpopulation.  Within this subsample, we detected
12 \xray\ point-sources with $L > 10^{39}$~\lum, when $\approx$4 were expected
from our global model.  From our Monte Carlo simulations, the probability of
obtaining 12 sources with $L > 10^{39}$~\lum\ is $\approx$0.2\%, suggesting
that the low-metallicity sample as a whole contains an excess of luminous point
sources.  For comparison, the total point-source luminosity $L_{\rm ps}$, and
number of sources with $L > 10^{38}$~\lum, are consistent with expectations
from the global model, $P(L_{\rm ps}) = 7$\% and $P(N_{\rm 38}) = 55$\%,
respectively, suggesting that the enhanced population is limited to the most
luminous sources.

A more detailed view of the low-metallicity XLF is displayed in Figure~14,
which shows the combined completeness-corrected, SFR-normalized XLF for the
five lowest-metallicity galaxies in our sample.  In Figure~14, we overlay our
best-fit global model XLF, which includes contributions from HMXBs, LMXBs, and
CXB sources ({\it faded blue, red, and green curves}).  The global model predicts
that the XLF of the low-metallicity galaxies is dominated by HMXBs above $L
\sim 10^{38}$~\lum.  A factor of $\approx$2--10 times excess of sources over
the global model is observed for $L \simgt 5 \times 10^{38}$~\lum\ for the
low-metallicity subset, with the largest and most significant excess measured
around $3 \times 10^{39}$~\lum.  Thus, the HMXB XLF of low-metallicity galaxies
takes on an enhanced ``hump'' above the global model at $L \simgt
10^{39}$~\lum.

Qualitatively similar enhancements were observed by Basu-Zych \etal\ (2016,
BZ16) in the $L \simgt 10^{40}$~\lum\ XLFs of low-metallicity Lyman-break
analog (LBA) galaxies Haro~11 and VV114, and the relatively nearby
low-metallicity galaxy NGC~3310 (e.g., Miralles-Caballero \etal\ 2014) appears
to show a similar excess of $L \simgt 10^{38}$~\lum\ sources compared to the
M12 relation (see, e.g., Fig.~14 of M12).  Using the LBA observations, combined
with measurements of $L_{\rm X}$/SFR versus metallicity from the literature
(Basu-Zych \etal\ 2013a, Brorby \etal\ 2014; Douna \etal\ 2015), BZ16
constructed two model scenarios for the low-metallicity XLF consistent with the
data.  These models include an HMXB XLF that (1) flattens or extends the
shallow high-luminosity slope to brighter limits ($\simgt$10$^{40}$~\lum;
hereafter ``bright-slope'') or (2) increases in normalization, as the
metallicity decreases.  Both scenarios result in a rise in $L_{\rm X}$/SFR with
decreasing metallicity consistent with the $L > 10^{40}$~\lum\ LBA XLFs, the
HMXB XLF of typical galaxies (based on M12), and the observed $L_{\rm X}$/SFR
versus metallicity correlation, which is also consistent with the Fragos \etal\
(2013b) population synthesis predictions for the $L_{\rm X}$/SFR versus
metallicity relation.

In Figure~14, we show both BZ16 predictions (i.e., varying bright-slope and
normalization with metallicity) for the $\approx$1/2~$Z_\odot$ HMXB XLF, with
model contributions from LMXB and CXB sources added for fair comparison with
our data.  The bottom-panel of Figure~14 shows the ratio of the low-metallicity
galaxy data from this study and BZ16 models compared to our best-fit global
model.  While the BZ16 models produce elevated HMXB XLF predictions, neither
scenario describes well our overall XLF constraints for the $\approx$1/2~$Z_\odot$ galaxies in our sample.  As noted above, the
excess of sources in the low-metallicity sample appears to begin at $L \simgt
10^{39}$~\lum, roughly an order of magnitude below that in the BZ16
bright-slope model ({\it magenta dot-dashed curve}).  Furthermore, the BZ16
enhanced normalization model nicely fits the enhanced $L > 10^{39}$~\lum\ hump,
but does not predict the return to the global XLF level at $L \simlt
10^{39}$~\lum.  It is currently not clear if the overall observed trend of
increasing $L_{\rm X}$/SFR with declining metallicity can be attributed to a
smooth development and enhancement of the XLF hump we observe here.  It is also
possible that more complex changes occur in the HMXB XLF shape with metallicity.
Despite this, a more systematic study of how the HMXB XLF varies as a function
of metallicity is tractable, but would require a sample of galaxies that span a
broader range of metallicity compared to those in this study.  Such an
investigation, and its implications for XRB population synthesis models, will
be the subject of future work.

\subsubsection{Enhanced LMXBs in Massive Elliptical Galaxies}

In addition to the statistically-significant enhancement of $N_{39}$ for HMXBs
in the lowest-metallicity galaxies in our sample, we also find enhancements in
the LMXB populations for some of the early-type galaxies.  Most notably,
NGC~4552, which has an E-type morphology, is observed to have a statistically
significant excess of low-luminosity LMXBs, $N_{38}$, compared to the global
model prediction (see bottom panels of Fig.~13).  For massive early-type
galaxies like NGC~4552, it has been shown by several authors (e.g.,
Harris~1991; Bekki \etal\ 2006; Peng \etal\ 2008; Harris \etal\ 2013) that the
number of GCs per unit stellar mass can be enhanced and vary significantly from
galaxy-to-galaxy.  In such galaxies, the contributions from dynamically formed
LMXBs coincident with GCs can dominate the XLF of the galaxy (see, e.g., Kim \&
Fabbiano~2004; Irwin~2005; Juett \etal\ 2005; Lehmer \etal\ 2014; Peacock
\etal\ 2017).  Although all galaxies in our sample are expected to contain some
contributions from GC LMXBs, and our global model will include an average
contribution from these GCs that is characteristic of the average number of GCs
per unit mass, our global model will not accurately predict the LMXB XLF for
galaxies with strong deviations from this average.  As previous studies have
shown, the galaxies that are most likely to show deviations are massive
ellipticals with relatively large dark-matter halos (see, e.g., Harris \etal\
2013).
%
%%%%%%%%%%%%%%%%%%%%%%%%%%%%%%%%%%%%%%%%%%%%%%%%%%%%%%%%%%%%%%%%%%%%%%%%%%%%%%%%%%
% Figure 15
%%%%%%%%%%%%%%%%%%%%%%%%%%%%%%%%%%%%%%%%%%%%%%%%%%%%%%%%%%%%%%%%%%%%%%%%%%%%%%%%%%
%
\begin{figure}
\figurenum{15}
\centerline{
\includegraphics[width=8.5cm]{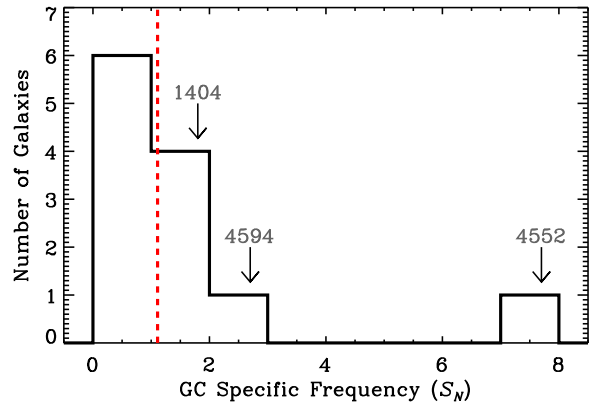}
}
\caption{
%%%
Distribution of GC specific frequncies, $S_N$, for 12 out of the \ngal\
galaxies in our sample, based on published values from Harris \etal\ (2013).
The locations and names of the three galaxies with the highest $S_N$ values
have been annotated.  NGC~4552 has the largest $S_N$ value of our sample and
has a statistically significant excess of LMXBs compared to our global model
expectation, suggesting that GC LMXBs dominate the XLF in this galaxy.
%%%
}
%%%
\end{figure}
%%%%%%%%%%%%%%%%%%%%%%%%%%%%%%%%%%%%%%%%%%%%%%%%%%%%%%%%%%%%%%%%%%%%%%%%%%%%%%%%%%

%
%%%%%%%%%%%%%%%%%%%%%%%%%%%%%%%%%%%%%%%%%%%%%%%%%%%%%%%%%%%%%%%%%%%%%%%%%%%%%%%%%%
% Figure 16
%%%%%%%%%%%%%%%%%%%%%%%%%%%%%%%%%%%%%%%%%%%%%%%%%%%%%%%%%%%%%%%%%%%%%%%%%%%%%%%%%%
%
\begin{figure*}
\figurenum{16}
\centerline{
\includegraphics[width=8.5cm]{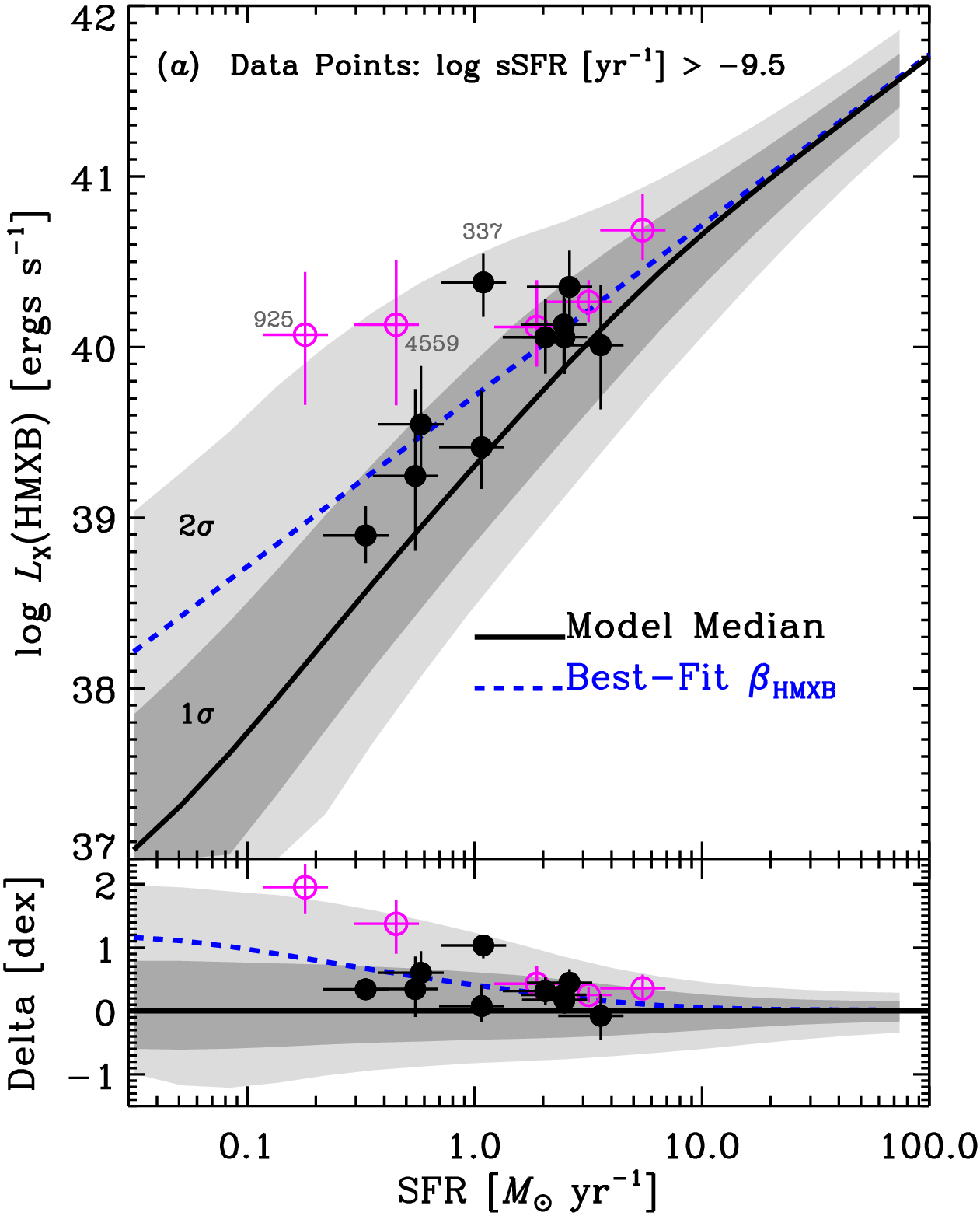}
\hfill
\includegraphics[width=8.5cm]{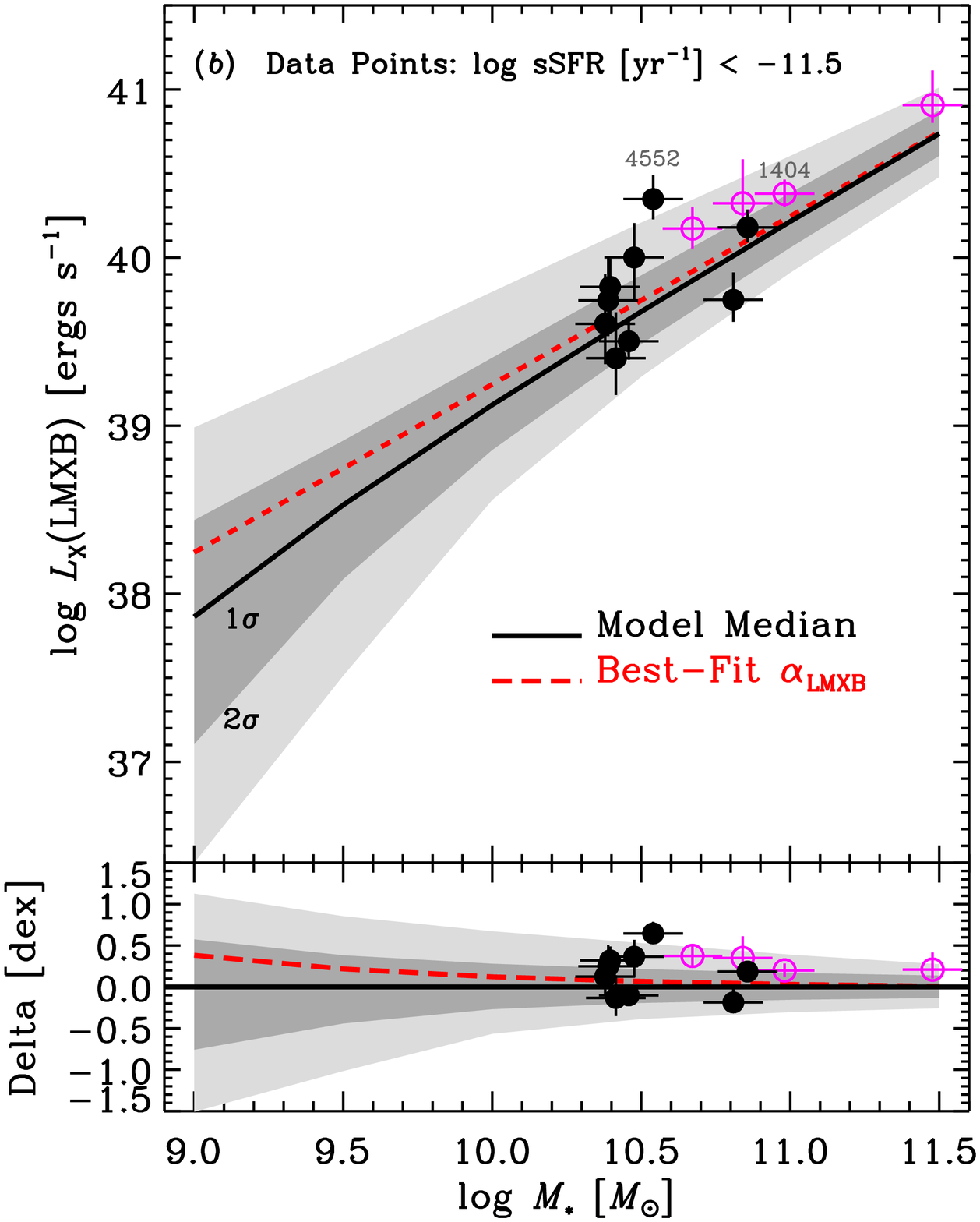}
}
\caption{
%%%
The $L_{\rm X}$(HMXB)--SFR ($a$, {\it top\/}) and $L_{\rm X}$(LMXB)--$M_\star$
($b$, {\it top\/}) relations based on our global model.  The blue dashed and
red long-dashed curves provide scaling-relation predictions based on
integration of the HMXB and LMXB XLFs, respectively: the $\beta_{\rm HMXB}$ and
$\alpha_{\rm LMXB}$ values calculated from our global model (see Table~4).  The
predicted median values are shown as solid curves with 16--84\% (1~$\sigma$; {\it dark gray}) and 2.5--97.5\% (2$\sigma$; {\it light gray}) confidence regions related to statistical
scatter indicated.
These
values were calculated using Monte Carlo simulations of data sets generated by
the global model, and are described in detail in $\S\S$5.2 and 5.3.  For
reference, galaxies in our sample that are predicted to be dominated by HMXBs
($\log$~sSFR/yr$^{-1} > -9.5$) and LMXBs ($\log$~sSFR/yr$^{-1} < -11.5$) are
shown, with known outliers annotated.
%%%
The bottom panels show the log-scale residuals of all quantities with respect
to the median relation, illustrating the level of scatter and relationship
with SFR and $M_\star$.  Note that the deviation of $\beta_{\rm
HMXB}$ and $\alpha_{\rm LMXB}$ with respect to the median grows to larger than
a factor of two for SFR~$\simlt$~2~\sfr\ and $M_\star \simlt
3 \times 10^9$~\msol.
%%%
}
%%%
\end{figure*}
%%%%%%%%%%%%%%%%%%%%%%%%%%%%%%%%%%%%%%%%%%%%%%%%%%%%%%%%%%%%%%%%%%%%%%%%%%%%%%%%%%

%%%%%%%%%%%%%%%%%%%%%%%%%%%%%%%%%%%%%%%%%%%%%%%%%%%%%%%%%%%%%%%%%%%%%%%%%%%%%%%%%%
% Table 6
%%%%%%%%%%%%%%%%%%%%%%%%%%%%%%%%%%%%%%%%%%%%%%%%%%%%%%%%%%%%%%%%%%%%%%%%%%%%%%%%%%
\begin{table*}
{\small
\begin{center}
\caption{Expected Statistical Scatter For Global Model}
\begin{tabular}{lcccccc}
\hline\hline
 & \multicolumn{6}{c}{$\log M_\star/M_\odot$} \\
 & \multicolumn{6}{c}{\rule{3.8in}{0.01in}} \\
 & (9.0) & (9.5) & (10.0) & (10.5) & (11.0) & (11.5) \\
\multicolumn{1}{c}{$\log$~sSFR} & $\log L_{\rm X}$ & $\log L_{\rm X}$ & $\log L_{\rm X}$ & $\log L_{\rm X}$ & $\log L_{\rm X}$ & $\log L_{\rm X}$ \\
\multicolumn{1}{c}{(yr$^{-1}$)} & (\lum) & (\lum) & (\lum) & (\lum) & (\lum) & (\lum) \\
\hline\hline
$-$12.5 & 37.86$_{-0.57}^{+0.76}$ & 38.53$_{-0.38}^{+0.44}$ & 39.13$_{-0.28}^{+0.27}$ & 39.68$_{-0.22}^{+0.19}$ & 40.22$_{-0.17}^{+0.15}$ & 40.74$_{-0.14}^{+0.13}$ \\
$-$12.3 & 37.86$_{-0.58}^{+0.75}$ & 38.53$_{-0.39}^{+0.44}$ & 39.13$_{-0.28}^{+0.27}$ & 39.68$_{-0.22}^{+0.20}$ & 40.22$_{-0.17}^{+0.15}$ & 40.74$_{-0.14}^{+0.13}$ \\
$-$12.1 & 37.86$_{-0.58}^{+0.76}$ & 38.53$_{-0.39}^{+0.44}$ & 39.13$_{-0.28}^{+0.27}$ & 39.69$_{-0.22}^{+0.19}$ & 40.22$_{-0.17}^{+0.16}$ & 40.75$_{-0.14}^{+0.13}$ \\
$-$11.9 & 37.86$_{-0.58}^{+0.75}$ & 38.54$_{-0.39}^{+0.44}$ & 39.13$_{-0.28}^{+0.27}$ & 39.69$_{-0.22}^{+0.19}$ & 40.23$_{-0.17}^{+0.15}$ & 40.75$_{-0.14}^{+0.13}$ \\
$-$11.7 & 37.88$_{-0.57}^{+0.76}$ & 38.54$_{-0.39}^{+0.43}$ & 39.14$_{-0.29}^{+0.27}$ & 39.70$_{-0.22}^{+0.20}$ & 40.24$_{-0.18}^{+0.16}$ & 40.76$_{-0.14}^{+0.13}$ \\
$-$11.4 & 37.88$_{-0.57}^{+0.75}$ & 38.55$_{-0.39}^{+0.43}$ & 39.14$_{-0.29}^{+0.27}$ & 39.71$_{-0.23}^{+0.19}$ & 40.25$_{-0.18}^{+0.16}$ & 40.78$_{-0.14}^{+0.13}$ \\
$-$11.2 & 37.89$_{-0.58}^{+0.75}$ & 38.56$_{-0.39}^{+0.43}$ & 39.16$_{-0.29}^{+0.27}$ & 39.72$_{-0.23}^{+0.19}$ & 40.27$_{-0.19}^{+0.16}$ & 40.80$_{-0.14}^{+0.13}$ \\
$-$11.0 & 37.90$_{-0.57}^{+0.73}$ & 38.58$_{-0.39}^{+0.42}$ & 39.18$_{-0.30}^{+0.26}$ & 39.75$_{-0.24}^{+0.20}$ & 40.30$_{-0.20}^{+0.16}$ & 40.84$_{-0.15}^{+0.14}$ \\
$-$10.8 & 37.93$_{-0.57}^{+0.72}$ & 38.61$_{-0.40}^{+0.41}$ & 39.22$_{-0.31}^{+0.27}$ & 39.79$_{-0.27}^{+0.20}$ & 40.35$_{-0.22}^{+0.17}$ & 40.89$_{-0.15}^{+0.14}$ \\
$-$10.6 & 37.97$_{-0.56}^{+0.68}$ & 38.66$_{-0.41}^{+0.40}$ & 39.27$_{-0.34}^{+0.27}$ & 39.85$_{-0.30}^{+0.21}$ & 40.43$_{-0.23}^{+0.18}$ & 40.97$_{-0.15}^{+0.14}$ \\
$-$10.4 & 38.04$_{-0.56}^{+0.64}$ & 38.72$_{-0.42}^{+0.38}$ & 39.34$_{-0.37}^{+0.27}$ & 39.95$_{-0.33}^{+0.22}$ & 40.53$_{-0.24}^{+0.19}$ & 41.07$_{-0.14}^{+0.15}$ \\
$-$10.2 & 38.14$_{-0.56}^{+0.59}$ & 38.82$_{-0.45}^{+0.37}$ & 39.45$_{-0.41}^{+0.28}$ & 40.07$_{-0.35}^{+0.24}$ & 40.67$_{-0.22}^{+0.20}$ & 41.20$_{-0.14}^{+0.15}$ \\
$-$10.0 & 38.27$_{-0.55}^{+0.53}$ & 38.96$_{-0.49}^{+0.35}$ & 39.61$_{-0.44}^{+0.29}$ & 40.25$_{-0.34}^{+0.26}$ & 40.83$_{-0.20}^{+0.21}$ & 41.35$\pm$0.14 \\
$-$9.8 & 38.44$_{-0.57}^{+0.48}$ & 39.13$_{-0.52}^{+0.35}$ & 39.81$_{-0.46}^{+0.32}$ & 40.45$_{-0.30}^{+0.28}$ & 41.01$_{-0.18}^{+0.20}$ & 41.52$\pm$0.14 \\
$-$9.6 & 38.65$_{-0.60}^{+0.44}$ & 39.35$_{-0.55}^{+0.37}$ & 40.04$_{-0.42}^{+0.33}$ & 40.66$_{-0.25}^{+0.27}$ & 41.20$_{-0.16}^{+0.18}$ & 41.70$_{-0.13}^{+0.14}$ \\
$-$9.3 & 38.89$_{-0.61}^{+0.43}$ & 39.61$_{-0.53}^{+0.38}$ & 40.28$_{-0.37}^{+0.33}$ & 40.87$_{-0.21}^{+0.24}$ & 41.39$_{-0.15}^{+0.17}$ & 41.89$_{-0.13}^{+0.14}$ \\
$-$9.1 & 39.16$_{-0.60}^{+0.42}$ & 39.88$_{-0.48}^{+0.38}$ & 40.53$_{-0.29}^{+0.31}$ & 41.08$_{-0.18}^{+0.21}$ & 41.59$_{-0.14}^{+0.15}$ & 42.09$\pm$0.13 \\
$-$8.9 & 39.45$_{-0.58}^{+0.42}$ & 40.15$_{-0.41}^{+0.37}$ & 40.77$_{-0.23}^{+0.28}$ & 41.29$_{-0.16}^{+0.19}$ & 41.79$_{-0.14}^{+0.15}$ & 42.29$\pm$0.13 \\
$-$8.7 & 39.75$_{-0.52}^{+0.41}$ & 40.42$_{-0.33}^{+0.34}$ & 40.98$_{-0.20}^{+0.23}$ & 41.50$_{-0.15}^{+0.17}$ & 42.00$_{-0.13}^{+0.14}$ & 42.50$\pm$0.13 \\
$-$8.5 & 40.04$_{-0.45}^{+0.39}$ & 40.67$_{-0.26}^{+0.30}$ & 41.20$_{-0.17}^{+0.20}$ & 41.71$_{-0.14}^{+0.15}$ & 42.21$_{-0.13}^{+0.14}$ & 42.71$\pm$0.13 \\
\hline
\end{tabular}
\end{center}
Note.---The expected integrated XRB luminosity for a variety of sSFR (by row) and $M_\star$ (by column) values.  Each quoted $L_{\rm X}$ value represents the median expected from our global model, with error bars representing the 16\% and 84\% confidence values that are expected for a given combination of sSFR and $M_\star$.  These values were obtained using the Monte Carlo simulations discussed in $\S$5.2.\\
}
\end{table*}

To investigate the relative levels that GC LMXBs are likely contributing to the
XLFs in each galaxy, we made use of the Harris \etal\ (2013) catalog of GC
specific frequencies for nearby galaxies.  The specific frequency, $S_N$, for a
given galaxy is defined as:
\begin{equation}
S_N \equiv N_{\rm GC} \times 10^{0.4(M_V^T + 15)},
\end{equation}
where $N_{\rm GC}$ is the number of GCs in the galaxy, and $M_V^T$ is the
galaxy-wide total $V$-band absolute magnitude.  In a broad sense, $S_N$, is a proxy
for the number of GCs per unit mass.  The Harris \etal\ (2013) catalog contains
measurements of the GC populations, including $S_N$, for a comprehensive sample
of 422 nearby galaxies.  We found entries for 12 of the \ngal\ galaxies in our
sample, and we have added the $S_N$ values of these galaxies to Table~1.  Not
surprisingly, measurements were available for the nearest and most massive
galaxies in the sample.  Given general trends of $S_N$ versus $M_\star$, we
would expect that the galaxies with available $S_N$ measurements would be
biased toward high-mass galaxies, which tend to have high-$S_N$.  In Figure~15, we display the distribution of $S_N$ values for the
sample, with the median value of $S_N^{\rm median} = 1.1$ indicated.  Ten out
of the 12 galaxies have $S_N < 2$, while the most significant outlier,
NGC~4552, has an $S_N = 7.7$, far above the next highest $S_N = 2.7$ for
NGC~4594.  

In terms of deviations from the global LMXB XLF, it is interesting to note that
the three galaxies with the highest $S_N$ values, NGC~1404, 4552, and 4594 all
have elevated values of $N_{38}$, with the most extreme galaxy (in $S_N$
terms), NGC~4552, having a statistically significant enhancement of
low-luminosity LMXBs.  Given the known enhancements in LMXB populations
generated by GC LMXBs, the above strongly implicates contributions from GC
LMXBs as being being responsible for the observed excess of LMXBs in NGC~4552
and possibly some of the other galaxies (e.g., NGC~1404 and 4594).  A more
detailed analysis involving direct identification of GC counterparts (see,
e.g., Kim \& Fabbiano~2010; Lehmer \etal\ 2014; Peacock \etal\ 2017) would be
required to quantify the level of influence GCs have on these galaxies.  Such a
paper is the subject of work currently in preparation (Ferrell \etal\ 2019,
in preparation).

\subsection{Characterizing the Statistical Scatter of the Global Model} 

The above analyses indicate that there are several galaxies that show
statistically significant deviations of their XRB populations compared to the
global model predictions; however, these deviations are strongly suggested to
be attributed to unmodeled dependencies in metallicity and GC LMXB population
contributions.  In spite of these examples, the global model provides a good
characterization of the XLFs for the majority of the galaxies in our sample
(see Table~5).  We can therefore use the global model to provide good estimates
of the typical emission, and scatter-related uncertainty, from XRB populations
in galaxies, given their SFR and $M_\star$ values.  However, we note that these
calculations are appropriate for galaxies with metallicities and GC specific
frequencies close to the average values of our sample: $\langle Z \rangle
\approx Z_\odot$ and $\langle S_N \rangle \approx 1.5$, respectively.

As a practical matter, for galaxies that are much more distant than those
studied here, only the integrated $L_{\rm X}$ can be measured.  In this
section, we make use of our global XRB XLF model to predict $L_{\rm X}$ values,
and their potential variations due to scatter, given only SFR and $M_\star$
values.  As discussed at the beginning of $\S$5.2, low-SFR or low-$M_\star$
populations are subject to large variations in measured $L_{\rm X}$ due to
poorly populated HMXB and LMXB XLFs.  For galaxies in these categories, the
average scaling relations, $\alpha_{\rm LMXB} \equiv L_{\rm X}({\rm
LMXB})/M_\star$ and $\beta_{\rm HMXB} \equiv L_{\rm X}({\rm HMXB})$/SFR, are
unlikely to give correct estimates of the integrated XRB population
luminosities, since these are only accurate when the XLFs are fully populated.  

To determine how $L_{\rm X}$ and its scatter would vary with SFR and $M_\star$,
we followed closely the Monte Carlo procedure outlined above in $\S$5.2.  We
first generated a grid of 15 sSFR values covering $\log$~sSFR
(yr$^{-1}$)~=~$-12.5$ to $-8.5$ and six $M_\star$ values ranging from $\log
M_\star (M_\odot) =$~\hbox{9--11.5}.  These ranges cover broader ranges of galaxy
properties than those found in our sample.  For a given pairing of sSFR and
$M_\star$, we ran our Monte Carlo simulation (see $\S$5.2 for details) to
generate simulated HMXB and LMXB source lists down to a luminosity limit of $L =
10^{35}$~\lum.  Here, we did not include completeness functions, as we had done in $\S$5.2
above, since we are interested in the total intrinsic luminosity.  Summing
the luminosities of the populations gives Monte-Carlo-based estimates of
$L_{\rm X}({\rm HMXB})$, $L_{\rm X}({\rm LMXB})$, and $L_{\rm X}$ (i.e., the
sum of HMXBs and LMXBs).  For a given pair of sSFR and $M_\star$, we generated
a total of 1000 $L_{\rm X}({\rm HMXB})$, $L_{\rm X}({\rm LMXB})$, and $L_{\rm
X}$ values each, and constructed probability distribution functions.

In Figures~16$a$ and 16$b$, we display the $L_{\rm X}({\rm HMXB})$~versus~SFR
and $L_{\rm X}({\rm LMXB})$~versus~$M_\star$, respectively, including the
expected median ({\it black solid curves\/}) and scatter (i.e., {\it gray shaded regions\/}) in the
relations, as well as the $\beta_{\rm HMXB}$ and $\alpha_{\rm LMXB}$ scaling
relations for fully populated XLFs.  For comparison, we include the locations
of galaxies that are expected to be HMXB and LMXB dominant, based on having
$\log$~sSFR (yr$^{-1}$)~$> -9.5$ and $\log$~sSFR (yr$^{-1}$)~$< -11.5$,
respectively.  As expected, the scatter and the deviations of the median
$L_{\rm X}$ from the respective relations grow with decreasing SFR or $M_\star$
due to the XLF becoming less populated.  These effects are larger in the
HMXB--SFR scaling than for the LMXB-$M_\star$ scaling, since the relatively
shallow-sloped HMXB XLF leads to large variations in $L_{\rm X}({\rm HMXB})$,
when the XLF is poorly populated.  For HMXBs, the median $L_{\rm X}({\rm
HMXB})$ is lower than that implied by $\beta_{\rm HMXB}$ by more than a factor
of two for SFR~$\simlt 2$~\sfr; all but seven of our galaxies have SFR values
in this range.  While for LMXBs, the median $L_{\rm X}({\rm LMXB})$ is a factor
of two lower than that implied by $\alpha_{\rm LMXB}$, only for galaxies with
$M_\star \simlt 3 \times 10^9$~\msol; only four of our galaxies have stellar
masses in this range.  The scatter itself ranges from $\approx$0.3--0.7~dex for
HMXBs across SFR=~0.1--10~\sfr\ and $\approx$0.2--0.4 for LMXBs across $\log
M_\star (M_\odot) =$~9.5--11.

In Table~6, we tabulate the results of our Monte Carlo simulations.  For a
broad range of sSFR and $M_\star$ combinations, we provide the median (50\%),
16\%, and 84\% confidence ranges for the total $L_{\rm X}$, which contains
contributions from both HMXBs and LMXBs.  In Figure~6, we display the 16\% and
84\% ranges of $L_{\rm X}$/SFR versus sSFR based on these results for the
median stellar mass of our sample $M_\star = 2 \times 10^{10}$~\msol\ ({\it gray
shaded region\/}) and for a low stellar mass bin at $M_\star = 3 \times
10^{9}$~\msol\ ({\it dotted curves}), above which 36 out of the \ngal\ galaxies
in our sample lie.  As we examined in $\S$5.2.1, the most significant outliers,
like NGC~337, 925, 4552, and 4559 are apparent due to their enhanced $L_{\rm
X}$/SFR values over these ranges.  Nevertheless, given values of $M_\star$ and
SFR (and thus sSFR), the tabulated values in Table~6 can be used on a
galaxy-by-galaxy basis to obtain a realistic estimate of the expected XRB
$L_{\rm X}$ and scatter-related uncertainty.  As alluded to throughout all of
$\S$~5, these parameterizations will be improved in the future with studies
of how the XLF varies with additional physical properties,
such as metallicity and $S_N$.

%
%%%%%%%%%%%%%%%%%%%%%%%%%%%%%%%%%%%%%%%%%%%%%%%%%%%%%%%%%%%%%%%%%%%%%%%%
\section{Summary}
%%%%%%%%%%%%%%%%%%%%%%%%%%%%%%%%%%%%%%%%%%%%%%%%%%%%%%%%%%%%%%%%%%%%%%%%
%

In this paper, we have utilized 5.8~Ms of \chandra\ data, combined with
UV--to--IR observations, for \ngal\ nearby ($D \simlt30$~Mpc) galaxies to
revisit scaling relations of the HMXB and LMXB XLFs with SFR and $M_\star$,
respectively.  We make novel use of local environment to isolate XRB
populations in a variety of sSFR bins, which allows us to cleanly determine the
HMXB and LMXB XLF shapes and normalizations.  In addition to providing new
details on XRB XLF scaling relations, which can be applied to a variety of
astrophysical problems, this work presents several new data products and results, which we
summarize below.

\begin{itemize}

\item We present publicly available \chandra\ data products and catalogs, as
well as SFR and $M_\star$ maps for all \ngal\ galaxies in our sample.  These products
are constructed carefully following the procedures detailed in $\S$3.  

\item We report new fits to the XRB XLFs of all \ngal\ galaxies in our sample, including
estimates of CXB sources and the intrinsic source populations.  We explore how
the XLF normalizations, slopes, and calculated XRB luminosities depend on
galaxy SFR and $M_\star$ (see Fig.~5; Table~3).  We find that the XLFs show a clear decline in
normalization per unit SFR and a decrease in the $L > 10^{38}$~\lum\ XLF slope
with increasing sSFR (i.e., SFR/$M_\star$), as the dominant XRB population
shifts from LMXBs to HMXBs.  As a corollary, the integrated XRB luminosity,
$L_{\rm X}$, per unit SFR declines with increasing sSFR (see Fig.~6).

\item When analyzing XRB XLFs from subgalactic regions, selected in bins of
sSFR, we clearly see the transition in XLF shape and normalization per SFR from
the almost ``pure'' HMXB XLF at sSFR~$\approx 5 \times 10^{-10}$~yr$^{-1}$ to
the nearly pure LMXB XLF at sSFR~$\approx 10^{-12}$~yr$^{-1}$ (see Fig.~7).  We
present a global model that characterizes the scaling of the HMXB XLF with SFR
and LMXB XLF with $M_\star$ that describes well the data for all \ngal\
galaxies (model curves in Figs.~7 and 9 and Table~4).  The parameters of these models and uncertainties are determined using an MCMC procedure and are reported (see Fig.~8 and Table~4).

\item We find basic agreement between the HMXB XLF shape and scaling with SFR,
as presented in past papers (e.g., M12); however, our HMXB XLF reveals new
complex features, beyond the previously reported power-law shape (see Fig.~12$a$)  These
features include a steep power-law slope between $L \approx
10^{36}$--$10^{38}$~\lum, a ``bump'' or ``flattening'' between $L \approx
10^{38}$--$10^{40}$~\lum, and rapid fall off at higher luminosities.  These
features are highly significant and are robustly identified in independent
subsets of our data.  Similar features have been reported in some XRB
population synthesis models of the HMXB XLF.

\item We further find qualitatively good agreement between our LMXB XLF with
the previously-reported LMXB XLF from Z12, which was based on elliptical
galaxies.  However, our fits to the data, which is mainly driven by late-type
galaxies, prefer a somewhat shallower slope at $L \simgt 10^{39}$~\lum\ and a
steeper slope at $L \simlt 10^{38}$~\lum.  We further find evidence that the
LMXB XLF in higher-sSFR subsets is shallower at $L \simgt 10^{39}$~\lum\ and
steeper at $L \simlt 10^{38}$~\lum\ compared with our total-sample average (see
Fig.~12$b$).  We speculate that this is plausibly due to a stellar age effect,
in which the LMXB XLF is dominated by older stellar populations at low-sSFR
compared to the high-sSFR.  This would imply that, compared to older LMXB XLFs,
the LMXB XLF for younger populations contains excesses of LMXBs at all
luminosities except $L \approx$~$10^{38}$--$10^{39}$~\lum.  Some features of
this trend (e.g., more high-$L$ sources) have been predicted in population
synthesis models.

\item We use our global model and Monte Carlo simulations to identify galaxies
that have outlier XLF populations that are statistically significant.  We
identify four such galaxies: NGC~337, 925, 4552 (M89), and 4559.  Scrutiny of these
objects indicates that NGC~337, 925, and 4559 are among the lowest metallicity objects
in our sample, and NGC~4552 contains a significant excess of GCs per unit
optical luminosity (i.e., specific frequency) over all other galaxies in our
sample ($\S$5.2).  

\item To examine the effects of metallicity on the XLFs, we constructed the XLF
for the lowest metallicity galaxies in our sample (NGC~337, 925, 3198, 4536, and
4559).  We find statistically significant evidence that the HMXB XLF in
low-metallicity ($\approx$0.5$Z_\odot$) galaxies contains an excess of $L
\simgt 10^{39}$~\lum\ sources, but comparable numbers of $\simlt 10^{39}$~\lum\
sources, compared to the global average HMXB XLF for our sample, which has a
median metallicity $\approx$$Z_\odot$ (see Fig.~14).  This result is in line
with other studies that characterize how the integrated \xray\ luminosity per
SFR is anticorrelated with metallicity (e.g., Basu-Zych \etal\ 2016; Brorby
\etal\ 2016).  Our result provides a first characterization of the
$\approx$0.5~$Z_\odot$ HMXB XLF from $\log L$~(ergs~s$^{-1}$)~=~37--41.

\item We conclude that our global model is appropriate for galaxies that are of
roughly solar metallicity and have low GC specific frequencies.  Finally, with
this caveat, we use the global model, along with Monte Carlo simulations to
calculate the scatter in the integrated \xray\ luminosities of HMXB and LMXB
populations as a function of SFR and $M_\star$.  Such a quantity is useful, for
example, for \xray\ data sets that detect only the total \xray\ emission from
the galaxy without resolving the XRB populations.  We show that the median HMXB
and LMXB integrated luminosities deviates substantially (by more than a factor
of two) from the XLF-integrated average scaling relations, $L_{\rm X}$(HMXB)/SFR and $L_{\rm X}$(LMXB)/$M_\star$, at SFR~$\simlt 2$~\sfr\ and $M_\star \simlt 3 \times 10^9$~\msol,
respectively (see Figure~16).  The corresponding 16--84\% scatter ranges from
$\approx$0.3--0.7~dex for HMXBs across SFR=~0.1--10~\sfr\ and $\approx$0.2--0.4
for LMXBs across $\log M_\star (M_\odot) =$~9.5--11.  Characterization of the XRB scatter is provided in Table~6.

\item Future investigations are underway to quantitatively assess how
metallicity, stellar age, and GC specific frequency affect the XRB XLFs.  These
studies will provide expansive new constraints on close-binary population
synthesis models that are used to understand a variety of close-binary
populations (e.g., XRBs, gravitational-wave sources, and millisecond pulsars),
and the role of XRBs in environments that are not-yet observable (e.g., during
the epoch of heating when HMXBs are thought to dominate the \xray\ emissivity
of the Universe).

\end{itemize}

\acknowledgements We thank the anonymous referee for their helpful
suggestions, which have improved the quality of this paper.  We gratefully
acknowledge support from the National Aeronautics and Space Administration
(NASA) Astrophysics Data Analysis Program (ADAP) grant NNX13AI48G (B.D.L.,
R.T.E., A.Z.) and \chandra\ X-ray Center grant GO8-19039X (B.D.L. and A.P.).
A.Z. acknowledges funding from the European Union's Seventh Framework Programme
(FP/2007--2013)/ERC Grant Agreement n.~617001. 

Our work includes observations made with the NASA {\it Galaxy Evolution
Explorer} (\galex). \galex\ is operated for NASA by the California Institute of
Technology under NASA contract NAS5-98034. This publication makes use of data
products from the Two Micron All Sky Survey, which is a joint project of the
University of Massachusetts and the Infrared Processing and Analysis
Center/California Institute of Technology, funded by NASA and the National
Science Foundation (NSF). This work is based on observations made with the
{\it Spitzer Space Telescope}, obtained from the NASA/IPAC Infrared Science Archive,
both of which are operated by the Jet Propulsion Laboratory, California
Institute of Technology under a contract with the National Aeronautics and
Space Administration.  

{\it Facilities:} {\it Chandra}, {\it GALEX}, Sloan, 2MASS, {\it Spitzer}, {\it Herschel}

\software{ACIS Extract (v2016sep22; Broos et al. 2010, 2012), MARX (v5.3.2; Davis et al. 2012), CIAO (v4.8; Fruscione~et al. 2006), xspec (v12.9.1; Arnaud~1996)}

%
%%%%%%%%%%%%%%%%%%%%%%%%%%%%%%%%%%%%%%%%%%%%%%%%%%%%%%%%%%%%%%%%%%%%%%%%

%%%%%%%%%%%%%%%%%%%%%%%%%%%%%%%%%%%%%%%%%%%%%%%%%%%%%%%%%%%%%%%%%%%%%%%%%%%%%%%%%%
%

\appendix

\section{X-ray Point Source Catalog}

In Table~A1, we provide the \xray\ point source catalogs, based on the analyses
presented in $\S\S$3.2 and 3.3.  The columns include the following: Col.(1):
Name of the host galaxy. Col.(2): point-source identification number within the
galaxy. Col.(3) and (4): Right ascension and declination of the point source.
Col.(5): Offset of the point source with respect to the average aim point of
the \chandra\ observations. Col.(6) and (7): 0.5--7~keV net counts (i.e.,
background subtracted) and 1$\sigma$ errors. Col.(8)--(9) and (10)--(11):
Best-fit column density $N_{\rm H}$ and photon index $\Gamma$, respectively,
along with their respective 1$\sigma$ errors, based on spectral fits to an
absorbed power-law model ({\ttfamily TBABS $\times$ POW} in {\ttfamily xspec}).
For sources with small numbers of counts ($<$20 net counts), we adopted
Galactic absorption appropriate for each galaxy and a photon index of $\Gamma =
1.7$.  Col.(12) and (13): the respective 0.5--8~keV flux and luminosity of the
source. Col.(14): Flag indicating the location of the source within the galaxy.
Flag=1 indicates the source is within the $K_s$-band footprint adopted in
Table~1, and outside a central region of avoidance, if applicable.  All XLF
calculations are based on Flag=1 sources.  Flag=2 indicates that the source is
within the $K_s$-band footprint, but has a luminosity of $L < 10^{35}$~\lum,
and was thus excluded from our XLF analysis.  Flag=3 indicates that the source
is outside the 20~mag~arcsec$^{-2}$ $K_s$-band ellipse of the galaxy, but
within the ``total'' $K_s$-band ellipse.  Flag=4 indicates that the source is
located in the central region of avoidance due to either the presence of an AGN
or very high levels of source confusion. Flag=5 indicates that the source is
outside the ``total'' $K_s$-band ellipse.

%%%%%%%%%%%%%%%%%%%%%%%%%%%%%%%%%%%%%%%%%%%%%%%%%%%%%%%%%%%%%%%%%%%%%%%%%%%%%%%%%%
% Table A1
%%%%%%%%%%%%%%%%%%%%%%%%%%%%%%%%%%%%%%%%%%%%%%%%%%%%%%%%%%%%%%%%%%%%%%%%%%%%%%%%%%
\begin{table*}
\renewcommand\thetable{A1}
{\footnotesize
\begin{center}
\caption{X-ray point-source catalog and properties}
\begin{tabular}{llccccccccc}
\hline\hline
 &  &  $\alpha_{\rm J2000}$ & $\delta_{\rm J2000}$ & $\theta$ & $N_{\rm FB}$ & $N_{\rm H}$  & & $\log F_{\rm FB}$ & $\log L_{\rm FB}$  & Location \\
 \multicolumn{1}{c}{\sc Galaxy} & \multicolumn{1}{c}{\sc ID} &  (deg) & (deg) & (arcmin) & (counts) & ($10^{22}$~cm$^{-2}$) & $\Gamma$ & (\flux) & (\lum)  & Flag \\
 \multicolumn{1}{c}{(1)} & \multicolumn{1}{c}{(2)} & (3) & (4) & (5) & (6)--(7) & (8)--(9) & (10)--(11)  & (12) & (13) & (14) \\
\hline\hline
NGC337 & 1 & 00 59 43.53 & $-$07 35 01.33 & 1.7 & 7.8$\pm$4.2 & 0.056 & 1.7 &  $-$14.1 & 38.6& 4 \\
  & 2 & 00 59 47.50 & $-$07 34 16.68 & 0.8 & 41.0$\pm$7.9 & 0.109$\pm$0.154 & $<$3.06 &  $-$13.7 & 39.1& 2 \\
  & 3 & 00 59 48.51 & $-$07 34 56.71 & 0.5 & 65.3$\pm$9.7 & 0.314$\pm$0.381 & 1.98$\pm$0.74 &  $-$13.2 & 39.5& 1 \\
  & 4 & 00 59 49.48 & $-$07 34 35.66 & 0.2 & 106.8$\pm$12.1 & 0.308$\pm$0.326 & 1.60$\pm$0.53 &  $-$13.0 & 39.8& 1 \\
  & 5 & 00 59 49.49 & $-$07 35 23.53 & 0.7 & 22.3$\pm$6.2 & 0.779$\pm$0.410 & $<$3.06 &  $-$13.8 & 38.9& 2 \\
\\
  & 6 & 00 59 50.40 & $-$07 34 45.67 & 0.1 & 4.7$\pm$2.2 & 0.056 & 1.7 &  $-$14.3 & 38.5& 1 \\
  & 7 & 00 59 50.40 & $-$07 34 54.18 & 0.2 & 42.8$\pm$8.1 & 0.647$\pm$0.796 & 1.57$\pm$0.90 &  $-$13.3 & 39.5& 1 \\
  & 8 & 00 59 50.56 & $-$07 34 58.08 & 0.3 & 300.5$\pm$19.5 & 0.136$\pm$0.155 & 1.40$\pm$0.29 &  $-$12.5 & 40.3& 1 \\
  & 9 & 00 59 51.90 & $-$07 34 57.71 & 0.5 & 14.4$\pm$5.2 & 0.056 & 1.7 &  $-$13.8 & 39.0& 1 \\
  & 10 & 00 59 52.29 & $-$07 34 47.38 & 0.6 & 43.3$\pm$8.1 & 0.405$\pm$0.510 & 2.17$\pm$0.97 &  $-$13.4 & 39.3& 2 \\
\\
  & 11 & 00 59 53.31 & $-$07 34 56.49 & 0.8 & 4.9$\pm$2.2 & 0.056 & 1.7 &  $-$14.3 & 38.5& 2 \\
  & 12 & 00 59 53.32 & $-$07 35 20.76 & 1.0 & 27.3$\pm$6.7 & 0.477$\pm$0.787 & 1.67$\pm$1.12 &  $-$13.5 & 39.2& 2 \\
NGC584 & 1 & 01 31 09.45 & $-$06 54 34.08 & 3.7 & 12.1$\pm$4.9 & 0.036 & 1.7 &  $-$13.8 & 38.9& 4 \\
  & 2 & 01 31 17.83 & $-$06 54 34.75 & 2.6 & 5.9$\pm$2.4 & 0.036 & 1.7 &  $-$14.1 & 38.6& 4 \\
  & 3 & 01 31 18.02 & $-$06 51 48.29 & 0.7 & 8.8$\pm$4.4 & 0.036 & 1.7 &  $-$13.8 & 38.9& 1 \\
\\
  & 4 & 01 31 18.73 & $-$06 52 06.49 & 0.5 & 1.9$\pm$1.4 & 0.036 & 1.7 &  $-$14.5 & 38.1& 1 \\
  & 5 & 01 31 19.28 & $-$06 51 50.26 & 0.4 & 3.9$\pm$2.0 & 0.036 & 1.7 &  $-$14.2 & 38.4& 1 \\
  & 6 & 01 31 19.54 & $-$06 52 03.87 & 0.3 & 2.9$\pm$1.7 & 0.036 & 1.7 &  $-$14.4 & 38.3& 1 \\
  & 7 & 01 31 20.00 & $-$06 52 07.07 & 0.2 & 19.6$\pm$5.9 & 0.036 & 1.7 &  $-$13.6 & 39.1& 1 \\
  & 8 & 01 31 20.14 & $-$06 51 41.03 & 0.4 & 3.9$\pm$2.0 & 0.036 & 1.7 &  $-$13.9 & 38.8& 1 \\
\\
\hline
\end{tabular}
\end{center}
Note.---The full version of this table contains \nx~sources.  An abbreviated version of the table is displayed here to illustrate its form and content.  A description of the columns is provided in the Appendix.\\
}
\end{table*}

\end{document}